\documentclass[aps,prx,showpacs,twocolumn,reprint,superscriptaddress, floatfix]{revtex4-2}

\usepackage{qcircuit}
\usepackage[dvips]{graphicx}
\usepackage{amsmath,amssymb,amsthm,mathrsfs,amsfonts,dsfont,mathtools,physics}
\usepackage{epsfig}
\usepackage{hyperref}
\usepackage{braket}
\usepackage{bm}
\usepackage{enumerate}
\usepackage{color}
\usepackage{graphicx}
\usepackage{textcomp}
\usepackage{floatrow}    
\usepackage[caption=false, position=top]{subfig}
\usepackage{ragged2e}
\DeclareCaptionJustification{justified}{\justifying}
\usepackage[version=4]{mhchem}
\usepackage[T1]{fontenc}

\usepackage{enumitem}
\usepackage[capitalize]{cleveref}
\usepackage[normalem]{ulem}
\usepackage{comment}
\usepackage{xcolor}
\hypersetup{
    colorlinks,
    linkcolor={blue!50!black},
    citecolor={blue!50!black},
    urlcolor={blue!80!black}
}

\usepackage{amsthm}

\Crefname{theorem}{Theorem}{Theorems}
\theoremstyle{remark}

\newcommand{\qmaddress}{\affiliation{Quantum Motion, 9 Sterling Way, London N7 9HJ, United Kingdom}}
\newcommand{\oxaddress}{\affiliation{Department of Materials, University of Oxford, Parks Road, Oxford OX1 3PH, United Kingdom}}
\newcommand{\ucladdress}{\affiliation{Dept of Physics \& Astronomy and London Centre for Nanotechnology, University College London, London WC1E 6BT, United Kingdom}}

\begin{document}

\title{Robustness of electron charge shuttling:\\ Architectures, pulses, charge defects and noise thresholds}
\author{Minjun Jeon}
\email[minjun.jeon@materials.ox.ac.uk]{}
\qmaddress
\oxaddress

\author{Simon C. Benjamin}
\qmaddress
\oxaddress

\author{Andrew J. Fisher}
\qmaddress
\ucladdress

\begin{abstract}
    In semiconductor-based quantum technologies, the capability to shuttle charges between components is profoundly enabling. We numerically simulated various ``conveyor-belt'' shuttling scenarios for simple \ce{Si}/\ce{SiO2} devices, explicitly modelling the electron's wave function using grid-based split-operator methods and a time-dependent 2D potential (obtained from a Poisson solver). This allowed us to fully characterise the electron loss probability and excitation fraction. Remarkably, with as few as three independent electrodes the process can remain near-perfectly adiabatic even in the presence of pulse imperfection, nearby charge defects, and Johnson-Nyquist noise. Only a substantial density of charge defects, or defects at `adversarial' locations, can catastrophically disrupt the charge shuttling. While we do not explicitly model the spin or valley degrees of freedom, our results from this charge propagation study support the conclusion that conveyor-belt shuttling is an excellent candidate for providing connectivity in semiconductor quantum devices.
\end{abstract}

\maketitle

\section{Introduction}

Semiconductor based quantum computers have received attention due to the long spin coherence times\cite{Tyryshkin_2011, Zwanenburg_2013}, high-fidelity single-qubit\cite{yoneda_2018, Veldhorst_2014, Laucht_2015, Muhonen_2015, Xue_2019} and two-qubit gates\cite{Zajac_2018, Xue_2019, Harvey_Collard_2022, Huang2019-lj, Watson_2018} above the quantum error correction (QEC) thresholds\cite{xue_2022, Noiri_2022} and the ability to mass-produce them using conventional semiconductor process technologies\cite{Zwanenburg_2013, Vandersypen_2017, burkard_2023}. Numerous device architectures\cite{Taylor2005-om, Li_2018, buonacorsi_2019, Boter_2022, Jnane_2022, patomaki2023pipeline, sato2024generating} and fault-tolerant schemes\cite{Veldhorst_2014, xue_2022, crawford2022compilation, siegel2024early} for semiconductor based quantum computers have been proposed, and many of them\cite{Li_2018, buonacorsi_2019, Boter_2022, Jnane_2022, patomaki2023pipeline} require mid-range interactions between distant qubits with the distance between them of, say, $10$\,$\mu$m\cite{langrock_2023}. This need arises partly because of the signal fan-out problem\cite{Vandersypen_2017, langrock_2023}: it is otherwise challenging to connect enough control wires to a dense array of quantum dots (QD) to control them individually. However, it is also beneficial to have high qubit connectivity to reduce the number of gates for a two-qubit operation between two arbitrary qubits and hence reduce the overall circuit depth\cite{brierley2016efficient, crawford2022compilation}. The range of the direct exchange interaction between localised spin qubits is too short to provide coupling at relevant distances\cite{Trifunovic_2012, langrock_2023}.

One way to connect two distant qubits is to use an electromagnetic cavity to enable an interaction via microwave photons\cite{Beaudoin_2016, Mi_2017, Stockklauser_2017, Mi_2018, Landig_2018, Samkharadze_2018, Borjans_2019, Borjans_2020, Benito_2019, Warren_2021, Young_2022, Harvey_Collard_2022}. Coupling of two spin qubits via a cavity has been experimentally demonstrated\cite{Samkharadze_2018, Mi_2018, Borjans_2019, Harvey_Collard_2022} and  simulations\cite{Benito_2019, Warren_2021, Young_2022} show that high-fidelity two-qubit gates can in principle be performed via a cavity. There are significant remaining challenges: engineering a sufficiently strong spin-photon coupling and tuning the spin qubits to the resonance frequency of the cavity~\cite{Borjans_2019}; ensuring that the link is usefully fast~\cite{Jnane_2022}; understanding how to integrate such a cavity with conventional semiconductor fabrication technologies\cite{langrock_2023}. Nevertheless, this is a promising option for long range links between spins, even potentially bridging between chips, but for ranges of order 10\,$\mu$m (say) it may not be the natural solution.

An alternative approach involves shuttling: physically moving the electron qubit from one place to another\cite{Hermelin_2011, Bertrand_2016, Fujita_2017, Mills_2019, Sigillito_2019, Buonacorsi_2020_simulated_shuttling, Yoneda_2021, Seidler_2022, Young_2022, langrock_2023, Zwerver_2023, Xue_2024, kunne2023spinbus}. One advantage of this approach is that high-fidelity two-qubit gates can be performed by local exchange interactions. Shuttling can be performed by using various physical means, and some interesting experimental results have been reported, e.g. by using acoustic wave in \ce{GaAs}/\ce{(Ga, As)Al} devices\,\cite{Hermelin_2011, McNeil2011-cj, Bertrand_2016, Fujita_2017} and  by sequentially applying SWAP gates on spins\cite{Sigillito_2019, Young_2022}. One of the most well-known approaches is by modulating voltages in a series of metal gates. In this context, the two most widely studied methods are conveyor-belt shuttling\cite{Taylor2005-om, Seidler_2022,langrock_2023} and bucket-brigade shuttling (see e.g. \cite{Mills_2019, Zwerver_2023, foster2024dephasingerrordynamicsaffecting}). The two are compared by Langrock et al. in Ref.~\cite{langrock_2023}, and the authors conclude that the conveyor-belt mode is superior. This approach transports the electron in a single moving quantum dot, while bucket-brigade shuttling moves the electron by a series of adiabatic Landau-Zener transitions (LZT) through an array of tunnel-coupled quantum dots. Both conveyor-belt and bucket-brigade shuttling require only a fixed number of voltage signals and solve the signal fan-out problem\cite{langrock_2023}. However, bucket-brigade shuttling suffers from two significant difficulties:  the voltages must be fine-tuned to trigger an adiabatic LZT in a specific QD and the shuttling direction will be reversed if only a single diabatic LZT occurs\cite{langrock_2023}. Conveyor-belt shuttling does not suffer from these drawbacks. It has been demonstrated as a proof-of-principle by Seidler et al\cite{Seidler_2022} for a distance of $420$\,nm. According to their failure analysis due to, for example, potential disorder caused by charge defects, they showed that the success probability of shuttling back and forth for $420$\,nm was $99.4\%$. They also found that using a signal amplitude above a certain threshold results in better success probability. Xue et al.\cite{Xue_2024} further pushed the success probability of shuttling to $99.7\%$ for the distance of $19$\,$\mu$m (in a back-and-forth trip). In the analysis by Langrock et al.\cite{langrock_2023} an exploration of the dephasing mechanisms for spin qubits including the effects of, e.g., potential disorder and atomic scale interface roughness, concluded that an optimal speed of $10$\,m/s for a \ce{Si}/\ce{SiGe} device. Losert et al.\cite{losert2024strategies} modelled the effect of \ce{Si}/\ce{SiGe} alloy disorder on the spin of shuttled electron through the valley degree of freedom and found that it is beneficial to have slower speeds and an elongated quantum dot near the valley splitting minimum. Coherent conveyor-belt shuttling of an electron spin was achieved by Struck et al.\,\cite{Struck2024-un} with estimated infidelity of only $0.7$\,\% for a distance of $560$\,nm. The authors characterised the shuttling process by separating and reuniting an Einstein-Podolsky-Rosen (EPR) spin-pair. Recently, De Smet et al.\cite{desmet2024highfidelity} made a comparative study of the bucket-brigade and conveyor-belt shuttling. They found that conveyor-belt shuttling allows faster transport of spin with higher fidelity than the bucket-brigade shuttling. Furthermore, they developed a novel pulse sequence, named two-tone conveyor, which further improves the escape probability with faster transport and higher fidelity, using $8$ gates instead of $4$ gates per unit cell.

While there are extensive experimental and theoretical studies on shuttling in \ce{Si}/\ce{SiGe} devices, shuttling in SiMOS devices hasn't been experimentally demonstrated and hasn't been thoroughly studied in the past. Nevertheless, SiMOS has a number of advantages compared to Si/Ge. A SiMOS device has higher valley splitting than a \ce{Si}/\ce{SiGe} device because the distance from the gate to the active silicon channel is smaller than Si/Ge, inducing stronger electric field and smaller dot lateral dot size\cite{saraiva2021materialssiliconquantumdots, Cifuentes2024}. This results in effectively sharper interface, breaking the Dresselhaus symmetry and creating an in-plane spin-orbit interaction, which is necessary to control spins electrically\cite{saraiva2021materialssiliconquantumdots, Ruskov_2018}. However, a SiMOS device may be more subject to charge defects than a Si/Ge device because \ce{SiO_{2}} is amorphous and its lattice matching to \ce{Si} is worse than that of \ce{SiGe}\cite{saraiva2021materialssiliconquantumdots}.

As this brief survey of the literature indicates, conveyor-belt shuttling (also called QuBus) is a prime prospect for enabling mid-range connectivity in silicon spin quantum computers. In this paper, we use desktop-scale and HPC computational resources to directly simulate quantum dynamics of an electron wave function in various conveyor-belt shuttling scenarios in a simplified SiMOS device. We calculate the probabilities of loss and excitation, thus determining the reliability and adiabaticity  of the process. Our intent is both to validate earlier studies via our independent approach, and to explore aspects of the process that have not previously been modelled explicitly.  We focus on the dynamics of the electron charge, rather than the spin qubit itself; in a final section we discuss the implications of the former for the latter.

First, in section \ref{sec: noiseless_shuttling}, we compare the results of noise-free shuttling scenarios with different parameters (target distance, target speed, number of electrodes in a unit cell). Second, in section \ref{sec:johnson-nyquist}, we simulate noisy shuttling cases in the presence of the Johnson-Nyquist noise in the gates.  We find significant differences from the inclusion of the quantum correction factor, and show that high-frequency noise is particularly detrimental to the quality of shuttling. In section \ref{sec:charge_defects}, we simulate shuttling in the presence of up to three charge defects due to impurities. In the adversarial case of three trapped charges near the interface forming a repulsive potential wall completely delocalized the electron wave function, making the shuttling impossible. Finally, in section \ref{sec:snap}, we propose a new ultra-fast non-adiabatic shuttling method: by a series of sudden changes in the gate potentials in the shuttling direction, we force the electron repeatedly to run down one side of the potential well and climb up the other side. Through these various analyses we conclude that the conveyor-belt shuttling process is remarkably robust and remains near-adiabatic over a broad range of scenarios (while we do find limits beyond which this fails).

\section{\label{sec:shuttling_device}Description of the Shuttling Device}

\begin{figure}
    \centering
    \includegraphics[width=\textwidth]{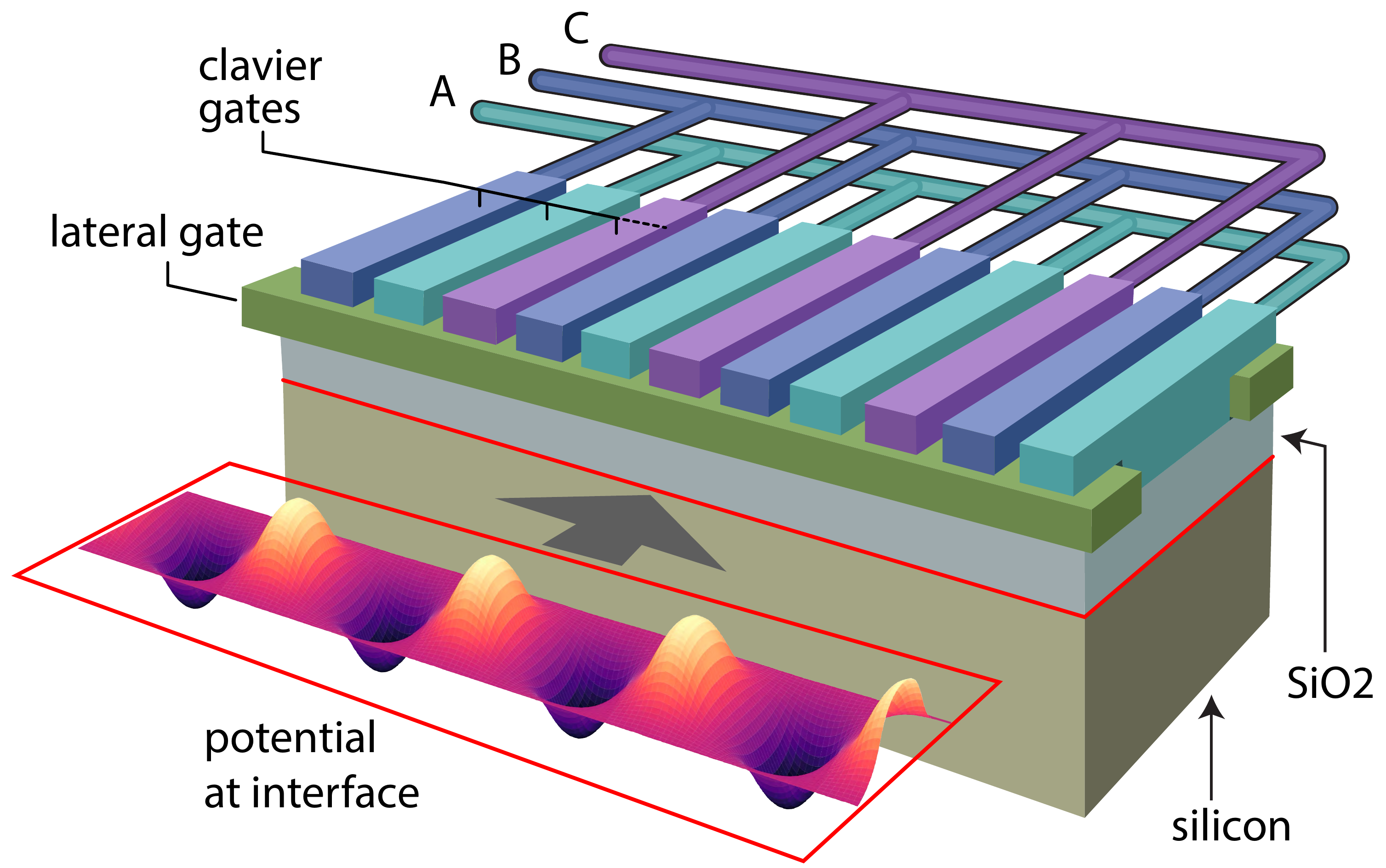}
    \caption{Conceptual illustration of the conveyor-belt shuttling device studied in this paper.  The electrons are confined near a silicon-oxide interface below a periodic array of gates to which external potentials can be applied via voltage lines A, B and C. In this paper we consider devices with three repeating electrodes, as depicted, as well as four\,\cite{langrock_2023} and five.}
    \label{fig:device_layout}
\end{figure}

\begin{figure}
\centering
\subfloat[]{%
	\includegraphics[width=\linewidth, trim={1cm 0.39cm 0 0.6cm},clip]{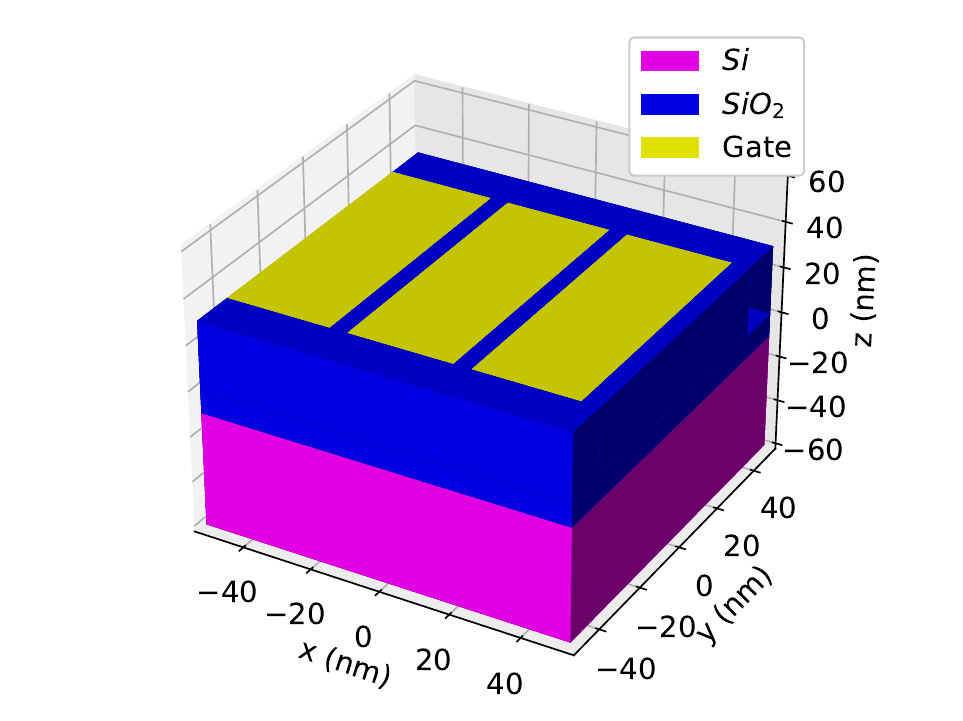}\label{fig:device_geometry_3D}
}%

\hfill

\subfloat[]{%
	\includegraphics[width=\linewidth,trim={0.25cm 0 0 1.25cm}, clip]{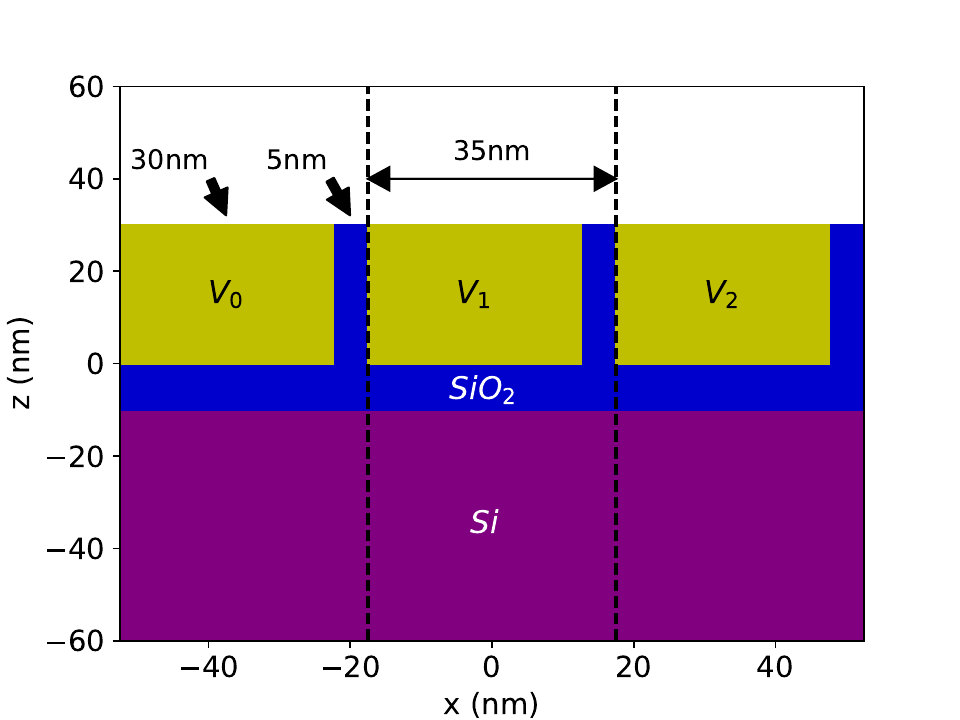}\label{fig:device_geometry_cross_section}
}%

\caption{Illustration of (a) the shuttling device in 3D and (b) its cross-section in the xz plane. The yellow boxes are clavier gates with width and height of $30$\,nm along the x and z direction and the length of $75$\,nm along the y direction. The gap between the gates is $5$\,nm. The size of this unit cell with three gates is $105$\,nm along the shuttling direction, i.e the x-direction. The clavier gates are implanted in the oxide layer (blue). Below the oxide layer, there is a \ce{Si} layer (magenta), whose base is grounded.}
\label{fig:device_geometry}
\end{figure}

Shuttling of an electron is achieved by creating a moving QD using a set of voltage pulses applied to metallic gates. Here, we explore shuttling devices broadly consistent with the Spin Qubit Shuttle (SQS) proposed by Langrock et al.\cite{langrock_2023}. A significant distinction is that while Langrock et al. focused on a silicon/silion-germanium device, here we consider devices consistent with silicon/silicon-oxide structures. Figure~\ref{fig:device_layout} shows a sketch of the device; Figure~\ref{fig:device_geometry} indicates the key dimensions. At the top surface, a periodic array of so-called clavier gates is deposited in the shuttling direction. The clavier gates are embedded inside the \ce{SiO2} layer. The electron moves in a channel, which is located near the interface of \ce{Si} and \ce{SiO2}. To confine the electron to move only in one direction, two lateral confinement gates are deposited just below the clavier gates and a large negative voltage (typically around $-1$\,V) is applied to these. Finally, the bottom of the device is grounded, i.e. $0$\,V. The voltages on the clavier gates form a periodic array of quantum dots along the channel; a single electron is initially loaded from a single-electron transistor (SET) into the leftmost dot, then shuttled along the channel by varying the clavier gate voltages until it reaches a second SET at the right-hand end. 

We made a number of simplifications to model this device. First, we assumed that the confinement gates have a sufficiently strong negative voltage that they act like hard walls at the sides of the channel. Second, since the electron moves below the confinement gates, we assumed that the effective potential it experiences is formed only by those parts of the clavier gates that are not screened by the confinement gates. Third, we assume there are infinitely many clavier gates lined up in a row in the shuttling direction. Finally, we assume that the confinement in the z-direction is very strong, and the quantum dots are formed nearly at the interface of the \ce{Si} and \ce{SiO2}\cite{burkard_2023}.

Given these assumptions, our device model is illustrated (for the case of three independent electrode voltages) in Figure~\ref{fig:device_geometry}. The yellow boxes denote the clavier gates, the blue area denotes \ce{SiO2}, and the magenta box represents the \ce{Si}. Figure~\ref{fig:device_geometry_3D} shows a 3D illustration of the full device while Figure~\ref{fig:device_geometry_cross_section} shows a cross-section of the device through the centre of the channel. The dimensions of the clavier gates are $(h,w,l)=30$\,nm $\times 30$\,nm $\times 75$\,nm , while the gap between the electrodes is fixed to $5$\,nm. Furthermore, the interface between the \ce{Si} and \ce{SiO2} is $10$\,nm below the bottom of the clavier gates as shown in Figure~\ref{fig:device_geometry_cross_section}.

The Hamiltonian of the electron in 3D is given by 

\begin{equation}\label{eqn:time_dependent_hamiltonian_noise_free_3D}
    H = \frac{1}{2} \mathbf{p}^{T}M^{-1}\mathbf{p} - e\Phi(V_{0}(t), V_{2}(t), ..., V_{N-1}(t)),
\end{equation}where $M$ is the anisotropic mass tensor in silicon, and $N$ is the number of gates per unit cell. Note that the potential, $\Phi$, is a function of time-dependent gate voltages. 

In this paper, we ran simulations in 2D, using only the transverse electron mass, i.e. $m^{*} = 0.19m_{e}$ (See section \ref{sec:numerical_simulation} and appendix \ref{appendix:subsec:airy_function}). This reduces the Hamiltonian to

\begin{equation}\label{eqn:time_dependent_hamiltonian_noise_free_2D}
    H = \frac{1}{2m^{*}} \mathbf{p}^{2} - e\Phi_{Si/SiO_{2}}(V_{0}(t), V_{2}(t), ..., V_{N-1}(t)),
\end{equation} where we sample the 2D potential $\Phi_{Si/SiO_{2}}$ from the 3D potential, $\Phi$, at the \ce{Si}--\ce{SiO2} interface (See section \ref{sec:numerical_simulation}). Aside from our investigation of charge defects in section \ref{sec:charge_defects}, the form of Hamiltonian in equation \ref{eqn:time_dependent_hamiltonian_noise_free_2D} doesn't change; However, the time-dependent gate voltages change due to, e.g., the Johnson-Nyquist noise in section \ref{sec:johnson-nyquist} or different pulse shapes in section \ref{sec:snap}.

\section{Voltage profiles for conveyor-belt Shuttling}\label{sec:conveyor_belt_shuttling}
\begin{figure*}
	\begin{minipage}[t]{0.495\columnwidth}
  	\centering
        \subfloat[]{%
	\includegraphics[trim={0 4.0cm 0 0}, width=\linewidth]{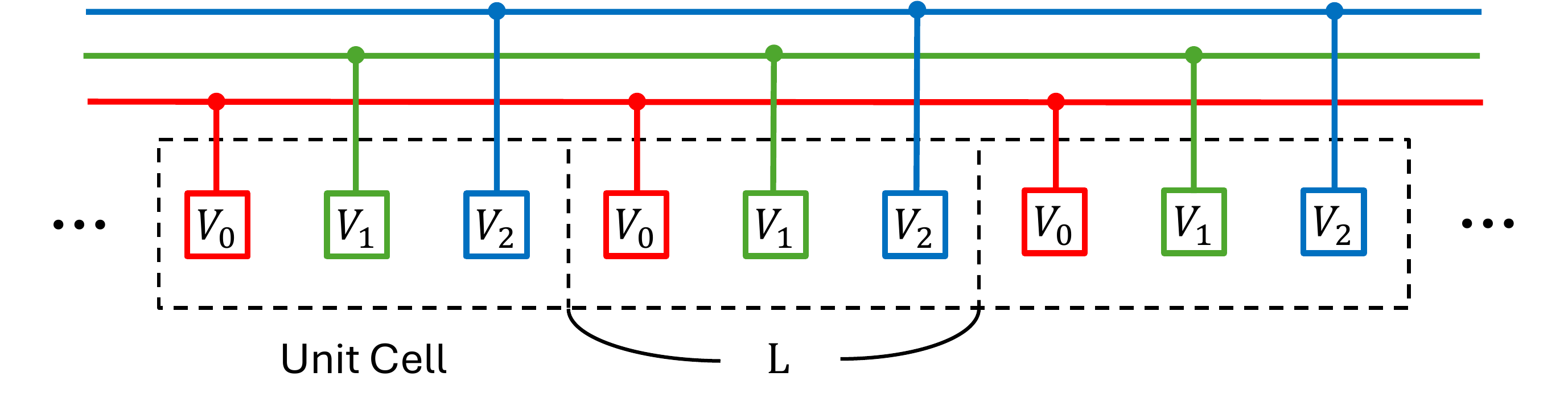}\label{fig:CB_mode_pulse_illustration_periodicity}}%

        \hfill
        
        \subfloat[]{%
	\includegraphics[trim={0 0.75cm 0 0},clip,width=\linewidth]{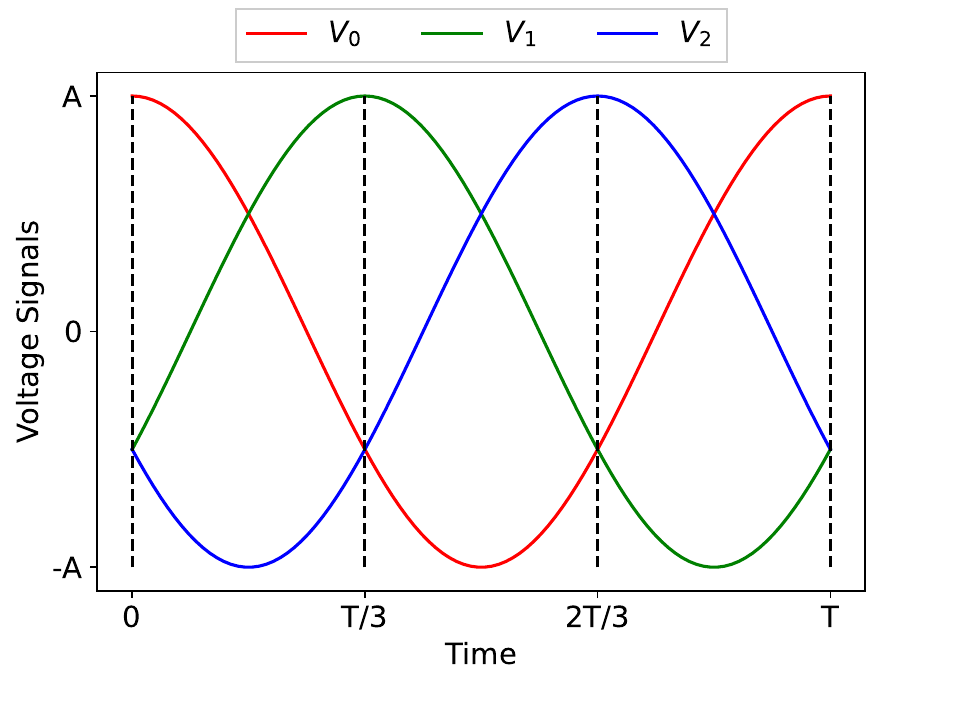}
          \label{fig:CB_mode_pulse_illustration_sinusoidal}}%
    
        \hfill
    
        \subfloat[]{%
	\includegraphics[width=\linewidth]{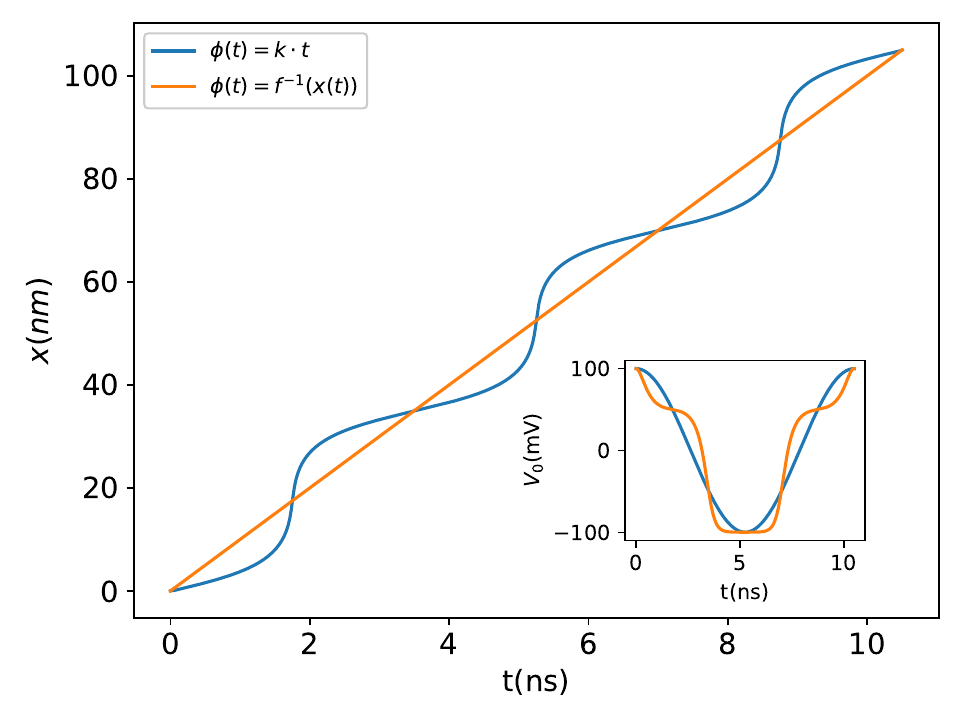}
          \label{fig:potential_max_x}}%
	\end{minipage}
	\hfill
    \begin{minipage}[t]{0.495\columnwidth}
        \subfloat[]{%
	\includegraphics[trim={0.70cm 0cm 0 0cm},clip, width=0.95\linewidth]{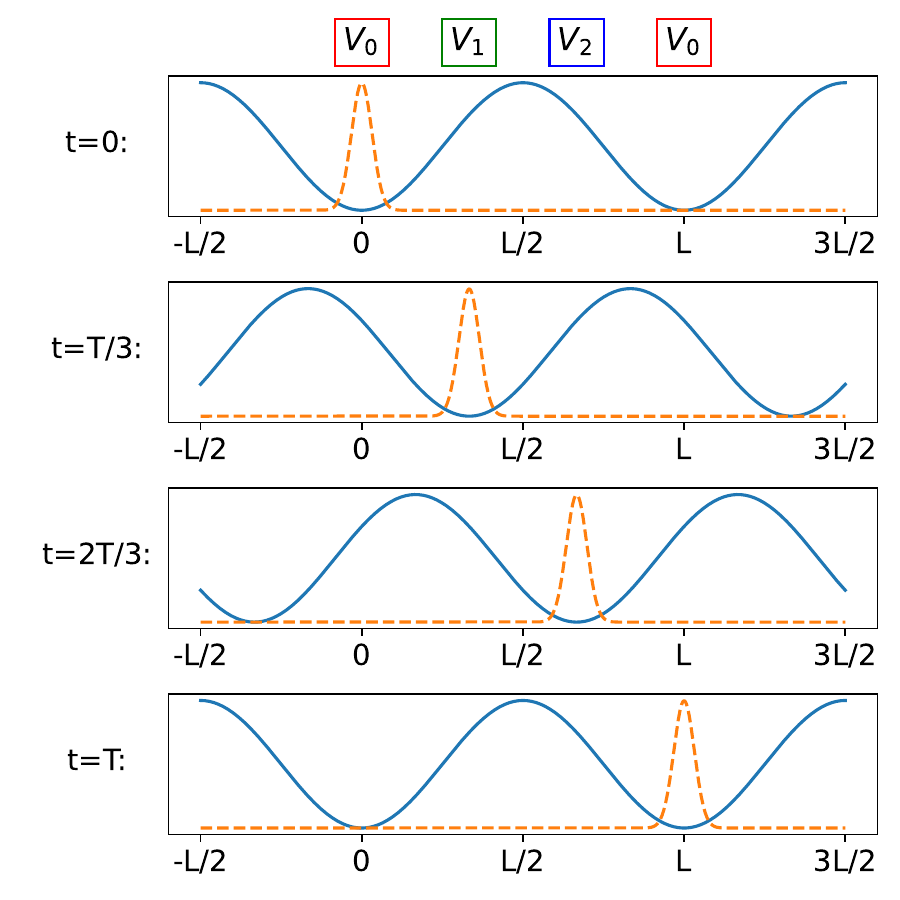}
    \label{fig:CB_mode_pulse_illustration_shuttling}}%
        
        \hfill

        \subfloat[]{%
	\includegraphics[width=\linewidth]{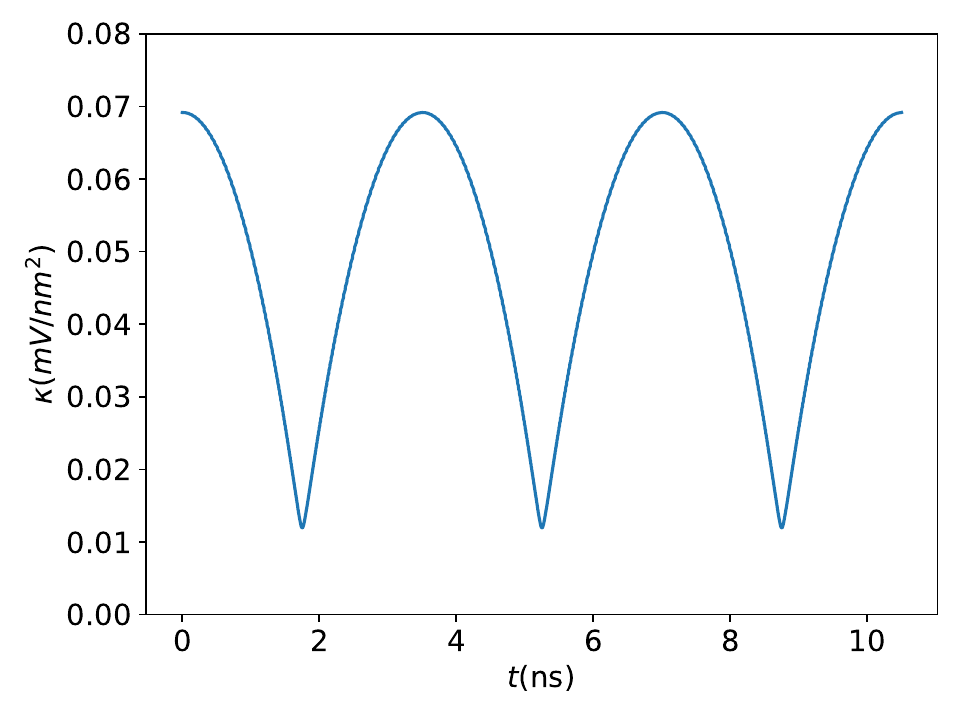}
          \label{fig:potential_max_curvature}}%
    \end{minipage}%
    
    \caption{Illustration of the voltage pulses applied to the gates in conveyor-belt shuttling for three gates per unit cell, i.e. $N=3$. (a) Every 3rd gate is connected to the same voltage source (denoted as red, green and blue lines). (b) The three independent pulses are sinusoidal with $2\pi/3$ phase difference. (c) The position, $x$, of the potential minimum as a function of time when the average shuttling speed is $10$\,m/s. The lines correspond to different choices of phase variation, $\phi(t)$: linearly varying phase $\phi(t)= k \cdot t$ (blue line), and phase obtained by using a look-up table, $f^{-1}$, $\phi(t)=f^{-1}(x(t))$ (orange line). The detailed arguments are given in Appendix~\ref{appendix:subsec:phase_variation}. (d) The resulting time-evolution of the potential and approximate position of the electron wave packet. The red, green, and blue boxes above denote the same clavier gates of the corresponding colour in (a) and (b). (e) Variation of the curvature at the potential minimum, $\kappa$, obtained by fitting the slice of potential energy at $y=0$ with a quadratic function (see also the plots of the full potential in Figure~\ref{fig:potential_3d_n_4}). Note that panels (a),(b), and (c) relate the panels (e), (f), and (g) of Figure 2 of  Langrock et al.\cite{langrock_2023}.}
    \label{fig:CB_mode_pulse_illustration}
\end{figure*}

\begin{figure*}
\subfloat[]{%
	\includegraphics[trim={1cm 1cm 1cm 1cm}, clip, width=0.495\textwidth]{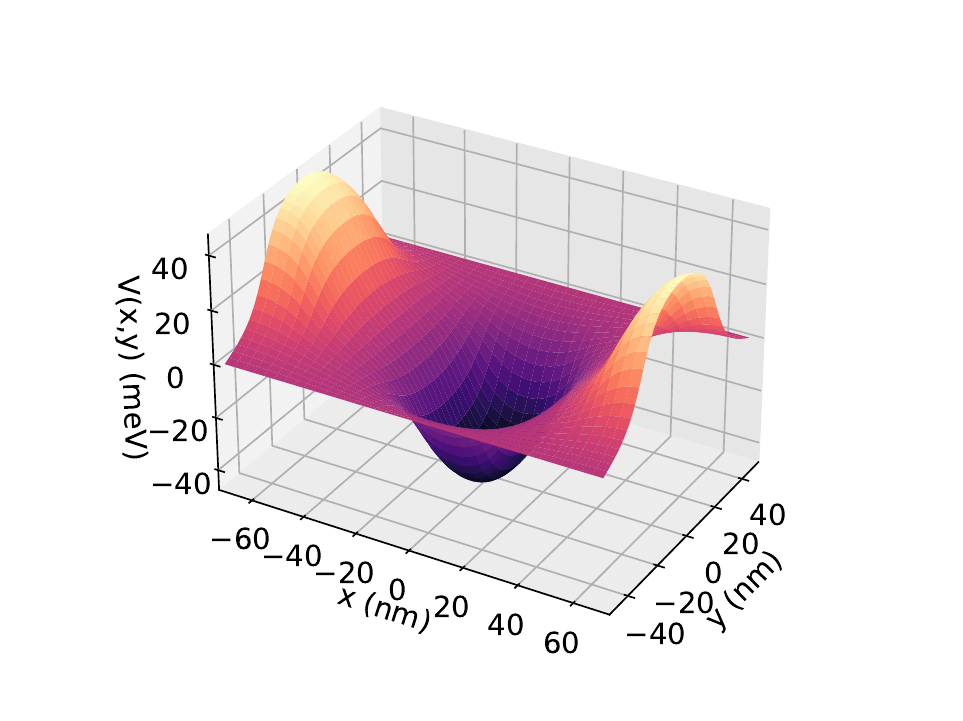}\label{fig:potential_3d_n_4_3d_1}
}%
~
\subfloat[]{%
	\includegraphics[width=0.495\textwidth]{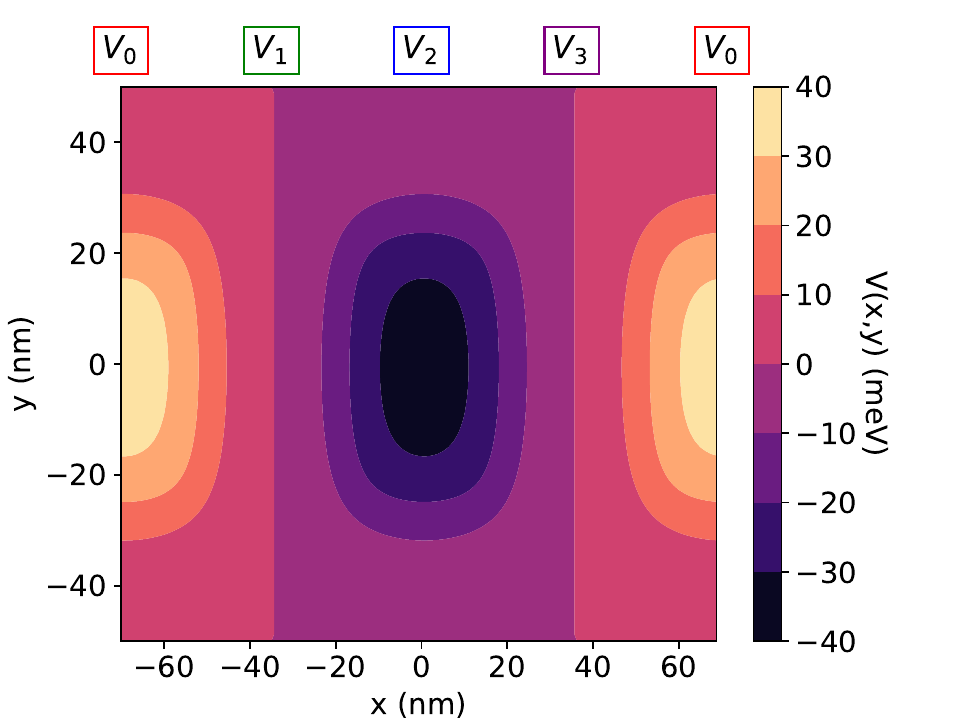}

  \label{fig:potential_3d_n_4_contour_1}
}%
\hfill
\subfloat[]{%
	\includegraphics[trim={1cm 1cm 1cm 1cm}, clip, width=0.495\textwidth]{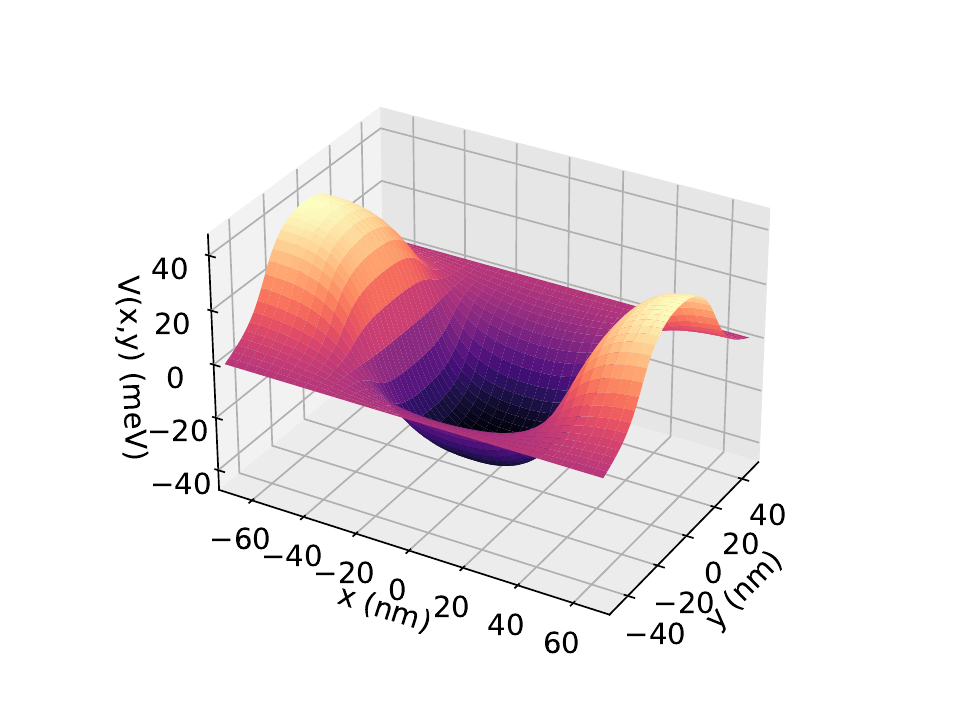}

  \label{fig:potential_3d_n_4_3d_2}
}%
~
\subfloat[]{%
	\includegraphics[width=0.495\textwidth]{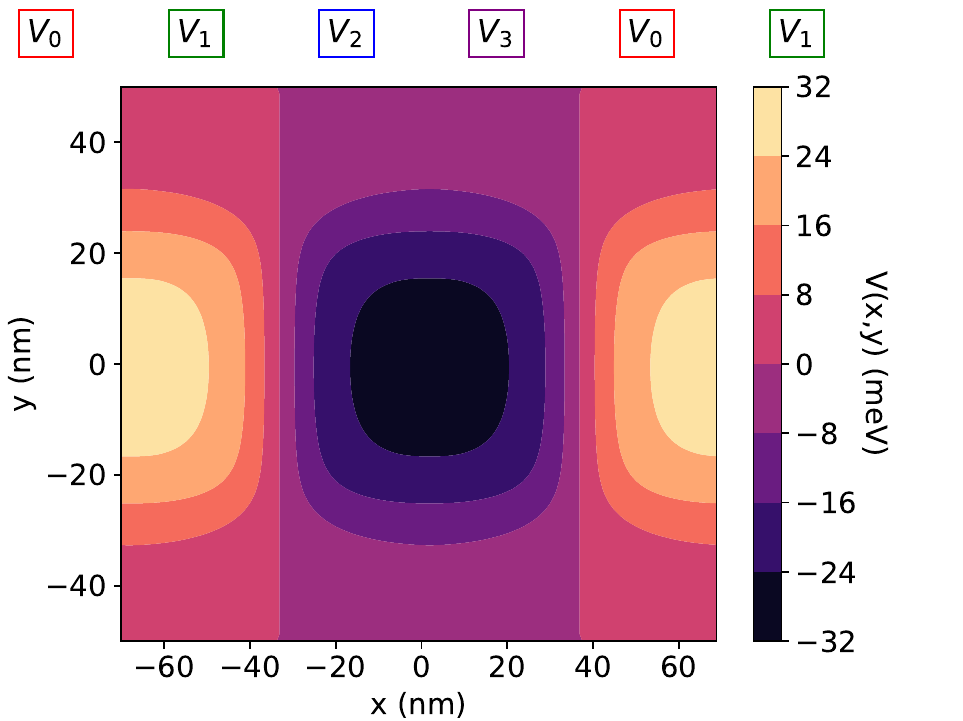}
  \label{fig:potential_3d_n_4_contour_2}
}%

\caption{3D and contour plots of the potential energy, $V(x,y)$, generated by the gates for $N=4$ with $A = 100$\,mV. (a) 3D  and (b) contour plots of the potential energy for one unit cell centred around the potential minimum when the curvature near the potential minimum, $\kappa(\phi)$, is maximum (corresponding to a potential well directly below an electrode). (c) 3D  and (d) contour plots of the potential energy for one unit cell centred around the potential minimum when $\kappa(\phi)$ is a minimum (corresponding to a potential well between two electrodes).}
\label{fig:potential_3d_n_4}
\end{figure*}

Conveyor-belt shuttling is achieved by creating a single QD moving in a desired trajectory (along the positive $x$ direction in our case). As explained in Langrock et al.\cite{langrock_2023}, such a potential can be created by applying sinusoidal voltage signals to the gates with a phase difference of $2\pi/N$ between successive gates; $N$ is then the number of independent voltage signals required:
\begin{equation}\label{eqn:sinusoidal_voltage_pulses}
    V_{i}(t) = A\cos(\phi(t) - \frac{2\pi i}{N}),
\end{equation}
where $A$ and $\phi(t)$ are the amplitude and phase of oscillation, respectively. Figure~\ref{fig:CB_mode_pulse_illustration_periodicity}  illustrates the repeating voltage signals; using an analogy from condensed-matter physics, we define a sequence of $N$ adjacent gates as a \textit{unit cell}. To be specific, the size of the unit cell along the shuttling direction, i.e. the x-direction, changes with the number of gate per unit cell, $N$. For example, the length of a unit cell is $105$\,nm for three gates per unit cell, but $175$\,nm for five electrodes per unit cell. Such a scheme solves the signal fan-out problem because the device only needs $N$ control lines regardless of the number of gates. For example, in the case of 4 gates in the unit cell, the applied pulses will be $\cos(\phi(t))$, $-\sin(\phi(t))$, $-\cos(\phi(t))$, and $\sin(\phi(t))$, where $\phi(t)$ is the phase as a function of time. The resulting evolution of voltage pulses at the clavier gates with time is illustrated in Figure~\ref{fig:CB_mode_pulse_illustration_sinusoidal}. To give a sense of direction, there must be at least three gates per unit cell.  Figure~\ref{fig:CB_mode_pulse_illustration_shuttling} shows the wave function propagating from left to right using the conveyor-belt shuttling at $4$ different times.

These voltage signals will successfully drive shuttling if the process proves to be adiabatic and thus the wave function closely follows the minimum of the potential energy. The instantaneous speed of shuttling is proportional to the first derivative of the phase $\phi(t)$ in the sinusoidal pulses. Hence, the shuttling trajectory depends on how the phase $\phi(t)$ is varied. We examined two possible ways to vary this phase: the first was a simple linear variation, while the second was designed to achieve a uniform propagation speed for the minimum of the quantum dot potential, the phases themselves being determined from a position-phase look-up table. Figure~\ref{fig:potential_max_x} shows the shuttling trajectories and voltage pulses (inset) for these two different methods of phase variation. 
However, a detailed comparison between the linearly increased phase and the phase obtained from the look-up table (in Appendix \ref{appendix:subsec:phase_variation}) showed that there is little practical difference between the two methods. Thus, we chose to update the phase linearly because it is easier to generate simple sinusoidal pulses on-chip than to apply more complicated pulses.

\section{\label{sec:numerical_simulation} Numerical Simulations}

For a given device geometry, specified as in section \ref{sec:shuttling_device}, it is necessary to solve the Laplace equation to obtain the QD potential, $\Phi(x,y,z,t)$, and then to solve the time-dependent Schr\"{o}dinger equation to simulate the dynamics of the shuttling. Periodic boundary condition was chosen along the shuttling direction, i.e. the x-direction, and $V=0$ was chosen for the bottom surface and sides of the shuttling track. In between the gates, Neumann boundary conditions of $\partial{\phi}/ \partial{z} = 0$ was imposed. At the \ce{Si}-\ce{SiO2} interface, the continuity of the displacement field was imposed, and the relevant relative permittivity for \ce{Si} ($11.69$) and \ce{SiO2} ($3.9$) were used. A detailed description of the boundary conditions is outlined in Appendix~\ref{appendix:subsec:boundary_conditions}. For the Poisson solver, we defined a uniform rectangular grid in 3D with finite difference approximation for the differential operators. We used successive over-relaxation (SOR)\cite{Young_1954, Frankel_1950} to obtain the time-dependent potential in the unit cell in Figure~\ref{fig:device_geometry}. For faster generation of the time-dependent potential, we used the superposition principle, based on the linearity of the Laplace equation, as noted in equation~\ref{eqn:superposition_principle_potential} in Appendix~\ref{appendix:subsec: numerical algorithms}. On the other hand, for the Schr\"{o}dinger solver, we used uniform rectangular grid in 2D on the plane defined by $z=-10$\,nm. We used the split operator method\cite{glowinski2017splitting} with symmetric Strang splitting\cite{strang1968construction, strang2012essays} to solve the time-dependent Schr\"{o}dinger equation. The convergence of the numerical methods were test in Appendix~\ref{appendix:subsec:convergence_studies}. Finally, the choices of unit systems and hyperparameters of the numerical methods are given in Appendix~\ref{appendix:subsec:Model}.

If we assume that the $z$-axis confinement is so strong that the electron only moves in the plane of the \ce{Si/SiO2} interface, modelling in 2D is enough to capture the relevant physics. The perpendicular extent of the wave-function is anyway reduced because the lowest-energy bound states are formed from the $\pm z$-valleys, so the motion in the $z$-direction is determined by the longitudinal (heavy) electron mass. Furthermore, \ce{SiO2} has large band gaps that allow strong electric fields to confine electrons in the z-direction without leakage out of the channel\cite{burkard_2023}. As a result, the typical confinement length of the QD in the $z$-direction is of order $1$\,nm\cite{Yang2013-wr, Veldhorst_2015, Ruskov_2018} while the oxide layer thickness is $10$\,nm. Thus, we sampled our 2D potential at the interface between \ce{Si} and \ce{SiO2}. A detailed comparison of the potential sampled at the interface and the potential averaged over the probability density of the ground state in the z-direction is given in Appendix~\ref{appendix:subsec:airy_function}.

We may further reduce the dimensionality to 1D if we assume that the voltages at the confinement gates are so high that the electron never undergoes excitation in the $y$-direction. A detailed comparison of 1D and 2D simulations is given in Appendix~\ref{appendix:subsec:1d_vs_2d}. 1D and 2D simulations yield different loss probabilities and excitation fractions with an order of magnitude difference for realistic parameters. This highlights the importance of simulating in 2D and that the potential is non-separable. We therefore report results of the more accurate 2D simulations in the remainder of the paper.

Note that the atomic scale interface roughness was neglected in our simulations.  Given that interface roughness has a similar nature to the Johnson-Nyquist noise, the results of section \ref{sec:johnson-nyquist} imply that it only affects the orbital excitation if there is a frequency component in the moving frame of the electron that is comparable to the energy gap in the orbital degree of freedom. For example, if the shuttling speed is $100$\,m/s and the smallest length scale of roughness is, say, $0.5$\,nm, the maximum change in frequency is $0.2$\,THz, which is much smaller than the frequency of characteristic energy gap, i.e. $\Delta E_{gs,2e}/h = 1.46$\,THz. 

We also neglect the effect of valley physics.  This is likely to have minimal effect on charge shuttling: valley-orbital anti-crossings are unlikely to occur because the four transverse valley states have much higher energy than the $\pm z$-valleys. In any device where the tensile strain in \ce{Si} exceeds $0.1$\,\%, the transverse valley states lie more than $20$\,meV above the $\pm z$-valleys~\cite{Euaruksakul_2008}, comfortably higher than our characteristic orbital energy gap of $6.02$\,meV ($A=100$\,meV and $N=4$ electrodes per unit cell). Furthermore, given strong confinement in the z-direction (confinement length $\sim 1$\,nm), there is an additional contribution to the splitting from the effective mass anisotropy (the longitudinal mass is around $5$ times bigger than the transverse mass).  On the other hand, the lowest valley excitation occurs within the $\pm z$-valleys and lies well below the orbital excitations (The energy scale of excited valley states is $\mathcal{O}$($10$-$100$\,$\mu$eV) while the energy scale of orbital states is  $\mathcal{O}$($1$\,meV)); it is not resolved in our calculations but does not significantly affect the location of the shuttled charge.  Therefore, valley-orbital anti-crossings are unlikely unless the tensile strain is unusually small. 

\section{\label{sec: metrics} Performance metrics}

\begin{figure}
    \centering
    \includegraphics[width=\columnwidth]{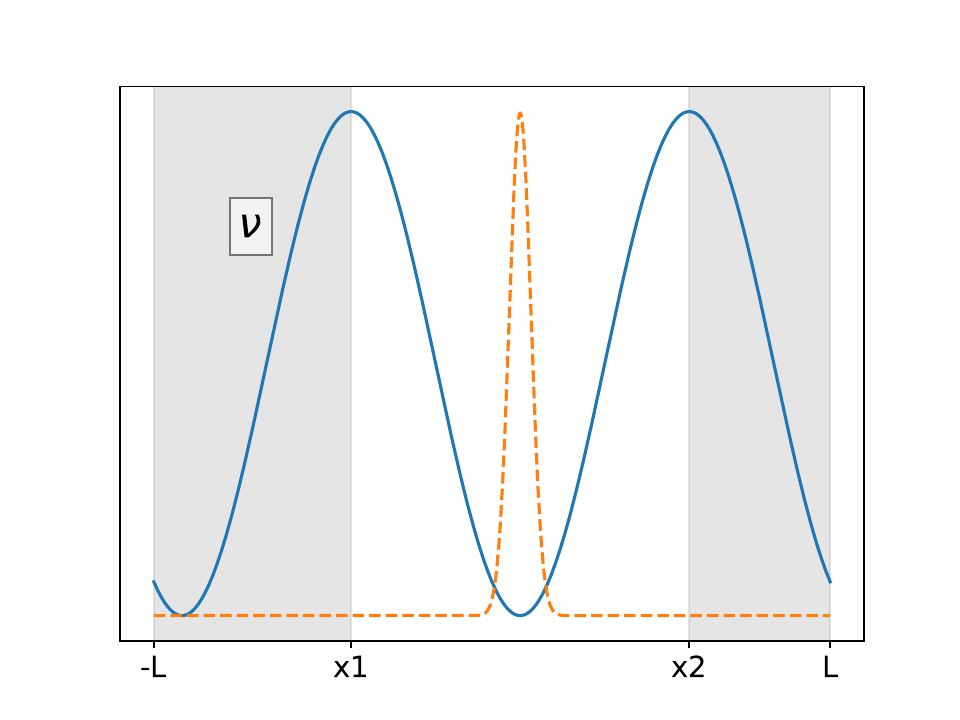}
    \caption{Illustration of the definition of the loss probability as the integrated probability density (orange line) over the shaded region  $\nu$, defined as the region outside the well in potential energy (blue line) being used to shuttle the electron.}
    \label{fig:loss_prob_metric}
\end{figure}

We will characterise the shuttling process by evaluating its capability to move the spin qubit to the target position (1) without losing the qubit and (2) with a good degree of adiabaticity. Specifically, the excitation in the orbital state is important as the $g$-factor of the electron in silicon depends both on its position and orbital state\cite{Kawakami_2014, langrock_2023}. Thus, the two most import imperfections to evaluate the shuttling scenarios are (1) the probability $P_L$ of losing the electron from the potential well and (2) the amount of excitation from the ground state. 

The loss probability is defined as the probability that the electron is found outside the single QD where it was initially loaded. Since we have periodic boundary conditions along the x-axis, we need at least two unit cells, i.e. two QDs, to calculate the loss probability. When there are two unit cells, the loss probability is equivalent to the probability of the electron to be in the other `wrong' QD (since we solve the TDSE only within the channel region, the electron cannot leave the channel). Figure~\ref{fig:loss_prob_metric} shows the illustration of calculation of loss probability. The loss probability is the probability in the shaded region, $\nu$:
\begin{equation}
    P_{L} = \int\limits_{\nu} dxdy \, |\psi(x,y)|^2 
\end{equation}where $\psi(x,y)$ is a 2-dimensional wave function.

The excitation fraction is a dimensionless measure of the level of excitation of the system due to non-adiabatic effects. It is defined as the ratio $\Delta E/\Delta E_{gs,2e}$ where $\Delta E$ is difference between the expectation value of the energy and the (instantaneous) ground state energy and $\Delta E_{gs,2e}$ is a characteristic energy gap; it should be interpreted as the excess energy relative to this characteristic energy gap. Since the excitation primarily occurs in the direction of shuttling, the characteristic energy gap was chosen as the energy gap of the ground to the second excited state, so that $\Delta E/\Delta E_{gs,2e} =  (E - E_{gs})/(E_{2e} - E_{gs})$. Figures~\ref{fig:fidelity_gamma} and  \ref{fig:fidelity_temperature} show that excitation primarily populates the second excited state, which is the excitation mode in the x-direction as shown in Figure~\ref{fig:excited_states_2d}.

Additionally, we calculated the probabilities of excitation to the $n$th eigenstate of the instantaneous Hamiltonian. This metric was used to compare the performance of the noisy shuttling cases, as the fidelity between the final state and the ground state deviates from $1$ by the order of only $10^{-7}$ for noise-free shuttling.

When we report metrics for the overall performance of the shuttling experiment, these are computed using the state of the system during the static phase after the shuttling procedure. The static phase involved an additional 5000 times steps ($\approx 13.5$\,ps) to evolve the state with the stationary potential at the end of the shuttling. While the excitation fraction remains constant up to a numerical precision during this period, the loss probability may vary if the potential barrier is too low, much lower than our default setting of $100$\,mV. Effectively, we sampled one loss probability value in this case; These scenarios, however, correspond to a failed shuttling, and any small fluctuation in loss probability is not of much interest.

\section{\label{sec: noiseless_shuttling}Noiseless Shuttling}
\begin{figure*}
	\subfloat[]{%
	\includegraphics[width=0.495\textwidth]{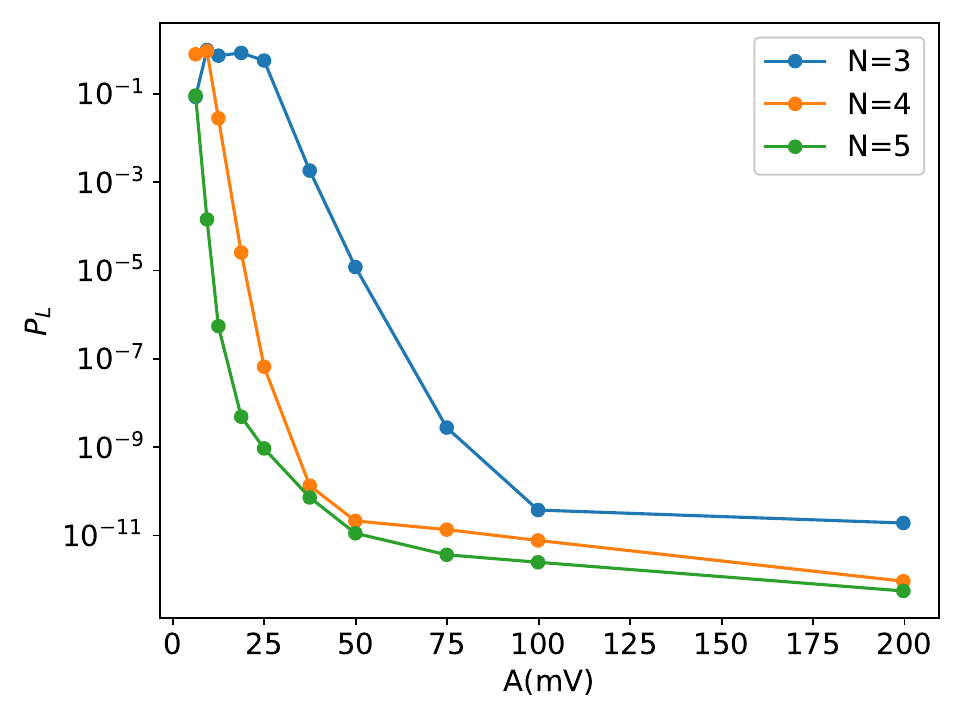}
    \label{fig:loss_probability_amplitude}}
    ~    
    \subfloat[]{%
    \includegraphics[width=0.495\textwidth]{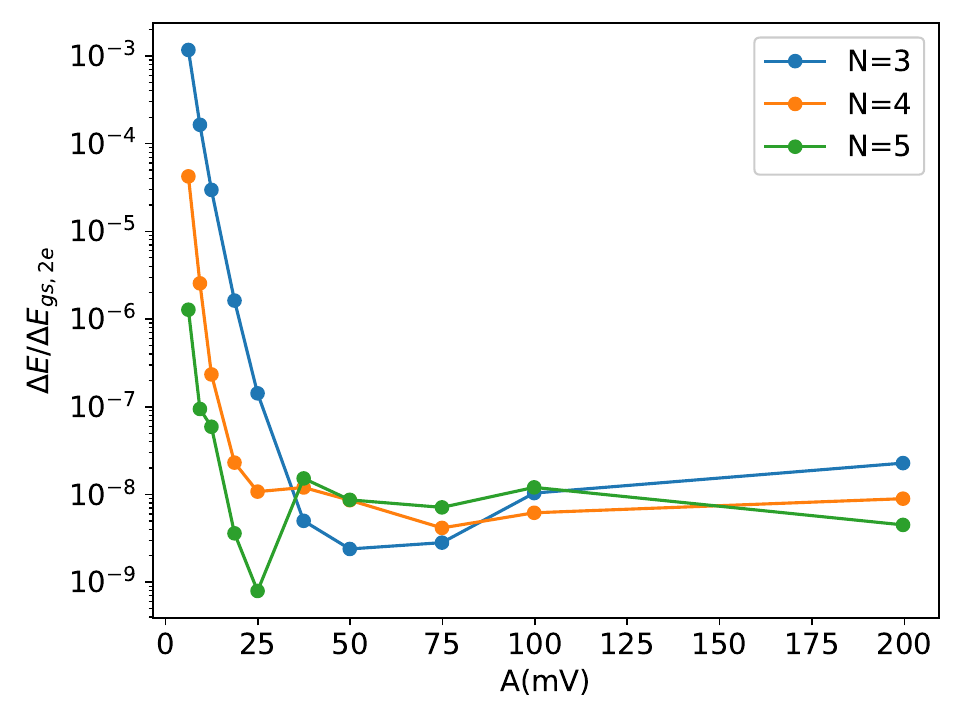}
	
    \label{fig:loss_probability_target_dist}}%

    \hfill	
    \subfloat[]{%
	\includegraphics[width=0.495\textwidth]{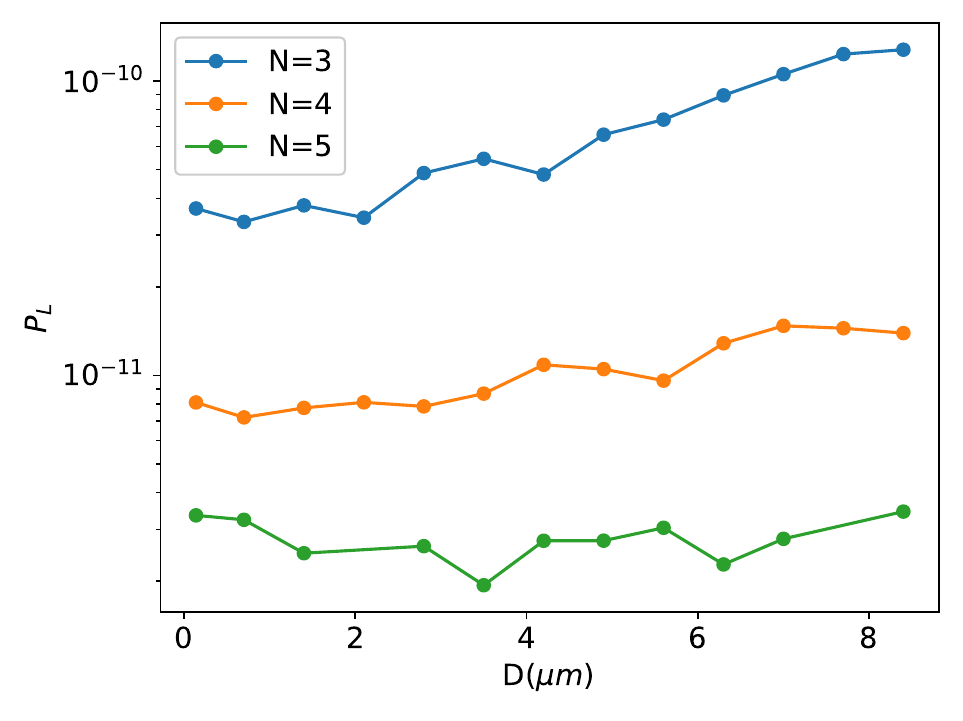}
    \label{fig:excitation_fraction_amplitude}}%
    ~    
    \subfloat[]{\includegraphics[width=0.495\textwidth]{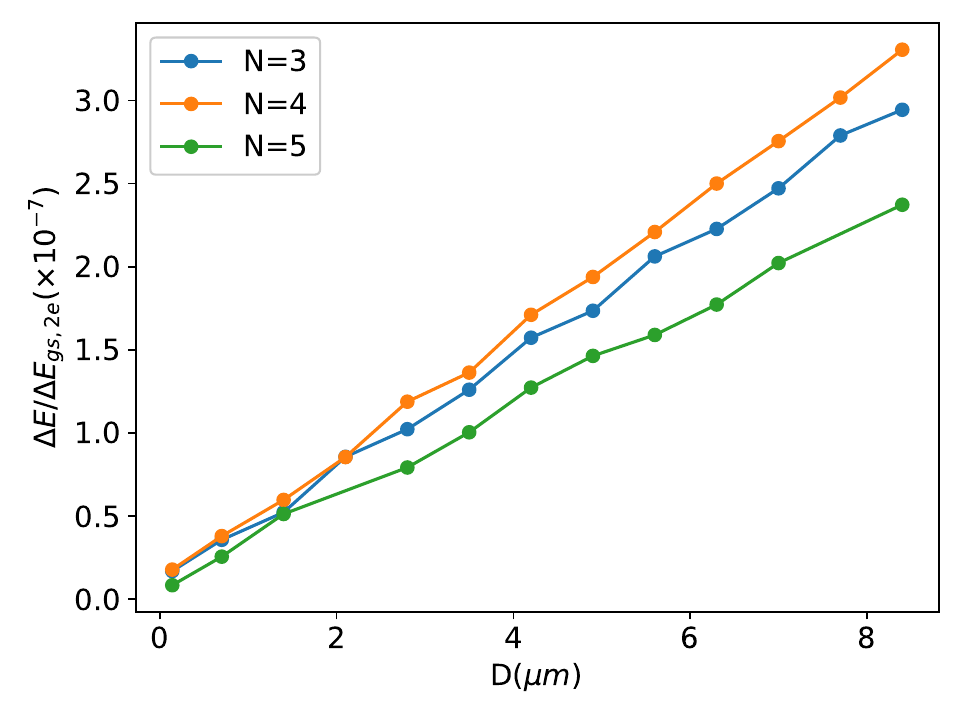
    }
    \label{fig:excitation_fraction_target_dist}}%

    \caption{The loss probability and excitation fraction for different noiseless shuttling scenarios: (a) loss probability and (b) excitation fraction as a function of voltage signal amplitude, $A$, for a shuttling distance of $1.4$\,$\mu$m; (c) loss probability and (d) excitation fraction as a function of target distance, $D$, for a signal amplitude of $100$\,mV.  Results for different numbers of electrodes per unit cell ($N=3,4,5$) are shown.}
    \label{fig:loss_and_excitation_target_dist}
\end{figure*}

In this section, we present the results from shuttling scenarios where there is no noise and no defect charges are present. While the quality of shuttling depends on many parameters, we selected three independent variables: (1) the target distance, (2) the amplitude of the voltage signals at the gates, and (3) the number of gates in a unit cell. 

Figures~\ref{fig:loss_probability_amplitude} and \ref{fig:excitation_fraction_amplitude} show the loss probability and excitation fractions for different amplitudes of sinusoidal oscillations at the gates. Larger signal amplitudes make a deeper QD, and thus the loss probability decreases. Our typical value of amplitude, $100$\,mV, resulted in a loss probability of $3 \times 10^{-11}$ even for $N=3$ electrodes. The loss probability reduces even further for larger numbers of electrodes as the depth of the QD and the inter-dot distance both increase. For example, when the amplitude is $50$\,mV, we see a loss probability of $10^{-5}$ for $N=3$ while we see the similar loss probabilities for $N=4$ and $N=5$ when the amplitudes are $25$\,mV and $12.5$\,mV.

Figures~\ref{fig:loss_probability_target_dist} and \ref{fig:excitation_fraction_target_dist} show the loss probability and excitation fraction with different target distances. The mean shuttling speed and the amplitude of voltage signals were fixed to $10$\,m/s and $100$\,mV, respectively. As the target distance increases, both the loss probability and excitation fraction increase. The worst case occurs when the number of electrodes is three and the target distance is $8.4$\,$\mu$m, which nevertheless results in near-ideal behaviour:  a loss probability of $1.3 \times 10^{-10}$ and an excitation fraction  of $2.7 \times 10^{-7}$.

Given these data, we conclude that noiseless shuttling is practically perfect when the default speed and amplitude were used with the target distance up to $8.4$\,$\mu$m. The quality of shuttling significantly depends on the amplitude of the voltage signal and the number of gates per unit cell. To reduce the loss probability, it is always beneficial to use more gates per unit cell; but for a broad range of cases we find that $N=3$ is quite sufficient for near-ideal performance.

\section{Sensitivity to Johnson-Nyquist noise} \label{sec:johnson-nyquist}

Since noise-free shuttling is nearly perfect, we further investigated the effect of discontinuities in the voltage signals. The full results are described in Appendix~\ref{appendix:subsec:step_changes} and we summarise here. Two extreme cases were studied: staircase-like potentials in time, with step-changes in the potential at defined intervals, and potentials formed from piece-wise linear functions connecting the midpoints of the steps of the staircase-like potential (See Figure~\ref{fig:voltage_signals_mode1_vs_mode2}). From Figure~\ref{fig:loss_and_excitation_step_vs_linear}, we concluded that staircase-like discontinuities in the voltage signals result in much more loss and excitation than continuous signals. 

Fortunately, these staircase-like voltage profiles  constitute an adversarial model that is somewhat unphysical, as in reality there is a finite response time for any change of voltage at the gates. However, rapid changes in the gate voltages on frequencies up to this cutoff can still arise from Johnson-Nyquist noise \cite{Johnson1927-ag, Nyquist_1928}, which is a thermal noise at the resistor caused by random thermal agitation. We therefore proceed to explore the impact of such noise when physically motivated. 

Figure~\ref{fig:circuit} shows a lumped-element model of a voltage source connected to clavier gates via a single bondwire. $L$ is the inductance of the bondwire, $R$ is the resistance of the metal connection from the bondpad to the gate, $C_{1}$ is the capacitance of the bondpad, and $C_{2}$ is the capacitance of clavier gate. (See appendix \ref{appendix:subsec:lumped_element_model}) The power spectral density of classical Johnson-Nyquist noise can be derived as
\begin{equation} \label{eqn:PSD_johnson_nyquist_classical}
    S_{C}(\omega) = 4k_{B}T \frac{N_{G}C_{2}}{C_{1}^2} \frac{\gamma}{\omega^{2} + \gamma^{2}},
\end{equation}where $\gamma=\frac{1}{RC_{2}}$ is the characteristic inverse time constant of the Lorentzian distribution and $N_{G}$ is the number of gates and metal connections connected to the same bondpad, as shown in the left side of Figure~\ref{fig:circuit}. The corresponding RMS voltage noise can be obtained as
\begin{equation}\label{eqn:rms_noise_temperature} 
    \Delta V_{\text{rms}} = \sqrt{2\pi k_{B}T \frac{N_{G}C_{2}}{C_{1}^{2}}}.
\end{equation}

\begin{figure}

\subfloat[]{%
	\includegraphics[width=\linewidth]{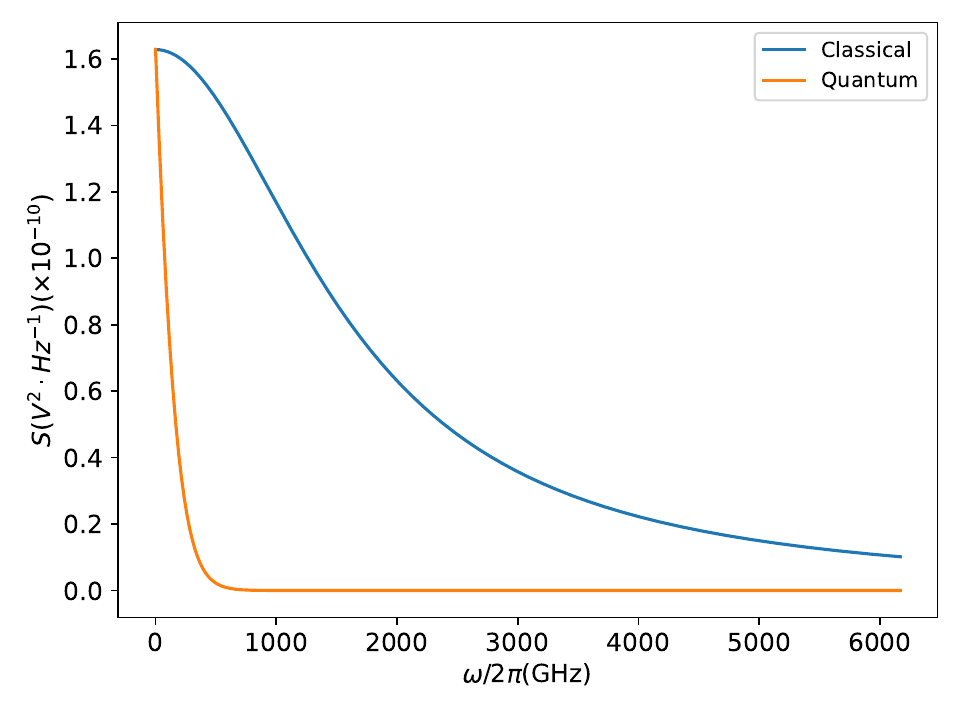}
  	\label{fig:PSD_Johnson_Nyquist}
}%

\hfill

\subfloat[]{%
	\includegraphics[width=\linewidth]{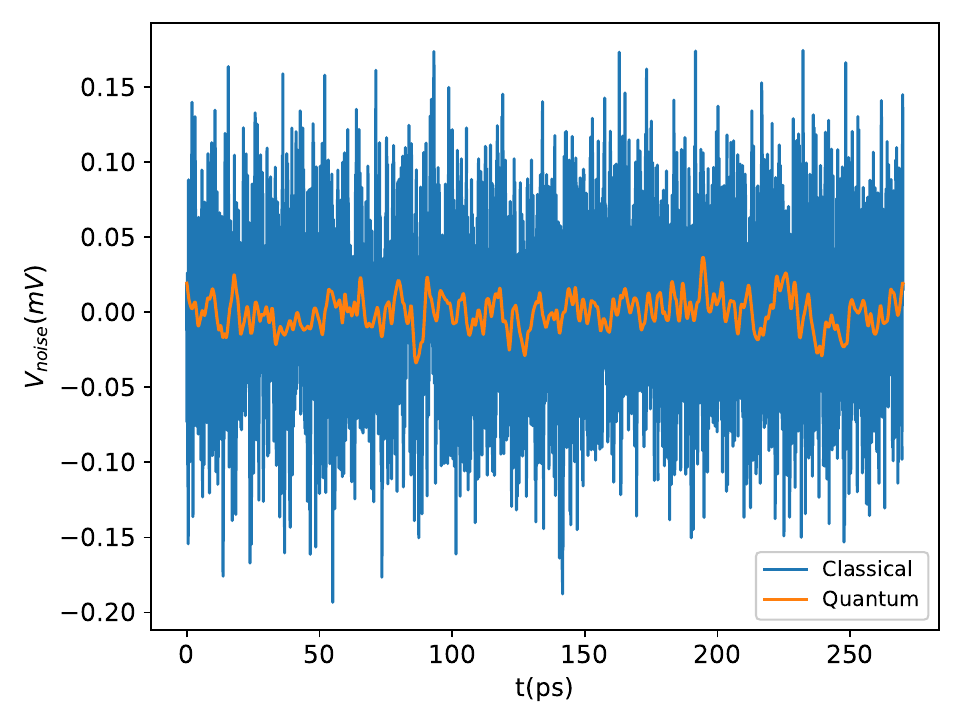}
  	\label{fig:quantum_classical_noise_comparison}
}%

\caption{(a) the power spectral density, $S$, of the classical (blue) and quantum (orange) Johnson-Nyquist noise, and (b) random instances of classical (blue) and quantum (orange) Johnson-Nyquist noise, $V_{noise}$, at $T=4$\,K, and $\gamma=10$\,THz}
\label{fig:eta_PSD_johnson-nyquist_noise}
\end{figure}

We first looked into the effect of classical Johnson-Nyquist noise. The complete results are described in Appendix~\ref{appendix:subsec:johnson-nyquist_classical}, and we summarise here. Figure~\ref{fig:loss_and_excitation_gamma_temp_classical} shows that both loss probability and excitation fraction significantly increase with higher cut-off frequency, $\gamma$, and higher temperature, $T$. We concluded that high frequency noise, especially the one that is comparable to the frequency corresponding to the characteristic energy gap, i.e. $\Delta E_{gs,2e/\hbar}$, is more harmful than low frequency noise. For example, for our default setting of $4$ gates per unit cell and the amplitude of the voltage signal of $100$\,mV, i.e. $N=4$ and $A=100$\,mV, the resulting characteristic energy gap is around $6$\,meV, which corresponds to the frequency of $1.46$\,THz, i.e. $\Delta E_{gs,2e}/h = 1.46$\,THz.

In reality, quantum effects have to be taken into account once the cut-off frequency reaches $\gamma \gtrsim k_BT/\hbar$. To account for this\,\footnote{There has been a debate about whether to include the zero-point fluctuations in the correction term\cite{Kish_2016}, which is an additive term of $\hbar\omega/2k_{B}T$ to the correction factor. We found that our situation is close to an example by Kish et al.\cite{Kish_2016}, a resistor connected to an antenna with a photon counter. This is because the amount of fluctuation in the voltage of the gates is \textit{measured} by the electron shuttled underneath the gates, which acts like a photon counter in the example. Furthermore, note that spontaneous absorption doesn't exist while spontaneous emission exists due to the zero-point energy. Since the electron is absorbing energy from the gates, zero-point energy cannot be transferred to the electron, and this implies the absence of the zero-point term in the correction factor. A more detailed discussion about the inclusion of the zero-point term can be found in Kish et al.\cite{Kish_2016}.}, we multiply the Lorentzian power spectral density in equation \ref{eqn:PSD_johnson_nyquist_classical} by a correction factor $\eta(\omega)$ corresponding to the ratio between the thermal mode populations in the classical and quantum cases:
\begin{equation}
    \eta(\omega) =\frac{\hbar \omega/k_{B}T}{e^{\hbar \omega/k_{B}T} -1},
\end{equation}
\begin{equation} \label{eqn:PSD_johnson_nyquist_quantum}
    S_{Q}(\omega) = S_{C}(\omega)\eta(\omega).
\end{equation}
The correction factor decreases from $\eta(0) = 1$ as the frequency increases, with an asymptotic value of $\eta(\omega) = 0$. Since $\eta(\omega) \leq 1$ for all frequencies, the power spectral density now deviates from the pure Lorentzian distribution with the higher frequency components more strongly suppressed. Figure~\ref{fig:PSD_Johnson_Nyquist} shows the classical and quantum power spectral density at a temperature of $4$\,K and cut-off frequency $10$\,THz. For a given cut-off frequency $\gamma$, we expect that the quality of shuttling will be improved relative to the corresponding classical case owing to the smaller PSD at higher frequencies. Figure~\ref{fig:quantum_classical_noise_comparison} shows instances of classical and quantum noise generated with the same circuit elements at $T=4$\,K; the reduction in high-frequency noise in the quantum case is evident, and the total noise power decreases to only $5.5$\,\%  of the classical value.

Simulations of the noisy shuttling process were performed by generating random noise profiles from the power spectral density, by the procedure given in Appendix~\ref{appendix:subsec:generation_of_noise}. The values of the circuit elements in Figure \ref{fig:circuit} are given in appendix \ref{appendix:subsec:lumped_element_model}. The shuttling distance was chosen to be $1.4$\,$\mu$m, which corresponds to $10$ unit cells; thus, we assumed $10$ gates are connected via a single bondpad to a single voltage source with the total capacitance of $N_{G} \times C_{2} = 1$\,fF.

\begin{figure*}
\centering
    	\subfloat[]{%
    	 \centering
    \includegraphics[width=0.495\textwidth]{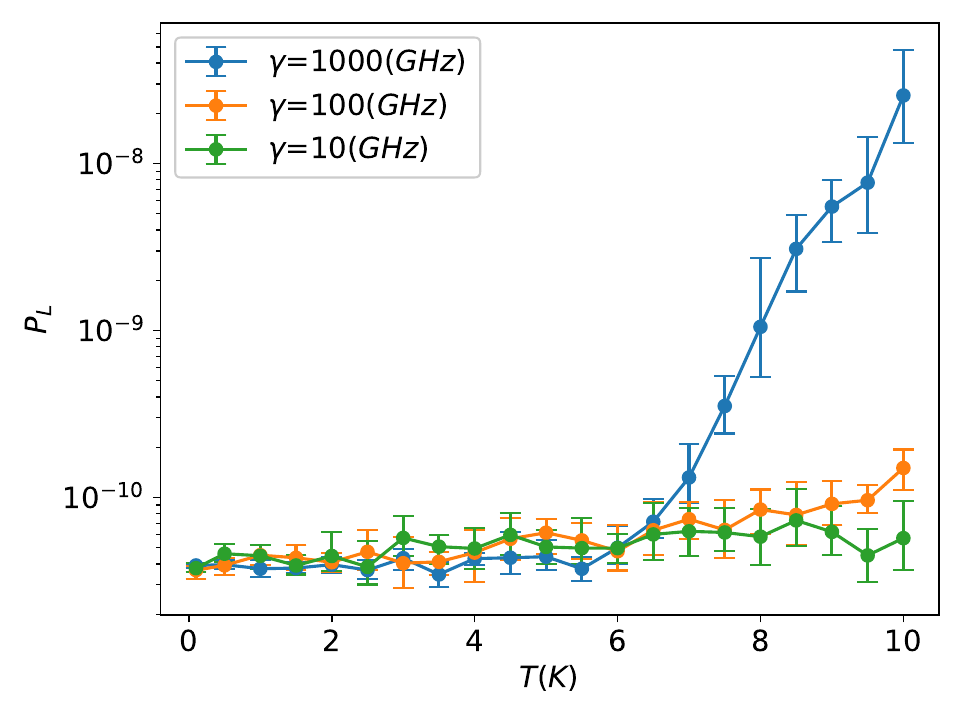}

          \label{fig:loss_quantum_noise_temperature}
    	}%
    	~
    	\subfloat[]{%
        \includegraphics[width=0.495\textwidth]{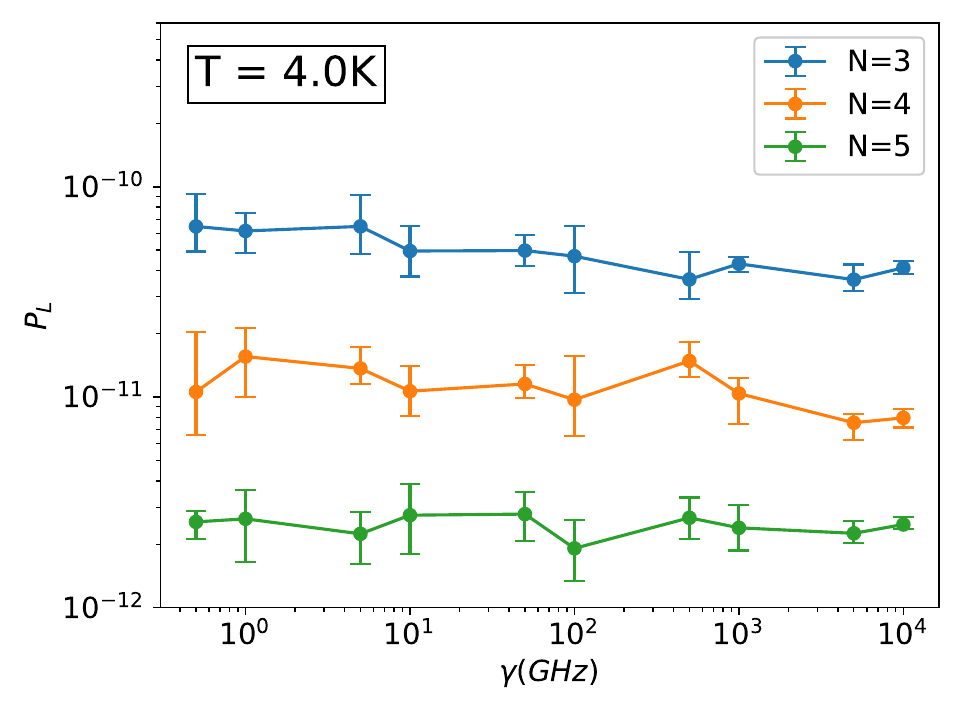}
			
    \label{fig:excitation_fraction_quantum_noise_temperature}    	
    	}%
    \hfill
    \subfloat[]{%
        \centering
    \includegraphics[width=0.495\textwidth]{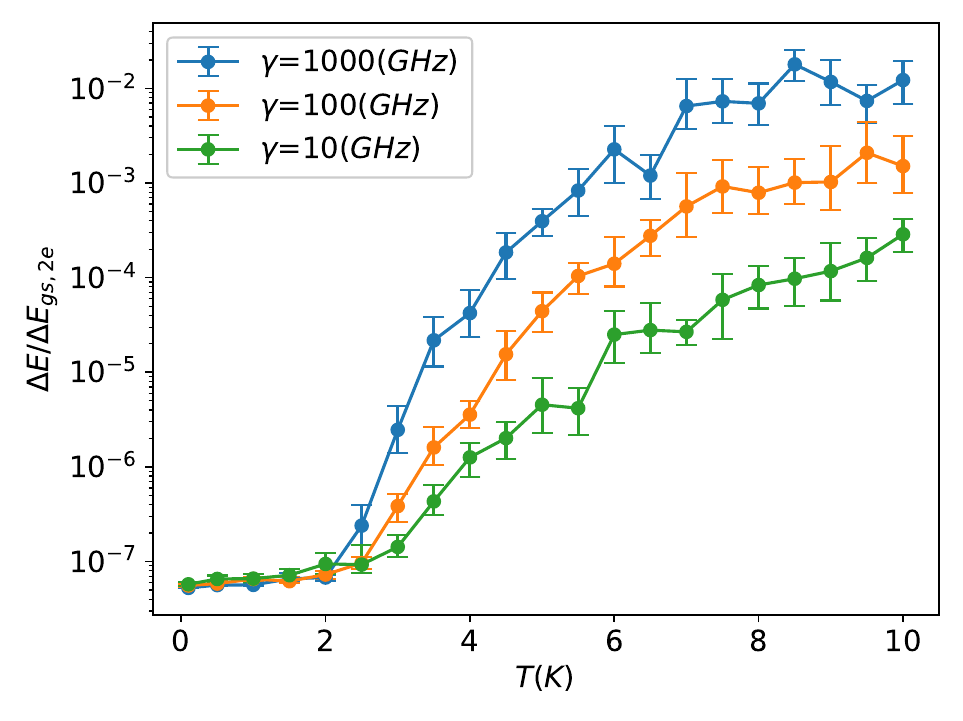}
          \label{fig:loss_quantum_noise_gamma}
    }%
    ~
    \subfloat[]{%
    \centering
    \includegraphics[width=0.495\textwidth]{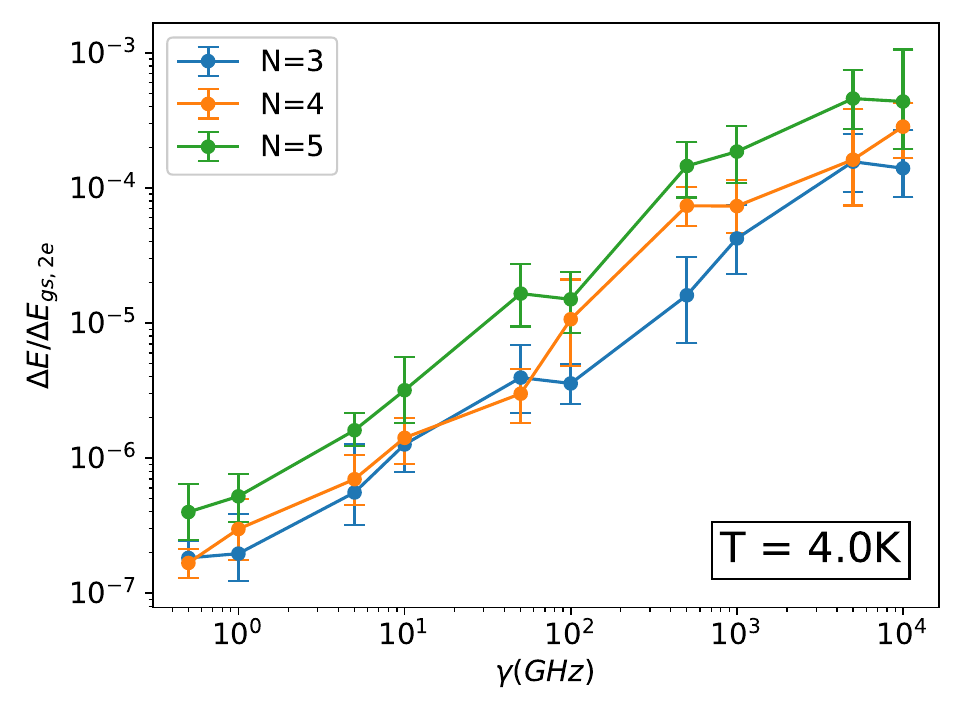}
    \label{fig:excitation_fraction_quantum_noise_gamma}
    }%
    
    \label{fig:loss_excitation_fraction_qauntum_noise}
    \caption{Loss probability and excitation fraction: (a, b) as a function of temperature with three different cut-off frequencies, i.e. $\gamma=10, 100, 1000 GHz$ for three gates per unit cell ($N=3$) and (c, d) as a function of cut-off frequency $\gamma$ with varying number of gates per unit cell, $N$.}
    
\end{figure*}

Figures~\ref{fig:loss_quantum_noise_temperature} and \ref{fig:excitation_fraction_quantum_noise_temperature} show the loss probability and excitation fraction for three gates per unit cell, i.e. $N=3$, at different temperatures $T$ ranging from $0.1$\,K to $10$\,K and with varying cut-off frequencies, $\gamma = 10,\ 100,\ 1000$\,GHz. For all cut-off frequencies, the excitation fraction tends to increase with temperature. However, there is highly significant increase only for $\gamma=1000$, where loss is seen to increase by three orders of magnitude (with an appreciable climb starting at lower temperatures). The excitation fraction also increases with both temperature and $\gamma$.

However, while both loss and excitation are finite and can rise severely with temperature, the primary conclusion is that they remain practically negligible. If we make the assumption that shuttling of qubits will not occur above a $4$\,K, we can confirm that at this temperature there is near-ideal behaviour. One observes that $P_L$ is always below $10^{-10}$ and the excitation fraction remains below $10^{-3}$ for all three architectural variants $N=3,\,4,\,5$.

\begin{figure}
    \centering
    \includegraphics[width=\textwidth]{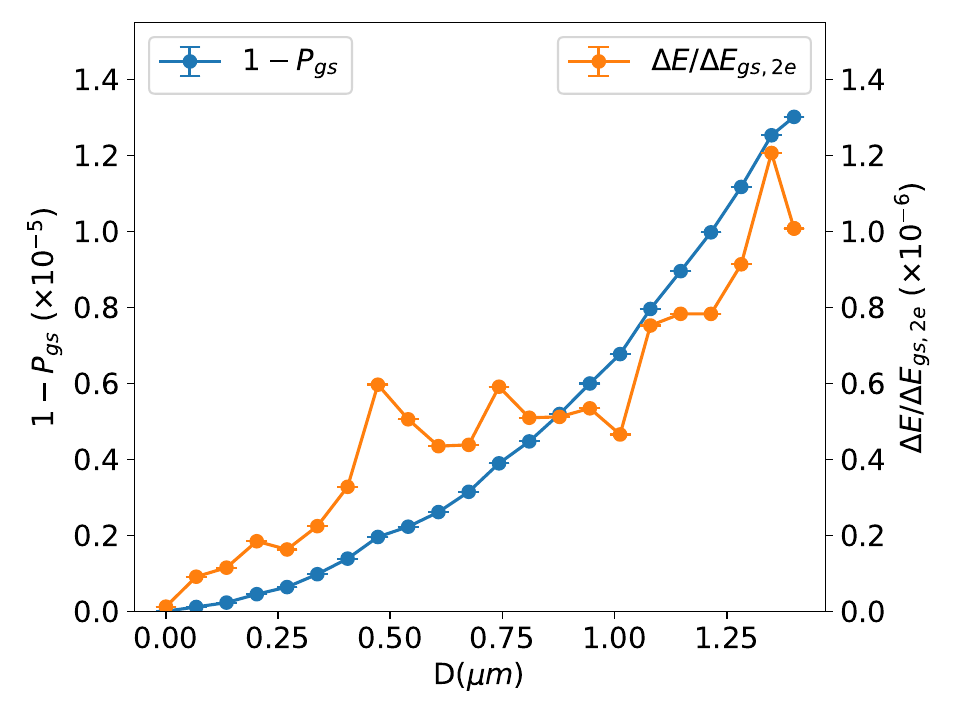}
    \caption{Probability of excitation outside of the ground state (blue) and excitation fraction during the shuttling (orange) for a target distance of $1.4$\,$\mu$m at $10$\,m/s. Other parameters were set as follows: The amplitude of voltage signals was $50$\,mV, i.e. $A=50$\,mV, the temperature was $2$\,K, i.e. $T=2$\,K, and there were three gates per unit cell, i.e. $N=3$.}
    \label{fig:infidelity_exc_frac_along_distance}
\end{figure}

Additionally, to confirm that there is no excitation during the shuttling process, we noted down the probability to remain in the ground state and excitation fraction in the middle of shuttling. Figure~\ref{fig:infidelity_exc_frac_along_distance} shows the probability of excitation outside of the ground state throughout the shuttling for a target distance of $1.4$\,$\mu$m at $10$\,m/s. During the shuttling, the probability of excitation was in the order of $10^{-6}$ to $10^{-5}$, the excitation fraction was in the order of $10^{-7}$ to $10^{-6}$ suggesting that the entire process of shuttling is largely adiabatic.

We conclude that, when high frequency components are suppressed by the correction factor, the effect of Johnson-Nyquist noise is negligible and the loss probability is comparable to the noiseless shuttling in the temperature ranges of practical interest. Furthermore, the entire process of shuttling remains adiabatic.

\section{Sensitivity to Charge Defects} \label{sec:charge_defects}

As Langrock et al.\cite{langrock_2023} pointed out, trapped charges due to impurities can affect the performance of shuttling if they occur near the interface defining the qubit layer. In this section, we investigate the effect of negative charge defects on the loss probability and excitation fraction. We used three unit cells and five electrodes per unit cell for these simulations, and the electron was shuttled across two unit cells in the presence of charge defects. Note that the trapped charges were placed in the oxide layer, and we used the permittivity of the oxide layer to compute the Coulomb peaks formed by the trapped charges. Since the oxide thickness is $10$\,nm (see section \ref{sec:shuttling_device}.), we chose the mid-point, i.e. $5$\,nm, as a default distance of defects from the interface. The Coulomb repulsion terms from the negatively charged defects are added to the Hamiltonian in equation \ref{eqn:time_dependent_hamiltonian_noise_free_2D}:

\begin{align}
    H &= \frac{\hbar^{2}}{2 m^{*}} \mathbf{p}^{2} - e\Phi(V_{0}(t), V_{2}(t), ..., V_{N-1}(t)) \\ \nonumber
    &+ \sum^{N_\text{\it defects}}_{i=1} \frac{e^{2}}{4\pi\epsilon_{0}\epsilon_{Si}|\mathbf{r} - \mathbf{r}_{i}|},
\end{align}where $\{r_{i}\}_{i = 1...N_\text{\it defects}}$ are positions of the charge defects in 3D space. Note that the motion of electron is still confined in a 2D space; The Coulomb repulsion from the static charges is calculated as if they are above (or below) the plane of motion in the 3D space. 

\begin{figure}
    \centering
    \includegraphics[width=0.9\textwidth,trim={0.40cm 0cm 0 0cm},clip]{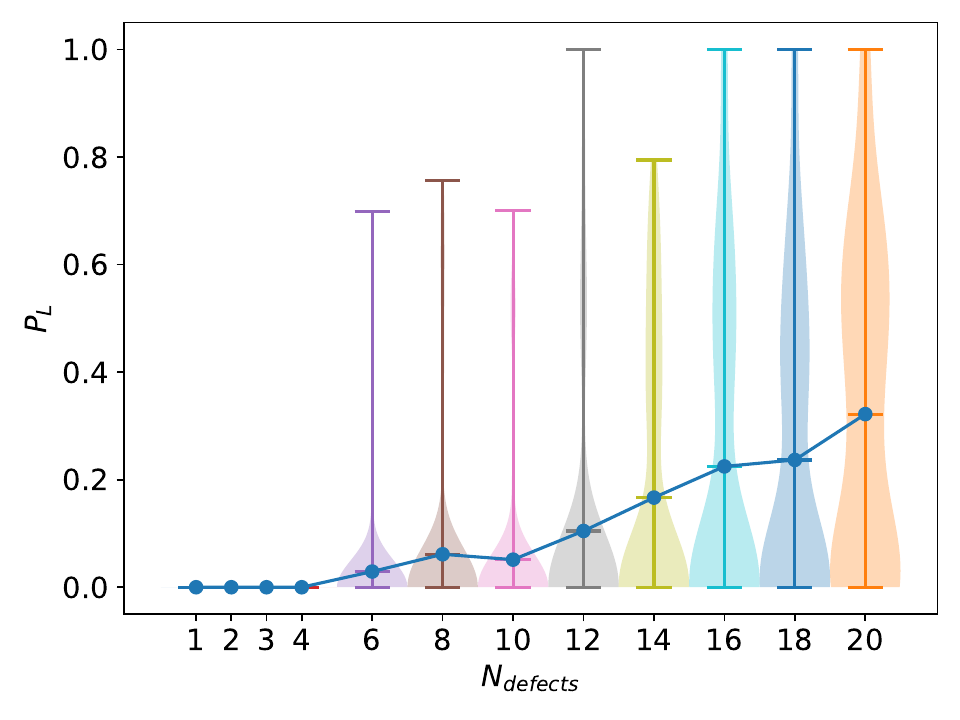}
    \caption{The loss probability of electron wave function with varying number of charge defects, $N_{defects}$, in the channel. Note that, for each number of charge defects, $100$ random configurations of charges were simulated. Up to $N_\text{\it defects}=4$, the loss probability remains nearly zero while we start to see cases with high loss probability with more than 6 defects present. The blue line connects the mean loss probability for each $N_\text{\it defects}$.}
    \label{fig:loss_prob_varying_number_defects}
\end{figure}

Figure~\ref{fig:loss_prob_varying_number_defects} shows the electron loss probability in the presence of varying number of charge defects. We considered a range of $N_\text{\it defects}$, the total number of charge defects, and for each case we simulated $100$ random configurations. For $4$ or fewer charge defects, the loss probability remains near to zero; but this probability climbs for higher defect counts. Notably, for as few as 6 defects, we did observe at least one case where the loss probability exceeds 50\% -- a catastrophic failure of the shuttling channel where the electron is likely to be ejected from the confinement region. 

To investigate further we explored `adversarial' scenarios where we seek the worst-case positioning for a small number of defect charge(s). We initially simulate scenarios with a single trapped negative electronic charge in the centre of the channel located either $2$\,nm or $5$\,nm away from the interface, with varying shuttling speeds. We also simulated cases where two and three trapped charges are aligned at a given x-coordinate, and so are liable to form a potential wall to repel the shuttled electron. In particular, two and three charges were positioned symmetrically around the centre axis of the channel, i.e. $y=0$, with the distance between two adjacent charges to be $1/3$ and $1/4$ of the full width of the channel ($100$\,nm), respectively.

\begin{figure}
	\subfloat[]{%
	\includegraphics[width=\linewidth]{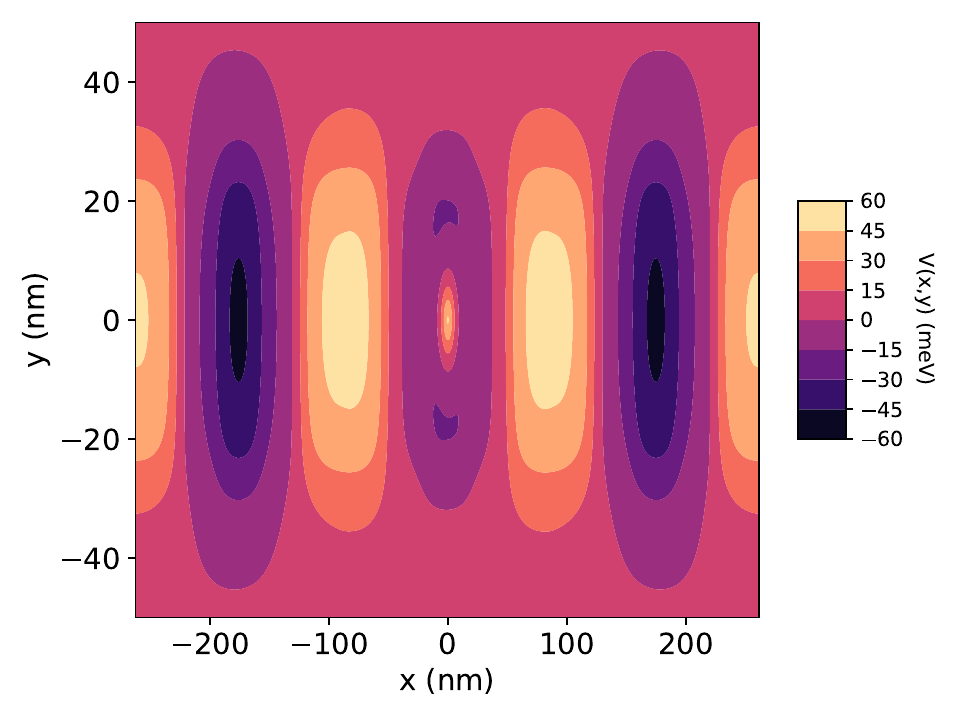}
        \label{fig:contour_plot_one_charge_defect}
}%
    \hfill
    \subfloat[]{%
    \includegraphics[width=\linewidth]{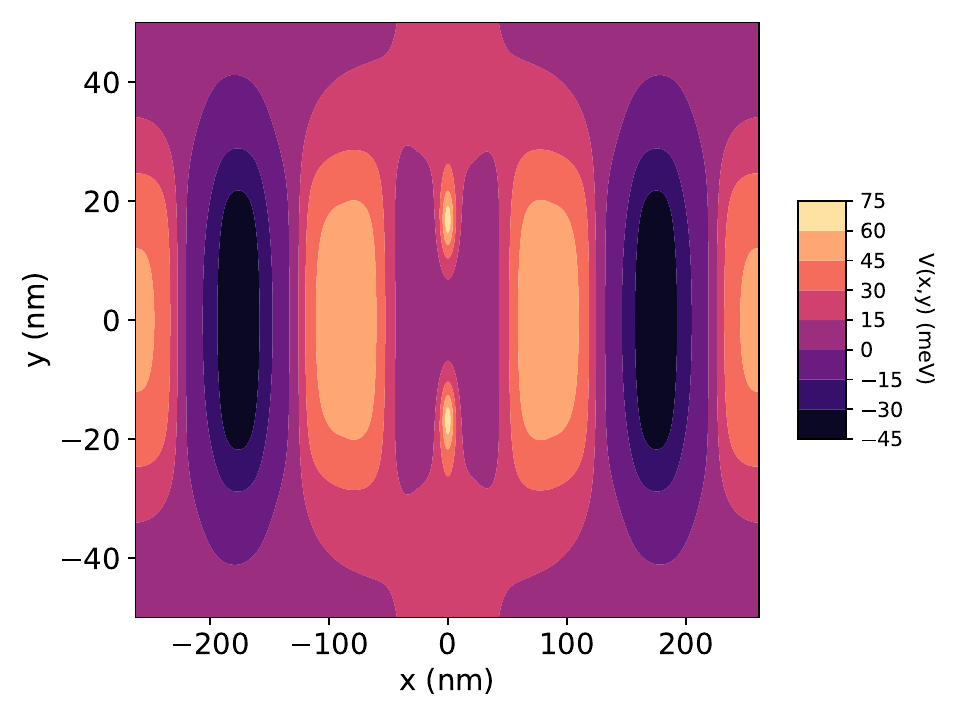}
        \label{fig:contour_plot_two_charge_defect}
    }%
    \hfill
    \subfloat[]{%
    \includegraphics[width=\linewidth]{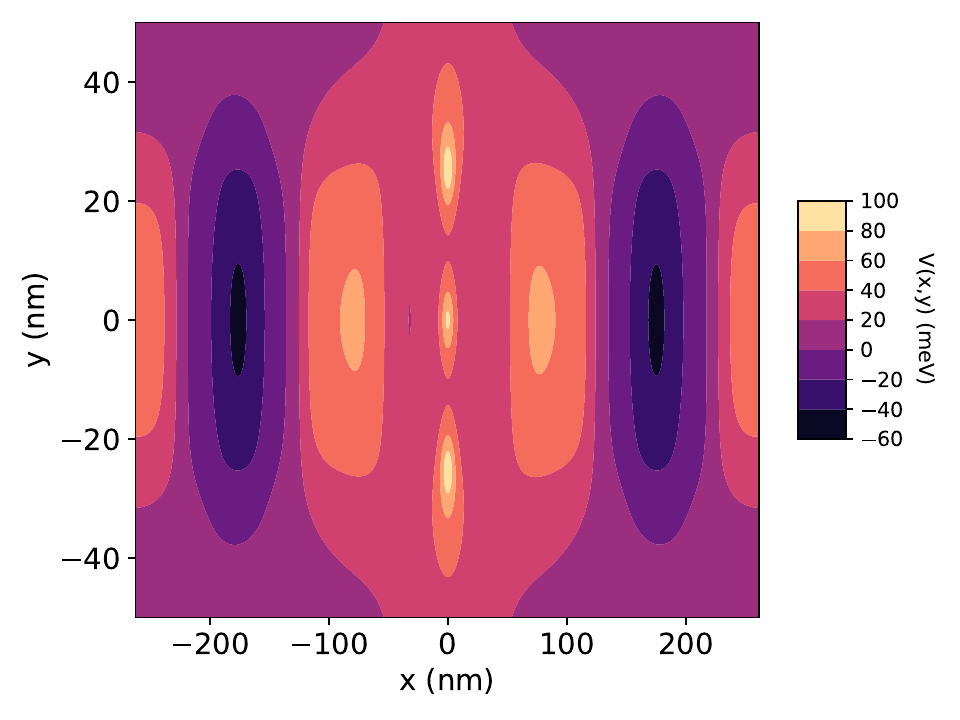}
        \label{fig:contour_plot_three_charge_defect}
    }%
    \caption{Contour plots of the potential energies with varying number of charge defects: (a) one, (b) two, and (c) three. Note that, for (b) and (c), the charges were aligned in the middle of the channel to form a wall to repel the electron. The distance from the interface to the charge defects was set to $5$\,nm.}
    \label{fig:countour_plots_charge_defects}
\end{figure}

Contour plots of the potential energies with one, two, and three charges placed at $5$\,nm above the interface are shown in Figure~\ref{fig:countour_plots_charge_defects} at a time where the phases of the gate voltages alone would produce a minimum in the potential energy at the charge location.

\begin{figure}[!ht]
	\subfloat[]{%
		\includegraphics[width=\linewidth]{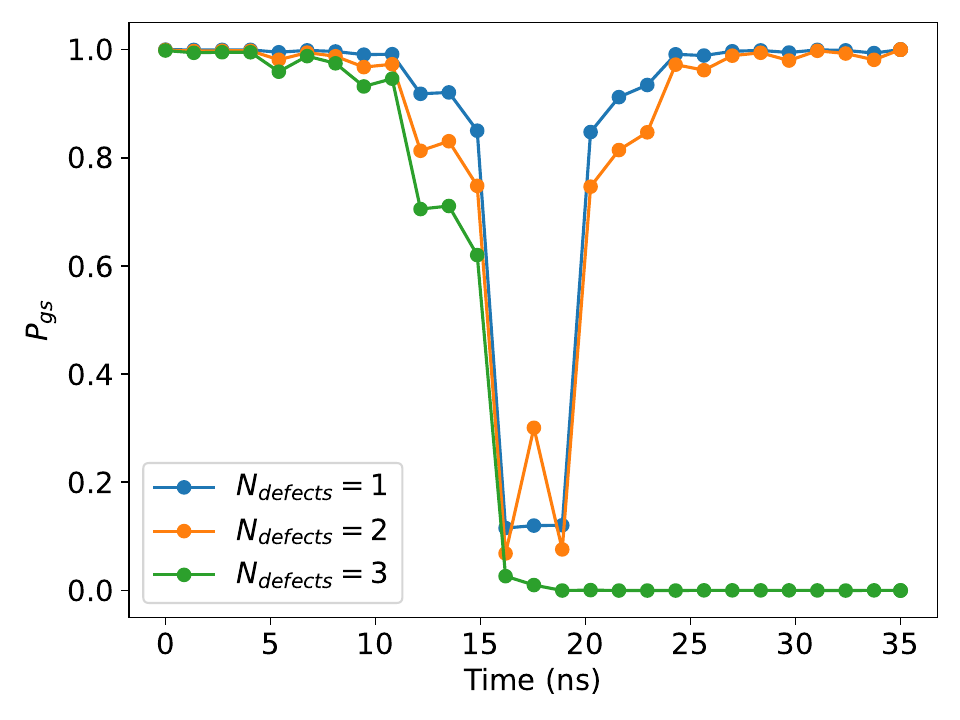}
        \label{fig:prob_gs_charge_defects}
	}%
    \hfill
    \subfloat[]{%
		\includegraphics[width=\linewidth]{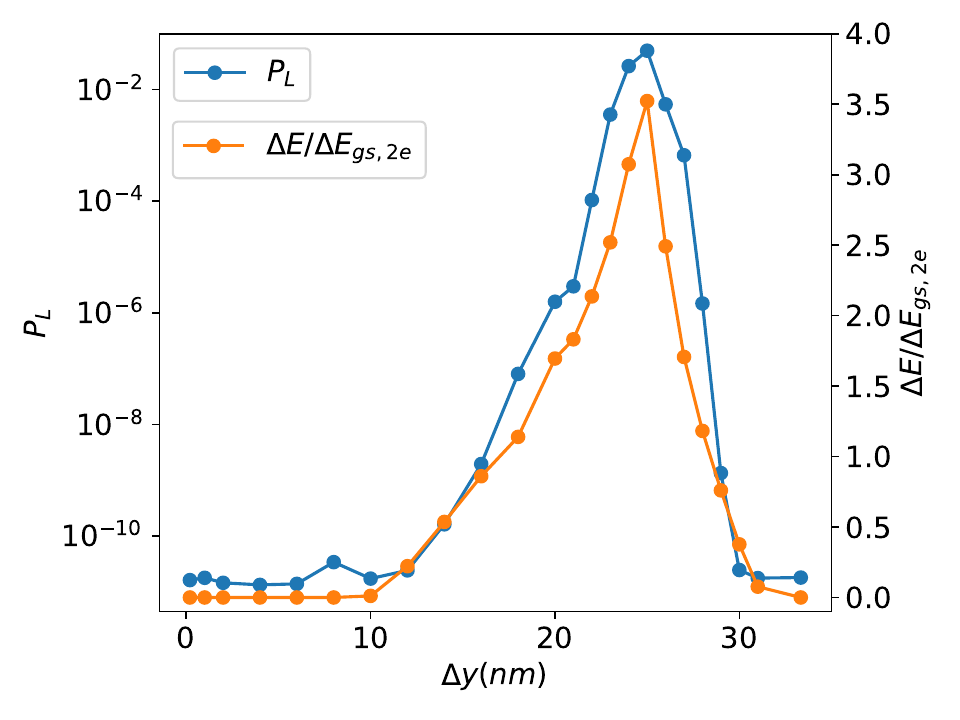}
        \label{fig:two_charge_defects_varying_distances}
	}%
    \caption{(a) The probability to remain in the instantaneous ground state of the potential formed by the gates alone for a shuttling speed of $10$\,m/s and varying numbers of charge defects, $N_{defects}$: one (blue), two (orange), and three (green). The electron is closest to the defects in the middle of the shuttling. Note that, for one and two charge defects, the probability decreases temporarily below $0.2$ as the electron passes the defects but rises back to approximately $1$; for three charge defects, shuttling fails as the probability decreases to almost zero at the end of shuttling.The distance from the interface to the charge defects was set to $5$\,nm. (b) The loss probability and excitation fraction of the electron wave function in the presence of two charge defects, separated by $\Delta y$, for a shuttling speed of $10$\,m/s. $\Delta y = 0$ corresponds to two charges at the same location, equivalent to one charge defect with twice the charge. As $\Delta y$ increases the peak loss probability and excitation fraction are about $0.04$ and 4 respectively. }
    \label{fig:prob_gs_123_and_two_charge_defects_varying_distances}
\end{figure}

Figure~\ref{fig:prob_gs_charge_defects} shows the probability of remaining in the ground state of the potential that would be formed by the gates alone (i.e., excluding the Coulomb potentials of the charge defects) when the shuttling speed is $10$\,m/s, and the shuttling distance is $350$\,nm (i.e., the length of two unit cells for $5$ electrodes). This means the electron was shuttled from one trough to the next, i.e. from one dark oval to the next in Figure~\ref{fig:countour_plots_charge_defects}. Thus, the electron is closest to the charge defects in the middle of the shuttling at around $17.5$\,ns. While the transfer is still almost adiabatic for one and two charge defects, for three defects the probability to remain in the instantaneous ground state of the gate potential drops to almost zero.  When the shuttled electron encounters the potential wall formed by the three charge defects, its wave functions becomes almost completely delocalized; it becomes unbound from its well in the shuttling potential. This is therefore a catastrophic failure of the shuttling process, and the device could not be used as a shuttling channel until/less the defect charges are moved.

It is unsurprising (indeed inevitable) that a sufficiently adversarial scenario involving multiple trapped charges will prevent shuttling. What is more remarkable is the protocol's robustness to the cases that might, intuitively, seem very problematic -- i.e. that it takes three trapped charges to `block' the channel with high probability. Figure~\ref{fig:prob_gs_charge_defects} shows that a single trapped charge or a pair of charges will imply a radical change to the instantaneous ground state, but the process can remain near-ideal. The contrast to the case of the three-charge `wall' is dramatic. 

\begin{figure}
    \centering
    \includegraphics[width=\textwidth]{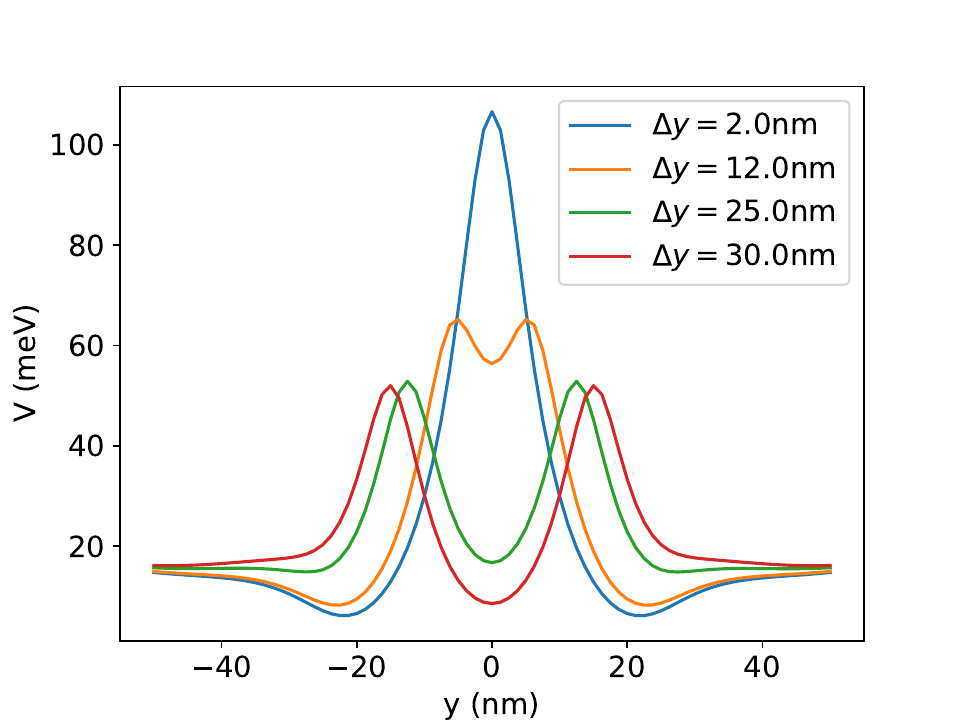}
    \caption{The cross-section of the potential in the $xz$ plane (at $x=0$\,nm in Figur~\ref{fig:contour_plot_two_charge_defect}) in the presence of two charge defects of varying separations $\Delta y= 2, 12 , 25, 30$\,nm. Each defect has charge of $-e$.}\label{fig:potential_cross_section_defects_varying_dist}
\end{figure}

We explored the worst case for the two-defect scenario. Figure~\ref{fig:two_charge_defects_varying_distances} shows the loss probability and excitation of the electron for varying defect separation, i.e. $\Delta y$. Both loss measures are at their most severe at about $\Delta y=25$\,nm. However, even at this point the loss is only $\approx4\%$; for separations outside a narrow $22$ to $27$nm range, the loss is again negligible. Figure~\ref{fig:potential_cross_section_defects_varying_dist} shows the cross-section of the potential on the $x=0$\,nm plane of figure \ref{fig:contour_plot_two_charge_defect} with varying defect separations. When the separation is small, e.g. $\Delta y=2$\,nm, the potential energy near the channel sides, e.g. $y \approx \pm 20$\,nm is low enough for the electron to move around the central barrier as in Figure~\ref{fig:two_charge_defects_varying_distances_states_2nm} in appendix~\ref{appendix:subsec:defects_varying_dist}. When the separation is large, e.g. $\Delta y = 30$\,nm, the potential energy in the middle is low enough for the electron to pass between the repulsive peaks as in Figure~\ref{fig:two_charge_defects_varying_distances_states_30nm} in appendix~\ref{appendix:subsec:defects_varying_dist}. However, at $\Delta y=25$\,nm, neither of these actions is easy: the local minima of potential energy($y=0,\pm ~ 25$\,nm) have roughly the same values as the potential energy at the channel edges ($y=\pm50$\,nm), and the electron requires higher energy to tunnel the barrier. Thus, Figure~\ref{fig:two_charge_defects_varying_distances_states_25nm} in appendix~\ref{appendix:subsec:defects_varying_dist} shows the high energy state of the electron tunneling through the barrier.

\begin{figure}
    \centering
    \includegraphics[width=\textwidth]{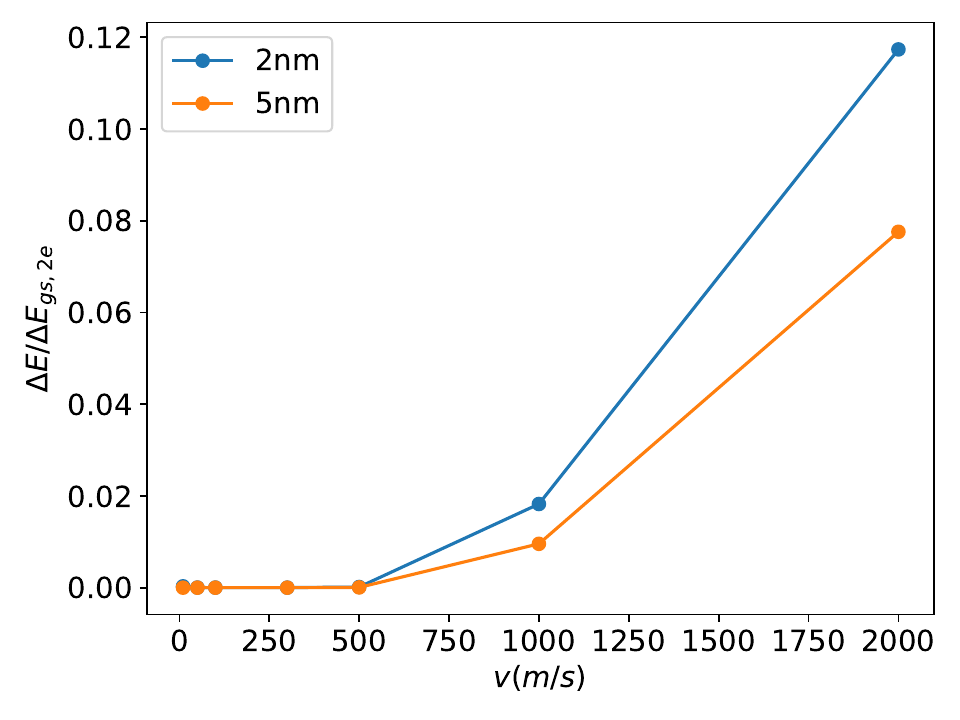}
    \caption{The excitation fraction with various shuttling speeds when one charge defect is present. The blue and orange lines correspond to excitation fraction when the charge defect is $2$\,nm and $5$\,nm away from the interface.}
    \label{fig:excitation_fraction_charge_defects}
\end{figure}

Our simulations also confirm that charge defects closer to the shuttling channel are more harmful to shuttling at varying shuttling speeds (see Figure~\ref{fig:excitation_fraction_charge_defects}).

\section{Advanced non-adiabatic ultra-fast shuttling} \label{sec:snap}

\begin{figure}
  \centering
  \includegraphics[trim={1.35cm 0 0 0}, clip, width=\linewidth]{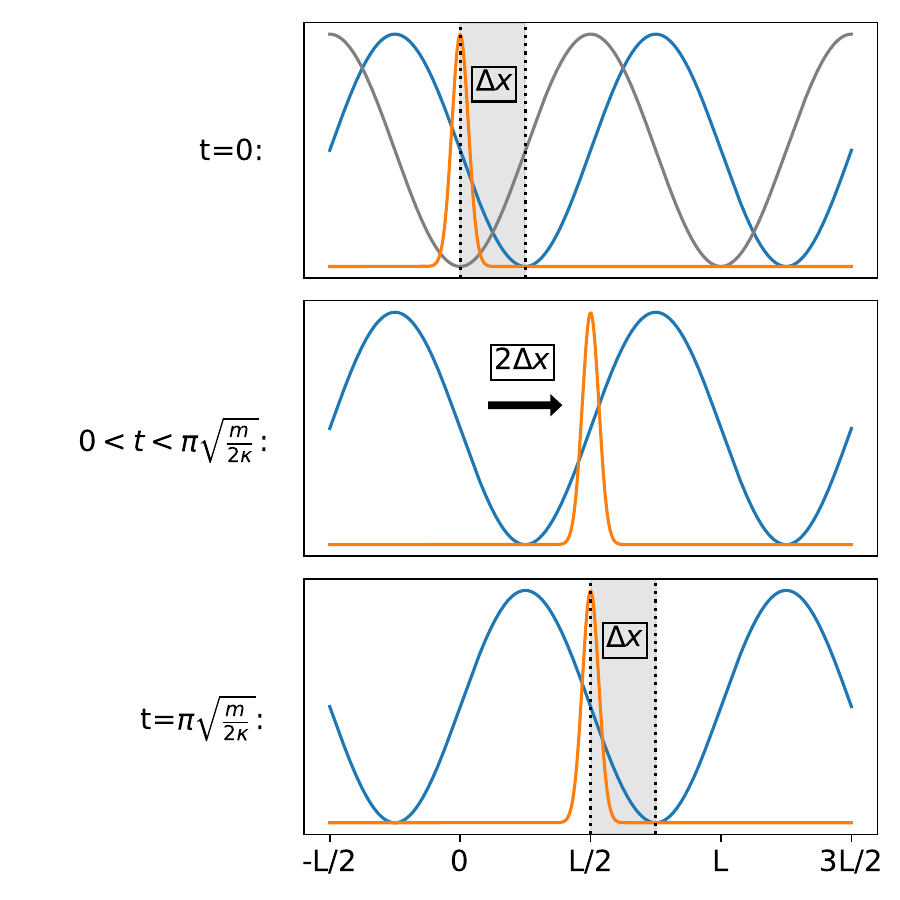}
  \caption{The illustration of snap method, where the blue line corresponds to the potential energy, the orange line corresponds to the probability density. Shaded regions were applied when instantaneous changes of the potential energy with displacement $\Delta x$ were made. The grey line at the $t=0$ panel represents the potential energy at $t < 0$. In the first panel, the potential is instantaneously shifted to the right by $\Delta x$. In the second panel, the potential is static until the wave function has evolved  to the other side of the well and comes to a halt. In the last panel, the potential is shifted again to the right by the same amount, $\Delta x$. This process continues until the target distance of shuttling is achieved. The time it takes for the electron wavefunction to curl up the other side of the well depends on the local curvature, $\kappa$, at the bottom of the well.}
  \label{fig:snap_method_illustration}
\end{figure}

Our analysis has confirmed the robustness of the conveyor-belt mode of shuttling: Over the range shuttling speeds that are likely to be targeted by near- or mid-term technologies, the method evidently excels. 

However, as quantum technologies mature it is possible that far greater shuttling speeds may be desirable. Therefore, in this final section we briefly explore the feasibility of an intentionally non-adiabatic shuttling method which could achieve extremely high shuttling rates, albeit with demands on voltage switching that are not practical at this time. 

In Appendix~\ref{appendix:subsec:step_changes}, we found that instantaneous changes in potential degrade the quality of shuttling. However, if we make such instantaneous changes at the right time, we can in principle perform shuttling with low loss and low probability of final excitation. The approach, which we informally call the `snap method', consists of four steps (see Figure~\ref{fig:snap_method_illustration}).
\bigskip
\begin{enumerate}
  \item Make an instantaneous change (or `snap') to the potential such that the minimum of potential energy is displaced to the right of the maximum of probability density by $\Delta x > 0$ (assuming that the shuttling direction is the $+x$ direction).
  \item Wait for the state to propagate across the potential well, climbing up the far side so that it effectively mirrors its initial position. The total distance travelled by the wave function is $2\Delta x$. The time taken can be approximated by $\Delta t=\pi \sqrt{m/2 \kappa}$, where $\kappa$ is the instantaneous local curvature at the bottom of the well (assumed approximately harmonic).
  \item Repeat 1 and 2 until the electron approaches the target position.
  \item Once the state approaches the target position, displace the potential such that the electron will have zero momentum at the target location; finally, displace the potential such that the minimum of the potential energy is at the target position. The shuttling is complete.
\end{enumerate}

Using the same numerical model, we simulated various scenarios of the snap method at different depths of the \ce{Si} layer. At the depth of $30$\,nm below the bottom of the clavier gates, we found the loss probability as low as $10^{-6}$ at the shuttling speed of $500$\,m/s with the excitation fraction of $2\times10^{-3}$. While the results suggest possibility of achieving full non-adiabatic transport with low loss and excitation, the method faces various challenges. For example, making instantaneous changes of potential is bounded by the maximum rate of change of voltage at the gates, which is around $14$\,mV/ps in current technology. Furthermore, the presence of charge defects will change the optimal timings of the instantaneous changes of the potential. Thus, we leave this fully non-adiabatic method as a future investigation: The results and detailed analysis of the snap method can be found in Appendix~\ref{appendix:sec:snap}.

\section{\label{sec: implication_spin}Implications for coherent qubit transport}

Our numerical studies have allowed us to analyse the orbital state of the electron in various shuttling scenarios. The loss probability measures how likely it is that the scheme will transport the electron to a target position, while the excitation fraction measures the adiabaticity of the overall shuttling process. In this section we consider the implications for the spin, i.e. the degree of freedom representing the qubit. 

The arguments and analysis in the study by Langrock et al.\cite{langrock_2023} are very relevant to the present section. There, the author's discusses various spin dephasing mechanisms. Spin dephasing still occurs when the shuttling is completely adiabatic in the spatial sector, because of variations in the local Zeeman splitting due to nuclear Overhauser fields and $1/f$ charge noise (which causes fluctuations in the local g-factor through spin-orbit interaction). While this effect can be mitigated by moving the electrons more quickly, averaging the local variations and leading to motional narrowing of the Zeeman splitting, fast movement makes shuttling non-adiabatic.

There are also potential sources of non-adiabaticity in shuttling: orbital excitation, both from the motion of electron and from electrostatic disorder in the QD potential.  Such disorder can arise, for example, due to Ohmic heating or charge defects at the  \ce{Si}--\ce{SiO2} interface. Furthermore, atomic-scale interface roughness makes the valley splitting in silicon, and the constitution of valley states, position-dependent. Shuttling the electron, so that it experiences different interface structure, therefore causes excitation in the valley degree of freedom. Such excitation into excited orbital and valley states gives rise to random phonon relaxations leading to a distribution of time spent in  the excited orbital and valley states. Because of spin-orbit coupling, the g-factor depends on the orbital and valley states, and the randomness of relaxation therefore becomes a source of spin dephasing.

The excitation fraction used in our paper directly measures the amount of orbital excitation arising from the motion of the electron. For realistic parameters of $A=50$\,mV and $T=2$\,K, Figure~\ref{fig:excitation_fraction_quantum_noise_temperature} suggests that the excitation in the orbital degree of freedom is minimal,  $\Delta E/\Delta E_{gs,2e} \sim 10^{-7}$. (Were it necessary, the Johnson-Nyquist noise can be further suppressed by using bondpads with higher capacitance and/or bondwires with higher resistance, at the cost of slowing down the control of the qubits). Thus, in the scenarios we have explored it is likely that the spin dephasing directly due to orbital excitation is minimal. Moreover, Figure~\ref{fig:infidelity_exc_frac_along_distance} shows the probability of excitation outside of the ground state throughout the shuttling for a target distance of $1.4$\,$\mu$m at $10$\,m/s. During the shuttling, the probability of excitation was in the order of $10^{-6}$ to $10^{-5}$, the excitation fraction was in the order of $10^{-7}$ to $10^{-6}$ suggesting that the entire process of shuttling is largely adiabatic.  This supports the claim that phonon relaxation from spatial excitation is unlikely to occur.

Spin dephasing can also occur via spin relaxation due to the motion of a QD\cite{Huang_2013}, whose rate is inversely proportional to the fourth power of the characteristic energy gap\cite{langrock_2023}, which is about $6$\,meV in our case for $N=4$ and $A=100$\,meV. Note that the spin relaxation rate is also proportional to the square of contribution of the potential from electrostatic disorder, $\delta V$, and inversely proportional to the third power of the correlation length $l^{\delta V}_{c}$\cite{langrock_2023}. Since the charge defects are closer to the shuttled electron in SiMOS than in \ce{Si}/\ce{SiGe} devices, the electrostatic disorder, $\delta V$, becomes bigger. Thus, there is a competition between the larger orbital splitting and stronger electrostatic disorder. Following the arguments of Langrock et al.\cite{langrock_2023}, we estimate the probability of a spin flip to be around $ 10^{-5}$ with the following parameters: energy gap $6$\,meV, $\delta V = 73.85$\,meV (the Coulomb potential of a charged defect at a distance of $5$\,nm), correlation length $l^{\delta V}_{c}=100$\,nm, shuttling speed $10$\,m/s and shuttling distance $1.4$\,$\mu$m. The probability of a spin flip is therefore negligible. However, our assumed correlation length, ($100$\,nm, following Langrock et al.\cite{langrock_2023}) may be an overestimate as the charge defects are closer to the interface in SiMOS than in \ce{Si}/\ce{SiGe}, and the spin relaxation may therefore be underestimated. Further calculations are needed to make a better estimate of spin relaxation in SiMOS; we leave this to future work, as we focus on the modelling of charge shuttling in \ce{Si}/\ce{SiO2}.


Our results agree with the one of the conclusions made by Langrock et al.\cite{langrock_2023}: spin dephasing is not significantly affected by the non-adiabatic effects in orbital degree of freedom. By approximating the channel as 1D, Langrock et al. calculated the excitation rate to the first excited state in the presence of electrostatic disorder and showed that the rate is suppressed by a Gaussian factor at low speeds, i.e. $v \ll 10^{-4}$\,m/s. This is in line with our results for the excitation fraction, which is no more than $3 \times 10^{-2}$ in the worst case of Johnson-Nyquist noise. Furthermore, Langrock et al. showed that the phonon relaxation is fast enough for the orbital state to relax without significant spin dephasing. Using realistic parameters, Langrock et al. estimated the amount of phase error due to random phonon relaxations during the entire shuttling, which is orders of magnitude smaller than the threshold error of $10^{-3}$.

We do observe more strongly non-adiabatic behaviours at higher shuttling speeds; for example peaks in loss probability and excitation fraction up to around $10^{-3}$ are observed at distances around $0.45$\,$\mu$m for shuttling speeds of $300\,\mathrm{ms}^{-1}$ (see Figure~\ref{fig:infidelity_exc_frac_against_gs_along_distance_all_speeds} in Appendix \ref{appendix:subsec:speed_adiabaticity}.).

\section{\label{sec: conclusions}Conclusions}

We investigated the feasibility of conveyor-belt shuttling of an electron's orbital state through numerical simulations. We have captured  the essential physics of a SiMOS shuttling device with a 3D solution of the Poisson equation but a 2D simulation of the electron wave-packet propagation near the Si/SiO$_2$ interface. We use periodic boundary conditions for both potential and wave-function along the row of `clavier gates' implementing the shuttling protocol, but force the electron wave function to vanish at a point within the confinement gates forming the sides of the channel. 
We introduced two important metrics in section \ref{sec: metrics} to evaluate the shuttling scenarios: the loss probability (effectively quantifying the failure of the electron to arrive at the desired location) and the excitation fraction (quantifying the non-adiabaticity of the shuttling). 

Shuttling in the absence of noise was described in section \ref{sec: noiseless_shuttling}. For target distance varying from $140$\,nm to $8.4$\,$\mu$m, we observed loss probabilities of the order of $10^{-10}$ or below and excitation fraction in the order of $10^{-7}$. Using three gates per unit cell rather than four, as previously proposed, would be enough to achieve this (albeit using 4 or more gates does lead to even more nearly-perfect operation).

We also simulated shuttling scenarios in the presence of Johnson-Nyquist noise. While the power spectral density of the classical Johnson-Nyquist is only valid when the energy corresponding to cut-off frequency, $\hbar \gamma$, is smaller than the energy of thermal excitation, $k_{B}T$, we tried both cases to clarify how much the quantum effect at low temperature and high frequency benefits the shuttling. 

The results of classical Johnson-Nyquist noise are given in Appendix~\ref{appendix:subsec:johnson-nyquist_classical}. The system becomes more resilient to the noise as the number of gates per unit cell increases. We found that high frequency noise, especially the one that is comparable to the frequency corresponding to the characteristic energy gap, i.e. $\Delta E_{gs,2e/\hbar}$, is more harmful than low frequency noise. The same behaviour is observed for the excitation fraction in Figure~\ref{fig:excitation_fraction_temp_classical}.

In section \ref{sec:johnson-nyquist}, the results of Johnson-Nyquist noise with the quantum correction factor are given. We found the loss probability and excitation fraction are greatly suppressed by quantum effects that reduce the high-frequency noise at low temperature as shown in Figure~\ref{fig:PSD_Johnson_Nyquist}. We find that the most important part of the noise is that occurring at the orbital excitation frequency in the direction of the shuttling; once this frequency is above the thermal frequency, changes in temperature have a limited effect. We conclude that the Johnson-Nyquist noise can be greatly suppressed by operating at low temperatures or with a low cut-off frequency, which is achieved by having bondwires with a large resistance, $R$ (given a fixed gate capacitance, $C_{2}$) or bondpads with high capacitance, $C_{1}$.  However, this will also limit the ability to produce fast voltage variations.

The effect of negative trapped charges near the interface on the shuttling was investigated, and the results are given in section \ref{sec:charge_defects}. We first looked into varying number of charge defects and found that loss probability was near zero up to four charge defects at random positions. To gain more insight, we looked into `adversarial' scenarios with a single trapped charge, two charges, or three charges positioned symmetrically around the centre of the channel. In the case of two and three charges, the Coulomb peaks can form a potential wall that repels the electron. We showed that the electron wave function was easily completely delocalized in the presence of three such trapped charges while the shuttling was remarkably successful in the presence of one and two trapped charges (with a small risk of loss in the two-defect case) . Furthermore, we confirmed that a shorter distance from the trapped charges from the interface and higher shuttling speeds harmed the shuttling more, as expected from the adiabatic theorem.

In section \ref{sec:snap}, we proposed a new non-adiabatic shuttling method, which allows fast shuttling with low loss probability. Despite our observation (see Appendix~\ref{appendix:subsec:step_changes}) that instantaneous changes of the potential are detrimental to conventional adiabatic shuttling, we leverage the idea that {\it properly timed} instantaneous changes can drive a coherent shuttling process. At the appropriate instant, the electron's wavefunction is a well-behaved coherent state in both the `before' and `after' potentials. The interval between the sudden changes, hence the shuttling speed, depends potential curvature near the minimum and the number of instantaneous updates per unit cell as shown in Figure~\ref{fig:average_speed_snap}. We found that there is a trade-off with this method as a function of depth below the electrodes: for deeper shuttling, the quantum dot becomes shallower, but also more nearly harmonic.  For our reference voltage pulse amplitude (100\,mV) we find the optimum depth is around 30\,nm.

In section \ref{sec: implication_spin}, we assess the impact of our results on the fidelity of spin transport. We find that the effect of random phonon-relaxation due to excitation of the orbital state would be minimal for the conveyor-belt shuttling.

Overall, we conclude that the Conveyor-belt shuttling is so adiabatic that the excitation fraction is smaller than $10^{-3}$ and loss probability is smaller than roughly $5 \times 10^{-9}$ even if we have only three gates per unit cell, in a reasonable temperature range $T \lesssim 4$\,K, at shuttling speeds up to $500$\,m/s, and in the amplitude range of $50$\,mV to $100$\,mV, in the presence of quantum Johnson-Nyquist noise. However, charge defects due to impurities can damage the shuttling if such defects form a repulsive potential wall in the middle of channel and delocalize the electron wave function.

Our approach is complementary to that of Langrock et al.\cite{langrock_2023} who solved Poisson-Schr\"{o}dinger equations to see the formation of quantum dots in the presence of charge defects, whereas we modelled the effect of charge defects by doing explicit time-dependent simulations of electron wave function on these potentials. While the \ce{Si}/\ce{SiGe} structure considered by Langrock et al. has the charge defect plane far apart from the shuttling channel (around $45$\,nm), the \ce{Si}/\ce{SiO2} structure considered in this paper is affected by charge defects more severely as the charge defect plane is closer to the channel ($5$\,nm in our typical simulations). As a topic for further study, we expect that $3$D simulations may result in improved loss probability and excitation fraction compared to the $2$D simulations performed in this paper as wave function has one more spatial direction to circumvent the potential barrier.

Similar charge modelling of bucket-brigade shuttling was performed by Buonacorsi et al.\cite{buonacorsi_2020} and Krzywda et al.\cite{krzywda2024decoherence}. Buonacorsi et al.\cite{buonacorsi_2020} reported the infidelity between the final state of shuttling and orbital ground state as low as $10^{-5}$ for a dot-to-dot transfer, and Krzydwa reported the charge transfer error as below as $10^{-7}$ for a dot-to-dot transfer. While the charge shuttling with bucket-brigade was shown to be still reliable for short distances of a few interdot distances, roughly a few $100$\,nm, shuttling of longer distances, e.g. a few micrometres, has not been simulated: Furthermore, bucket-brigade suffers from reversal of the shuttling direction once the dot-to-dot transfer fails. In contrast, we simulated conveyor-belt shuttling in the presence of Johnson-Nyquist noise for a distance of $1.4$\,$\mu$m and found nearly perfect loss probability below $5 \times 10^{-9}$ and excitation fraction below $10^{-3}$. However, we do acknowledge that charge defects can hinder the reliable conveyor-belt shuttling of electron in some adversarial cases.

Further investigation for the charge modelling could be made to find the effect of trapped charges randomly placed near the interface due to impurities. Even though we looked into the worst case of three charges forming a repulsive potential wall, we have not yet determined the statistical probability that a given charge {\it density} will disable a shuttling channel of a given length. It may also be interesting to ask whether any modification of the shuttling protocol can improve the robustness versus the charge defect environment. Finally, it would be helpful to investigate the effect of positive charge defects near the interface. Positive charge defects may trap the shuttled electron and form a bound state.

In this paper we have explored the dynamics of charge shuttling directly through numerically-intensive granular grid-based modelling of the 2D wavefuction, and then argued how the observed behaviour can be expected to affect qubit integrity. The natural next stage for a further study is to equip the model with spin and valley degrees of freedom so as to observe the qubit's evolution directly. For this to be meaningful, numerical models of state/position-dependent g-factor and atomic scale interface roughness are necessary and this is an exciting challenge.  

\acknowledgments{We give a special thanks to Christian Binder, Hamza Jnane, Richard Meister, James Williams, Tyson Jones, and Guido Burkard for valuable discussions.The authors would like to acknowledge the use of the University of Oxford Advanced Research Computing (ARC) facility \cite{Richards2015-fx}  in carrying out this work and specifically the facilities made available from the EPSRC QCS Hub grant (agreement No. EP/T001062/1). The authors also acknowledge support from EPSRC’s Robust and Reliable Quantum Computing (RoaRQ) project (EP/W032635/1), and the SEEQA project (EP/Y004655/1). 
 }

\appendix

\section{Details of Conveyor-Belt Shuttling}\label{appendix:details_CB_shuttling}

\subsection{Energy Evolution}\label{appendix:subsec:energy_evolution}

\begin{figure}
\centering
\includegraphics[width=\linewidth]{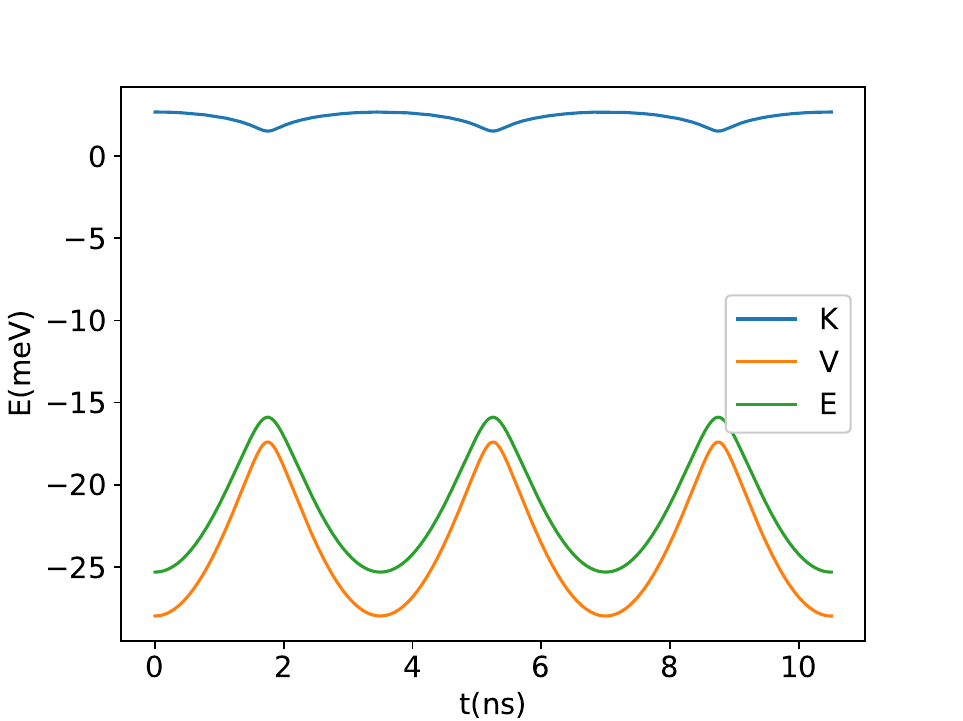}
\caption{For $N=3$, the kinetic energy (blue), the potential energy (orange), and the total energy (green) of electron during conveyor-belt shuttling for a distance of one unit cell, i.e. $105$\,nm for $N=3$. with linearly varying phase.}
\label{fig:energy_vs_time_cb_shuttling}
\end{figure}

Figure~\ref{fig:energy_vs_time_cb_shuttling} shows the evolution of kinetic, potential, and total energy when the electron is shuttled over one unit cell length for $3$ electrodes per unit cell. The energy curves are periodic with the period $L/(Nv)$, where $L$ is the length of the unit cell, $v$ is the shuttling speed, and $N$ is the number of gates per cell: this is the time taken for the electron to travel from one gate to the next.

The change of local potential energy results from the change of local curvature as shown in Figure~\ref{fig:potential_max_curvature}. The curvature a maximum when the electron is directly underneath one of the gates and a minimum when the electron is beneath the gap between two gates.

\subsection{Two Possible Phase Variations} \label{appendix:subsec:phase_variation}

\begin{figure}
	\subfloat[]{%
 	\includegraphics[width=\linewidth]{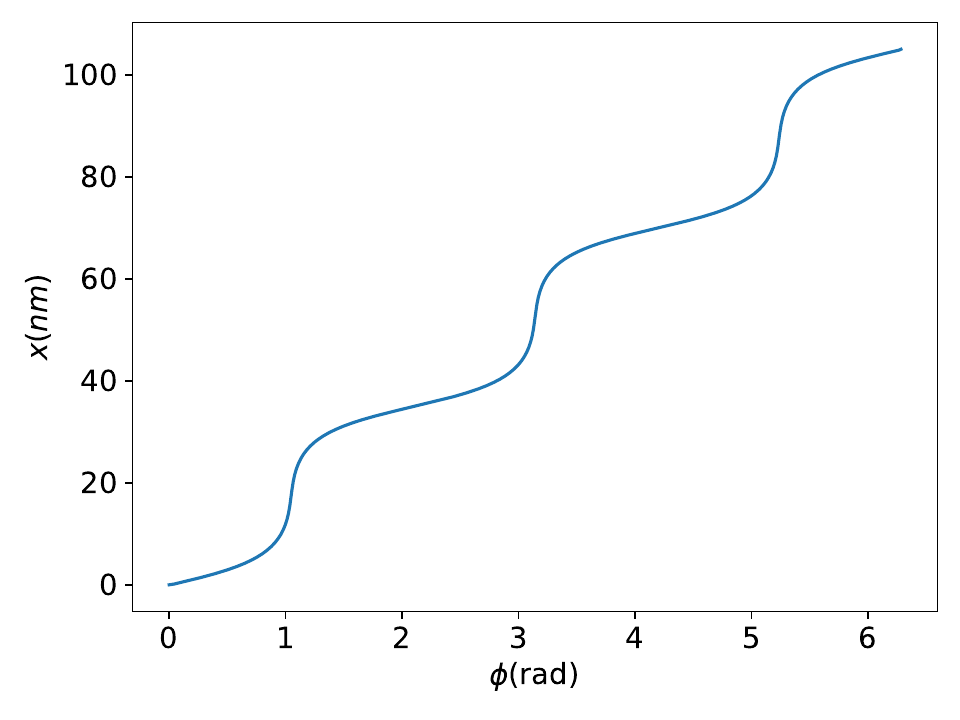}
  	\label{fig:potential_max_x_phi}
	}%
	\hfill
	\subfloat[]{%
	\includegraphics[width=\linewidth]{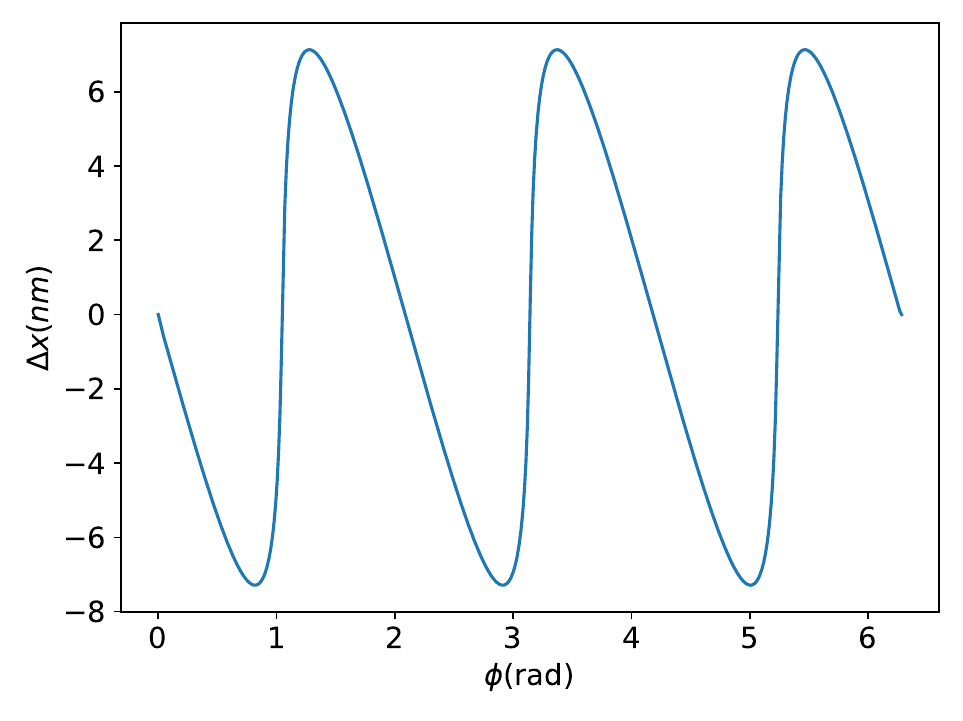}
  	\label{fig:potential_max_x_perturbation}
	}%
	\hfill
	\subfloat[]{%
		\includegraphics[width=\linewidth]{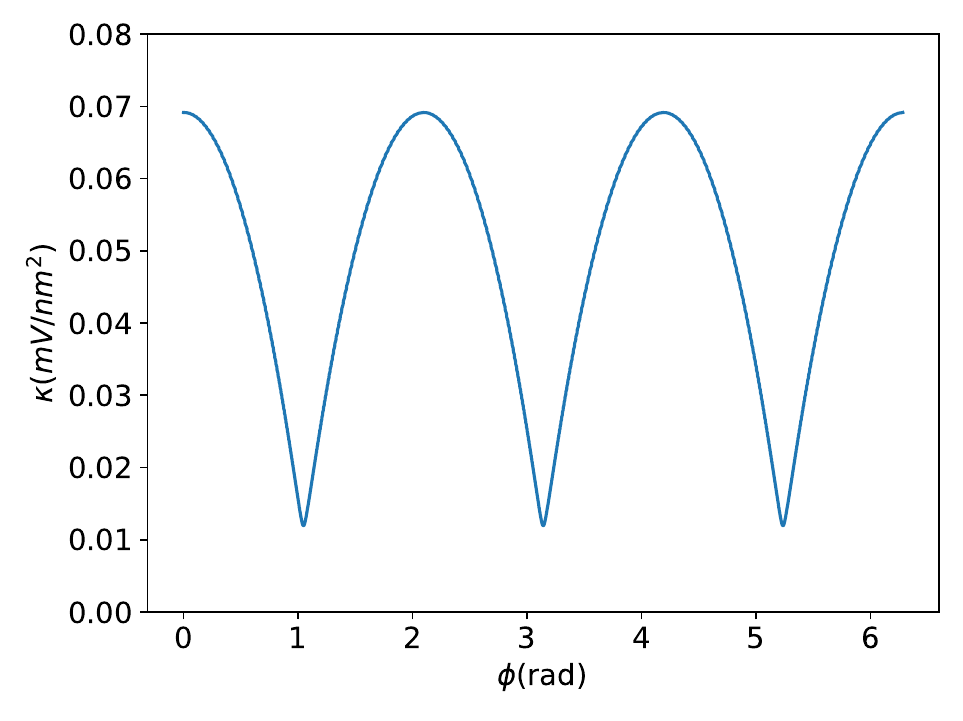}
  	\label{fig:potential_max_x_perturbation_phi}
	}%
\caption{(a) The position, $x(\phi)$, (b) the perturbation of the position, and (c) the local curvature of the potential energy minimum, $\Delta x(\phi)$, was obtained by excluding the linear function from the $x(\phi)$ in Figure~\ref{fig:potential_max_x}. As the electron is shuttled for a distance of one unit cell, there are $N$ repeating patterns, where $N=3$ in this figure. The position curve in (a) and the curvature curve in (b) are the same as the position and curvature curve in Figure~\ref{fig:potential_max_x} as a function of the phase, $\phi$, instead of time.}
\label{fig:potential_max_x_and_x_perturbation_phi}
\end{figure}

As mentioned in \ref{sec:conveyor_belt_shuttling}, the trajectory and speed of shuttling depend on how we vary the phase $\phi(t)$ of the sinusoidal pulses in equation \ref{eqn:sinusoidal_voltage_pulses}.

The simplest way is to update the phase linearly with time, i.e. $\phi(t) = k\,t$, where $k$ is the rate of change of phase. The periods of the sinusoidal pulses in the time domain are then $2\pi/k$ and the average shuttling speed is $v_{avg} = k \, \frac{L}{2\pi}$, where $L$ is the length of the unit cell. 

Nevertheless, the instantaneous speed is not constant, as the position of the minimum of the potential energy does not depend linearly on the phase but instead varies as shown in Figure~\ref{fig:potential_max_x_phi}, which can be further decomposed as a sum of a linear function and a periodic perturbation (with period $2\pi/N$ for $N$ electrodes) in Figure~\ref{fig:potential_max_x_perturbation}. 

We can use a non-linear phase function $\phi(t)$ to shuttle the electron along an arbitrary trajectory, $x(t)$. Suppose that $f$ is a function mapping the phase to the position of the minimum of potential energy, i.e. $x=f(\phi)$  with its shape presented in Figure~\ref{fig:potential_max_x_phi}. For any position $x$, one can use the inverse function $f^{-1}$ as a look-up table to find the corresponding phase; thus, given an arbitrary trajectory $x(t)$ one can  find the corresponding phase at each time step $\phi(t)=f^{-1}(x(t))$. Here, we implement this approach for shuttling with constant speed: $x(t) = v\, t$.

\begin{figure}
    \centering
    \includegraphics[width=\textwidth]{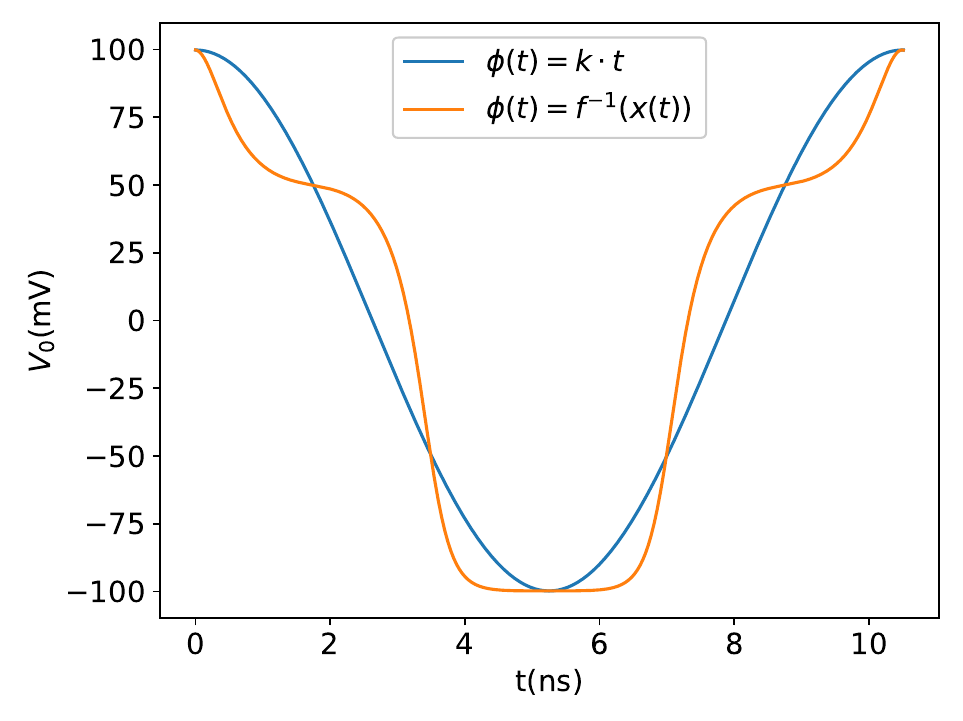}
    \caption{An example of voltage signal on the 1st gate in the unit cell for different choices of phase variation $\phi(t)$, i.e. linearly increasing phase, and phase generated by the look-up table(orange line). This is the same as the inset of Figure~\ref{fig:potential_max_x}}
    \label{fig:voltage_signals_look_up_table}
\end{figure}

Figure~\ref{fig:voltage_signals_look_up_table} shows the voltage signal at the first gate in the unit cell for the two phase variations $\phi(t)$: the blue line represents linear phase variation and the orange line represents shuttling at a strictly constant speed (with $v_{avg} = 10$\,m/s in both cases). Signals can also be generated for more general $x(t)$, such as trajectories involving either constant or time-dependent accelerations. 

For the noiseless shuttling up to the distance of $8.4$\,$\mu$m ($A=100$\,mV, $v=10$\,m/s), both phase variation methods yielded loss probability below $1.5 \times 10^{-11}$ and excitation fraction below $5 \times 10^{-7}$. Furthermore, we found no qualitative difference in these important metrics and no clear trend as to which profile gives better results.
Since it is easier to generate sinusoidal pulses than more complicated pulses with several frequency components, we chose the linearly varying phase as the default for the rest of the paper.

\subsection{\label{appendix:subsec:airy_function} The Effect of Finite Extent of the QD in the perpendicular direction}

Our shuttling model is perfectly two-dimensional and therefore ignores the finite extent of the quantum dots in the direction perpendicular to the interface.  In this section we explore the averaging of the electrostatic potential as a result of this finite thickness.

\begin{figure}
  \centering
  \includegraphics[width=\textwidth]{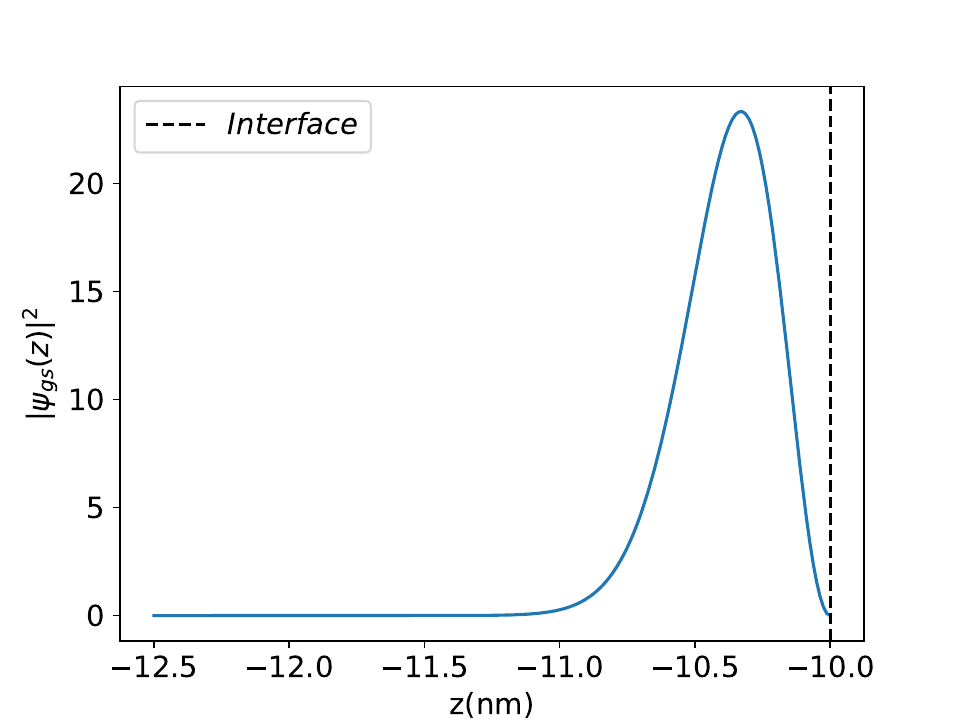}
  \caption{Probability density of the ground state in the z-direction (blue line). Probability density exists only in $z< -10$\,nm, where there is a silicon layer. The probability density tails off at $-11$\,nm, which makes the confinement length $1$\,nm below the interface (dotted line).}
  \label{fig:prob_gs_z}
\end{figure}

\begin{figure}
\centering
\subfloat[]{%
\includegraphics[width=\linewidth]{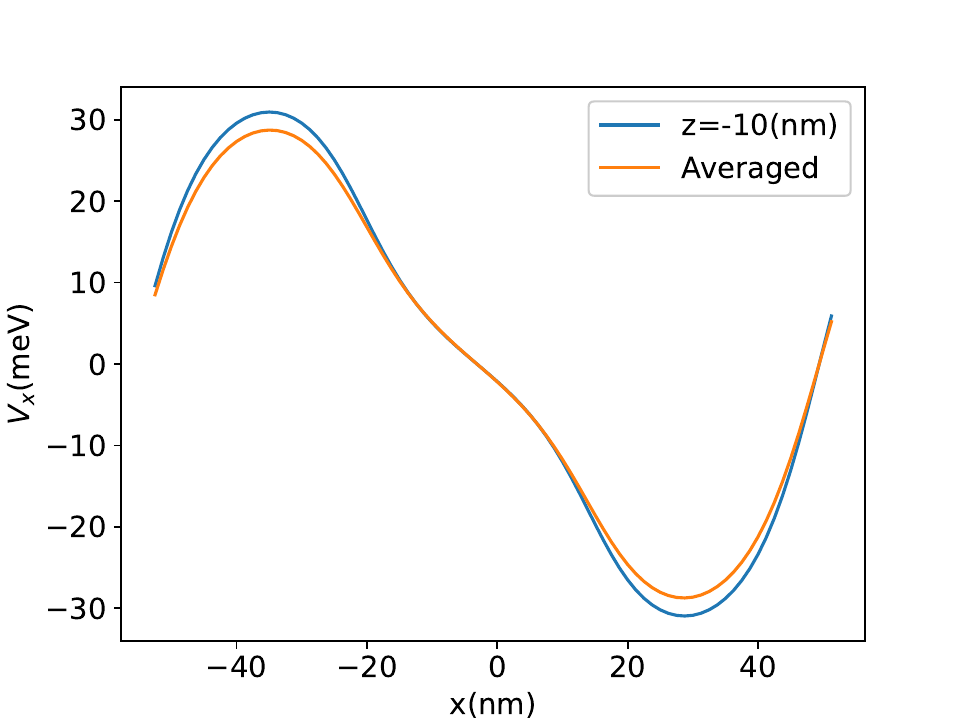}
  \label{fig:V_x_airy}}%
\hfill
\subfloat[]{%
\includegraphics[width=\linewidth]{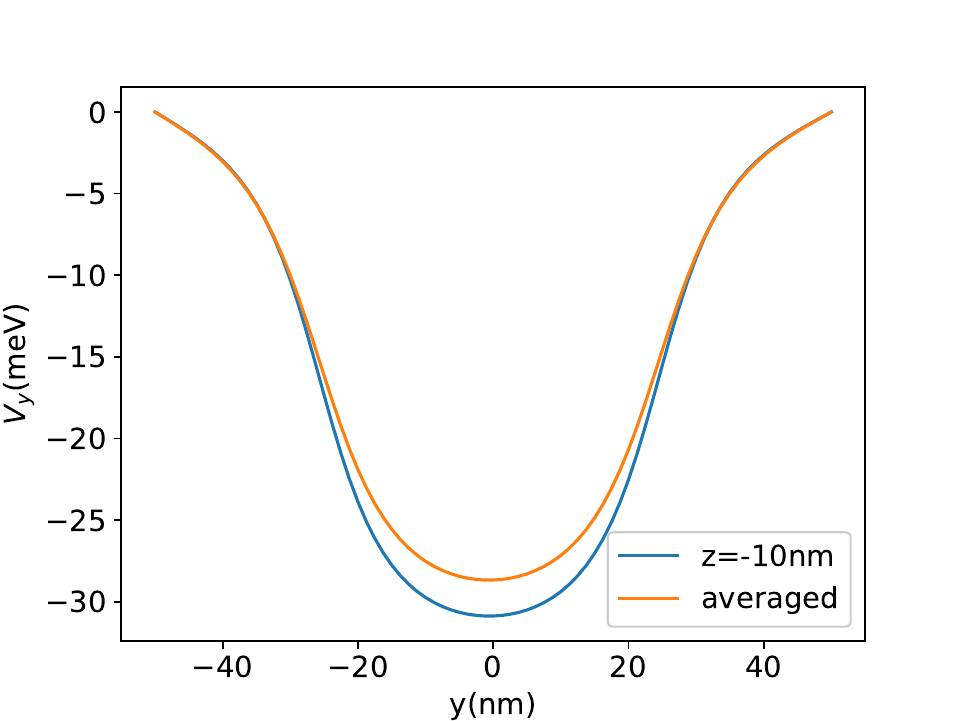}
  \label{fig:V_y_airy}
}%
\caption{Cross-section of potentials on (a) the $y=0\,\text{nm}$ and (b) $x=27.5\,\text{nm}$ planes for the potential sampled at $z=-10$\,nm (blue line) and the potential averaged over the probability density of the ground state in the z-direction for the thickness of $1$\,nm (orange line).}
\label{fig:V_x_y_airy}
\end{figure}

Figure~\ref{fig:V_x_y_airy} shows the spatial variation of two potentials, both along the channel and across it.  One potential is sampled directly at the \ce{Si}/\ce{SiO2} interface and the other is obtained by averaging over the probability density of the ground state in the z-direction, assuming that the confinement length of the QD is $1$\,nm, using an Airy function of the first kind, truncated up to the last x-intercept, as the ground state as shown in Figure~\ref{fig:prob_gs_z}.

Note that the difference between the depths of the potential is about $7$\,\% of the well depth of potential, and the effect of the averaging is similar to a rescaling of the gate voltages at the gates. Given these relatively small differences, we  use the potential sampled directly at the interface in the remainder of the paper.

\subsection{\label{appendix:subsec:1d_vs_2d} The Comparison of 1D and 2D Simulations for the Shuttling}

\begin{figure}
	\subfloat[]{%
	\includegraphics[width=\linewidth]{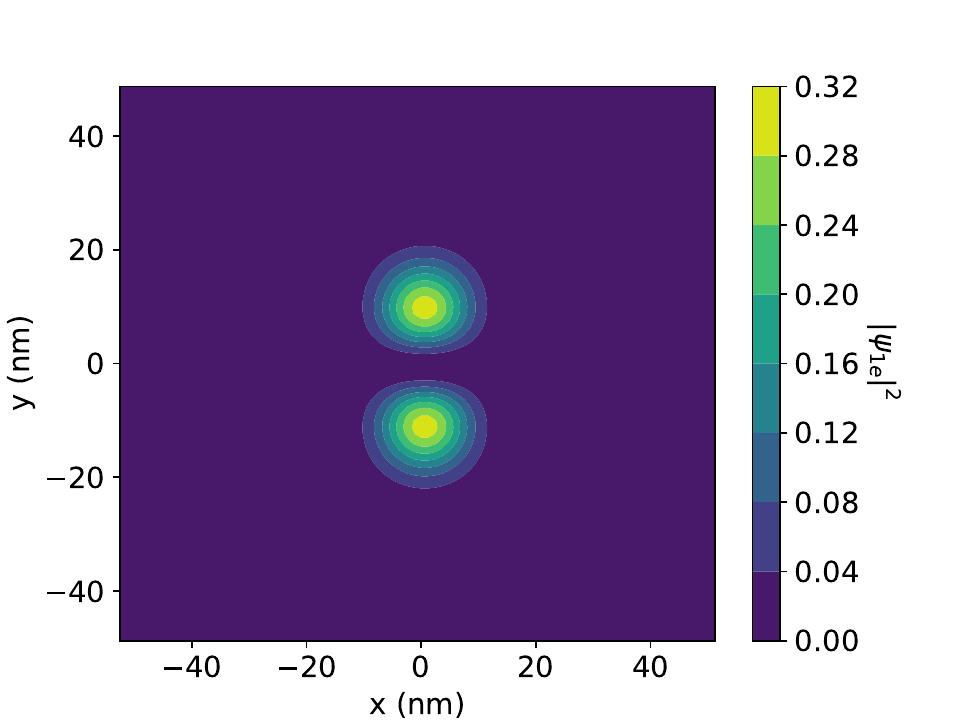}
  	\label{fig:1st_excited_state_2d}
	}%
	\hfill
	\subfloat[]{%
	\includegraphics[width=\linewidth]{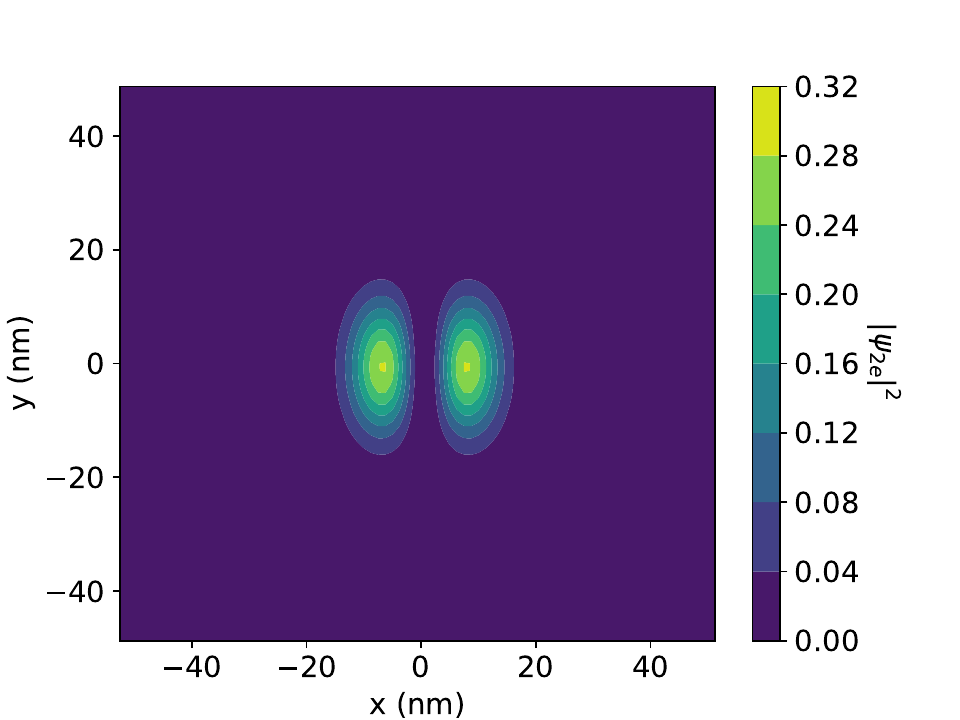}
  	\label{fig:2nd_excited_state_2d}	
	}%
	\hfill
	\subfloat[]{%
	\includegraphics[width=\linewidth]{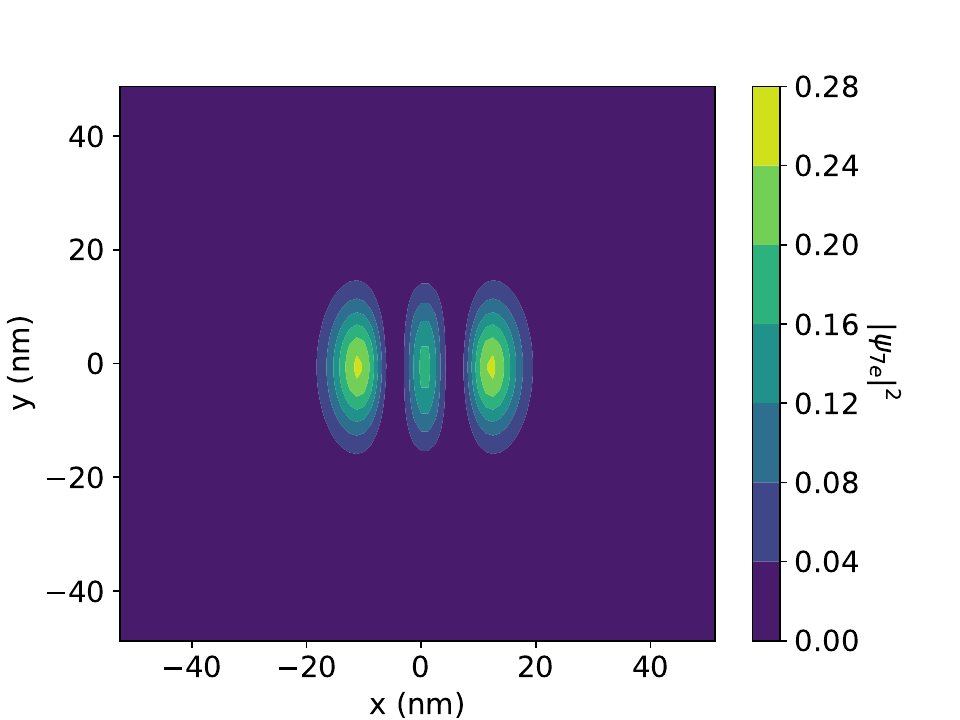}
  	\label{fig:7th_excited_state_2d}	
	}%
\caption{Contour plots of the probability densities of (a) the first, (b) second, and (c) 7th excited states of the 2D potential when $N=3$ and $A=100$\,mV.}
\label{fig:excited_states_2d}
\end{figure}

We found both the loss probability and the excitation fraction has an order of magnitude difference. For the noiseless shuttling for a distance of $1.4$\,$\mu$m ($A=100$\,mV, $v=10$\,m/s), the loss probability in the case of  Note that the energy gap of the 1D potential ($6.38$\,meV), is smaller than the energy gap of the 2D potential ($3.837$\,meV). This is because the first excited state of the 2D potential is in the y-direction as shown in Figure~\ref{fig:excited_states_2d}.

\section{Details of Numerical Simulations}\label{appendix:sec:numerical_simulations}

\subsection{Boundary Conditions}\label{appendix:subsec:boundary_conditions}

In this section, we outline the boundary conditions to solve the Laplace and time-dependent Schr\"{o}dinger equations, explain the numerical methods and techniques for both cases, and, finally, state the set of hyper-parameters in the simulations, which we define as the \textit{model}.

As noted in section \ref{sec:shuttling_device} and section \ref{sec:conveyor_belt_shuttling}, we make two assumptions that allow us to use periodic boundary condition in the shuttling direction (the $+x$ direction): (1) there is an infinite line of clavier gates along the axis and (2) there are $N$ independent voltage signals applied to the gates. These allow us to limit our domain to solve the Laplace equation to one unit cell of the device. Thus, the boundary conditions of the system are as follows:

\subsubsection{Periodic Boundary Conditions} Since the unit cell repeats along the x axis, we used periodic boundary condition along this axis, i.e. $\Phi(x,y,z) = \Phi(x+d, y, z)$, where $d$ is the length of the unit cell in the x-axis.

\subsubsection{Dirichlet Boundary Conditions} On the $y=\pm 50$\,nm planes, Dirichlet boundary conditions were imposed such that $\Phi(x,y=\pm 50\,\text{nm}, z) = 0$ for the Laplace solver. These boundary conditions are reasonable because the potential energy barrier from the confinement gates is an order of magnitude greater than the characteristic energy gap in the $y$-direction, which is the energy gap between the ground and first excited states (see Figure~\ref{fig:1st_excited_state_2d}), so the precise form of the top of the barrier is not critical: specifically, for our default setting of $A = 100$\,mV at the gates, the height of the potential energy is $8.5$ times bigger that the characteristic energy gap. When the gate voltage amplitude is smaller, the height of the potential barrier becomes only twice as big as the characteristic energy gap for $A = 6.24$\,mV. However, such small amplitudes were only used for the simulation of noiseless shuttling cases in section \ref{sec: noiseless_shuttling}. In these cases, the excitation fraction, defined in section \ref{sec: metrics}, is as low as approximately $5 \times 10^{-3}$ in the worst-case scenario (see section \ref{sec: noiseless_shuttling} for more details). In addition, hard-wall boundary conditions were imposed in the time-dependent Schr\"{o}dinger solver, i.e. the wave function is always zero on $y=\pm 50\,\text{nm}$ planes, so that there is no loss of probability outside the well in the y-direction.

For the bottom surface of the device, i.e. $z=-60$\,nm, Dirichlet boundary condition was imposed such that $\Phi(x, y, z=-60\,\text{nm}) = 0$. The position of the bottom surface doesn't change the overall physics: We found that the depth of potential energy changed about 1-2\% when the bottom surface of the device was $540$\,nm below the gates instead of $60$\,nm. Moreover, the electrode regions, represented as yellow boxes in Figure~\ref{fig:device_geometry}, have Dirichlet conditions applied at their boundaries fixing the potential at the voltages applied to individual gates.

\subsubsection{Neumann Boundary Conditions}
In Figure~\ref{fig:device_geometry_cross_section}, at $z=15$\,nm, in the gaps between the electrodes, Neumann boundary conditions were imposed such that $\frac{\partial{\Phi}}{\partial{z}}=0$. This is to reflect that the electric fields between two clavier gates should be parallel to the X-Y plane. At the interface between the \ce{Si} and \ce{SiO2}, i.e. $z=-10$\,nm, the displacement field should be continuous, and, thus, a Neumann boundary condition was imposed such that $\epsilon_{\text{ox}}\frac{\partial{\Phi}_{\text{ox}}}{\partial{z}}=\epsilon_{\text{Si}}\frac{\partial{\Phi}_{\text{Si}}}{\partial{z}}$.

\subsection{\label{appendix:subsec: numerical algorithms} Numerical Methods}

The numerical methods consist of two parts: a Laplace solver to obtain a time-dependent potential from the gates and a Schr\"{o}dinger solver to simulate the dynamics of the electron state in the shuttling device.

We used successive over-relaxation (SOR)\cite{Young_1954, Frankel_1950} to obtain the time-dependent potential in the unit cell in Figure~\ref{fig:device_geometry}. Instead of solving the Laplace equation to obtain the potential at every time step, we used the fact that any potential can be expressed as a linear combination of individual contributions from the gates\cite{Gurtner_2017}:
\begin{equation}\label{eqn:superposition_principle_potential}
    \Phi(x,y,z, t) = \sum^{N}_{i=1}u_{i}(t)\phi_{i}(x,y,z),
\end{equation}
where $N$ is the number of gates in the unit cell, and $\{\phi_{i}\}_{i=1..N}$ are the solutions to the Laplace equation when only one of the gates in the unit cell is turned on and the potentials on the others are zero. $u_{i}(t)$ is the voltage applied to the $i$th gate as a function of time. Moreover, the periodic boundary condition along the $x$-axis means the solutions $\phi_i$ for different electrodes can be generated from one another by simple translations.  Figure~\ref{fig:potential_3d_n_4} shows two examples of the potential energy obtained by the Laplace solver, at the points of maximum and minimum curvature $\kappa(\phi)$ near the potential minimum.

We used the split operator method\cite{glowinski2017splitting} with symmetric Strang splitting\cite{strang1968construction, strang2012essays} to solve the time-dependent Schr\"{o}dinger equation. By using the fact that kinetic energy operator is diagonal in k-space and the potential operator is diagonal in position space, the state of an electron was propagated as follows:
\begin{align}
    \hat{U}_{r}(\Delta t) &= e^{\frac{i}{\hbar}e\Phi(\Vec{r}, t) \Delta t}  \nonumber \\
    \hat{U}_{k}(\Delta t) &= e^{-\frac{i\hbar k^2 \Delta t}{2m}} \nonumber \\
    \psi(\Vec{r}, t+\Delta t) &= \hat{U}_{r}(\frac{\Delta t}{2}) F^{-1}[\hat{U}_{k}(\Delta t)F[\hat{U}_{r}(\frac{\Delta t}{2})\psi(\Vec{r}, t)]],
\end{align}where $F$ and $F^{-1}$ are Fourier and inverse Fourier transform to move from position space to the k-space, and $\hat{U}_{r}(\Delta t)$ and $\hat{U}_{k}(\Delta t)$ are propagators in position space and k-space, respectively.

Since we imposed hard-wall boundary condition on $\psi$ at the planes $y=\pm50$\,nm to mimic the effect of the confinement gates, we used a discrete sine transform (DST) to perform the Fourier transform in the $y$-direction. The DST allows us to impose the boundary condition, i.e.  $\psi(x, y=\pm50\,\text{nm}, t) = 0$, at the boundaries.

\subsection{\label{appendix:subsec:Model} Definition of Model}

The model is defined as a set of hyper-parameters with which the full simulation can be reproduced. These include: (1) the choice of the unit system  to map the simulation results to real systems; (2) parameters used in the numerical algorithms to simulate the dynamics, such as the step sizes of the grid; (3) device-specific parameters, such as the dimensions of the gates and permittivity of the device materials; (4) parameters to specify the shuttling scenarios, such as the distance, speed, acceleration of the electron as a function time.

In the Schr\"{o}dinger solver, we set the reduced Planck constant, electric charge constant, and the mass of the free electron to one, i.e. $\hbar=e=m_{e,0}=1$. With the remaining degree of freedom, we chose our length unit to be $10$\,nm. From the constraint $\hbar=1$, the time unit and energy unit are uniquely determined to be $0.8637$\,ps and $0.762$\,meV. Below are the equations to derive the time unit and energy units:

\begin{align}
    1 \text{(time unit)} &=\frac{m_{e} \times 1\text{(length unit)}^2}{\hbar} = 0.8637\,\text{ps} \nonumber \\
    1 \text{(energy unit)} &= \frac{\hbar}{1 \text{(time unit)}} = 0.762\,\text{meV}.
\end{align}

The spatial grid spacing used in the Laplace solver and Schr\"{o}dinger solver were $0.125$ for both $x$ and $y$ directions while the time step was $0.003125$. The convergence of the solvers was tested as described in Appendix~\ref{appendix:subsec:convergence_studies}. The relaxation parameter of the SOR in Laplace solver, $\omega$, was $1.9$, which controls the rate of convergence. The dimensions of the gates are outlined in section \ref{sec:shuttling_device} while $3.9$ and $11.69$ were taken as permittivity of \ce{Si} and \ce{SiO2}, respectively. Furthermore, the transverse electron mass in \ce{Si}, $m^{*}_{e,t}=0.19m_{e,0}$, was used for the motion in the 2D plane of the device (corresponding to population of the $\pm z$ valleys).

The default target distance and speed were set to $1.4$\,$\mu$m and $10$\,m/s. These choices were made because the length of one unit cell in the $x$-direction is an integer multiple of $35$\,nm, and the optimal speed of shuttling suggested by previous analytical calculations is around $10$\,m/s\cite{langrock_2023}. 

All our models approximate reality by the discretization of space and time; remaining sources of significant error are the Trotterization error of the split operator method\cite{glowinski2017splitting} in the Schr\"{o}dinger solver, and the error below the tolerance threshold of SOR in the Laplace solver. As we reduce the step sizes in our model, it becomes a better representation of reality but the computational cost grows by $\mathcal{O}(N_{p}\, N_{iter})$\cite{SKOTNICZNY2024102216} for the SOR and $\mathcal{O}(N_{p}\log N_{p})$ for the split operator method, where $N_{p}$ is the number of spatial grid points, and $N_{iter}$ is the number of SOR loops, which grows with increasing $N_{p}$. For both SOR and split operator method, the complexity grows only linearly with the number of grid points in time, i.e. $\mathcal{O}(N_{t})$.  For the specific choices made in our model, the numerical artefacts are explained in more detail in Appendix~\ref{appendix:subsec:numerical_precision}.

\subsection{Convergence Studies of the model} \label{appendix:subsec:convergence_studies}

\begin{figure}
	\subfloat[]{%
	\includegraphics[width=\linewidth]{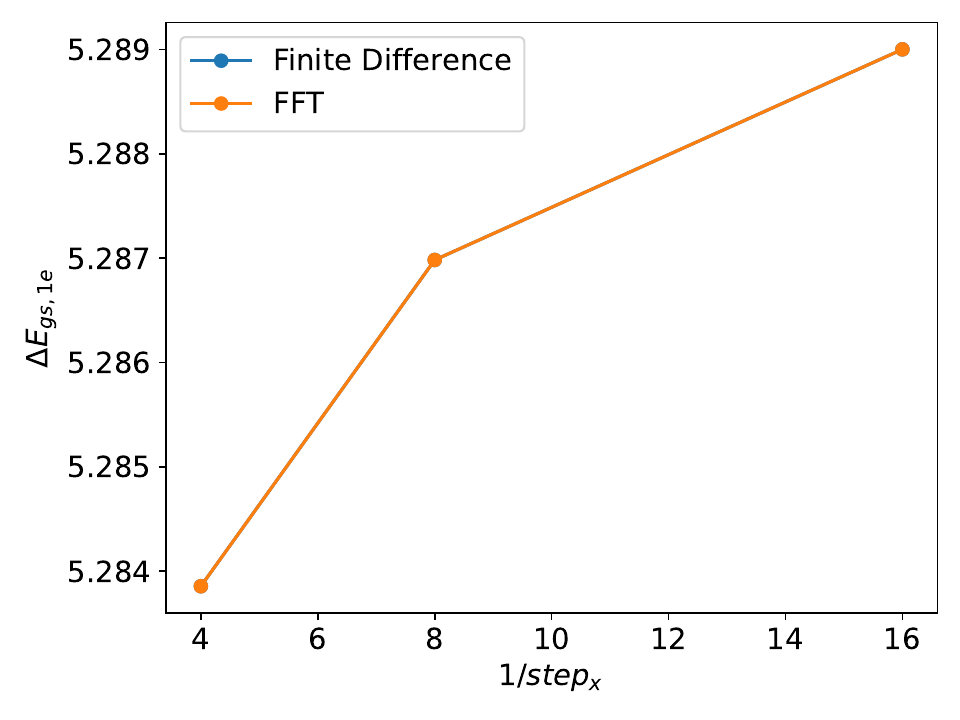}
  	\label{fig:convergence_step_x}
	}%
	\hfill
	\subfloat[]{%
	\includegraphics[width=\linewidth]{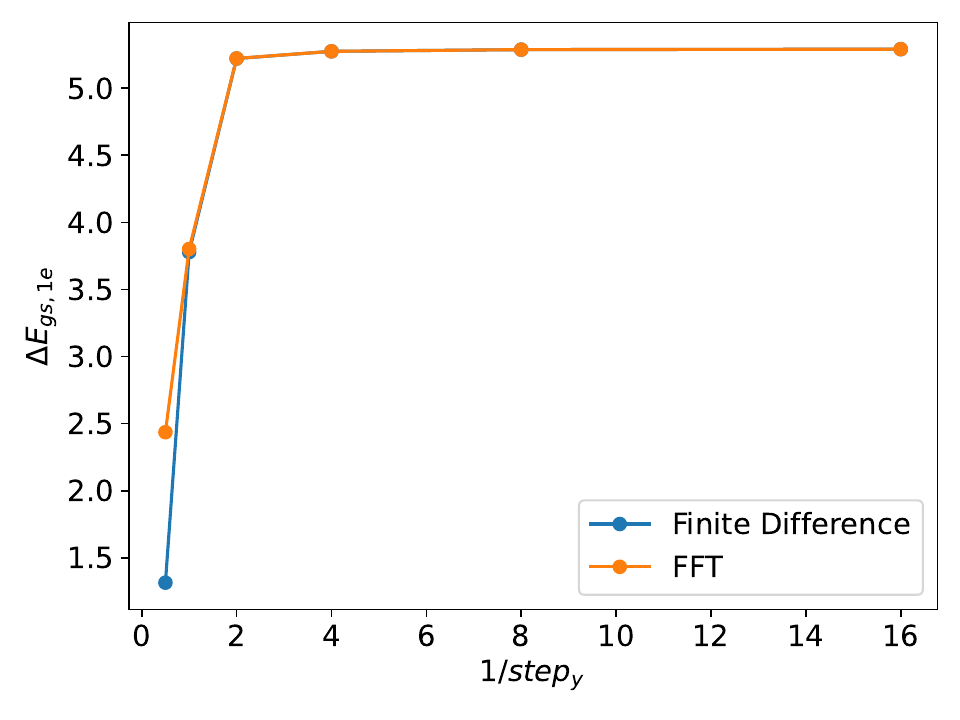}
  	\label{fig:convergence_step_y}
	}%
\caption{Convergence studies for the spatial resolution: The energy gap of the initial Hamiltonian obtained with different step sizes in (a) x and (b) y direction for $N=5$.}
\label{fig:convergence_spatial}
\end{figure}

\begin{figure}
\subfloat[]{%
\includegraphics[width=\linewidth]{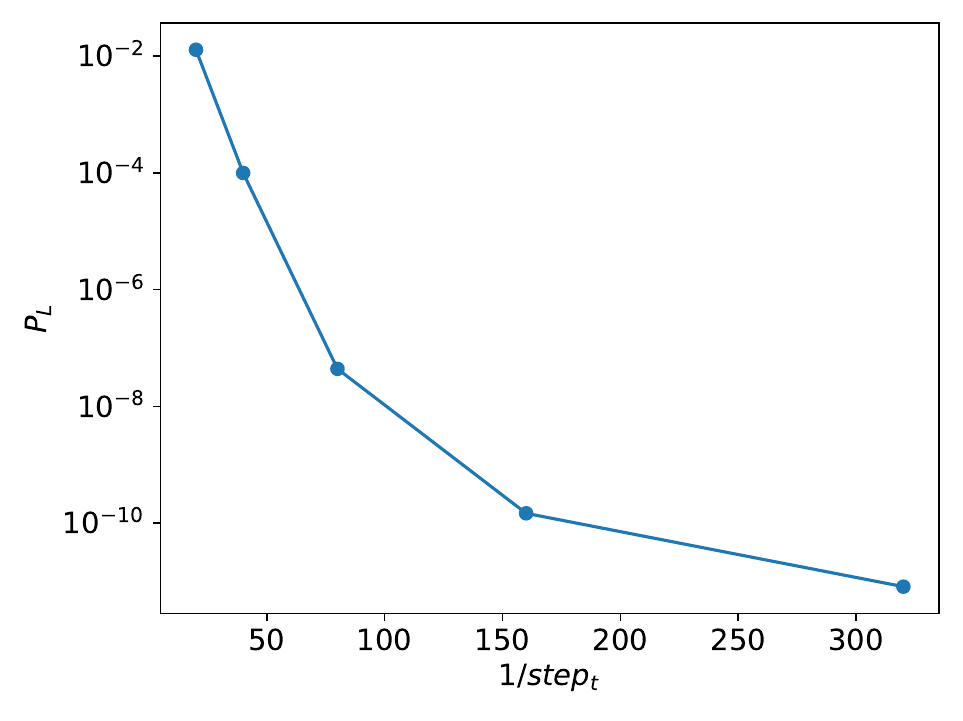}
  \label{fig:converegence_step_t_loss}
}%
\hfill
\subfloat[]{%
 \includegraphics[width=\linewidth]{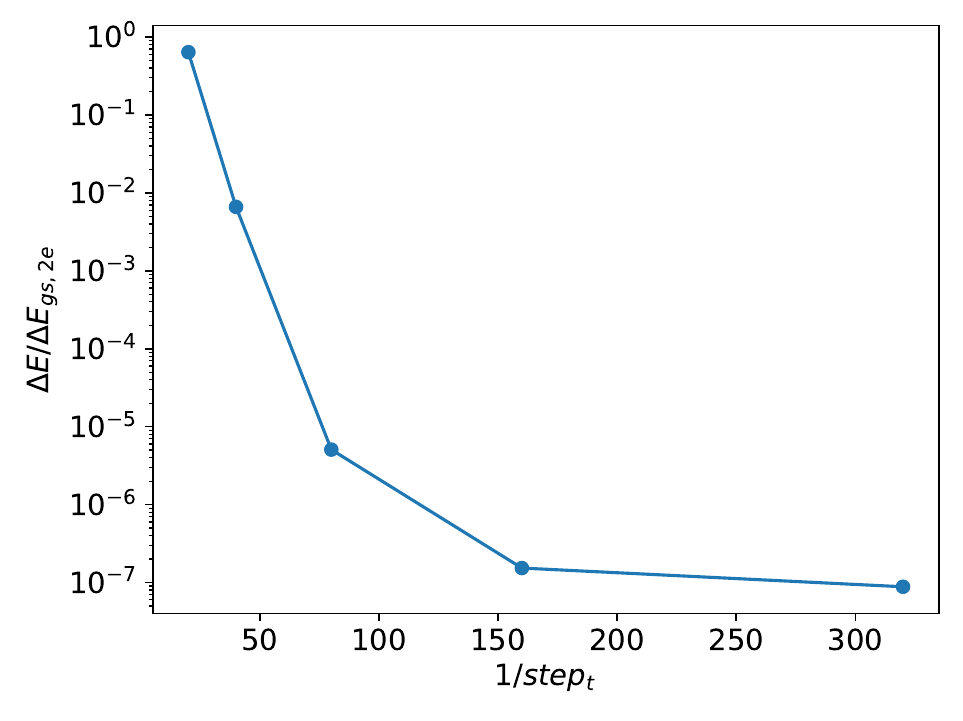}
  \label{fig:converegence_step_t_energy_excitation}
}%
\caption{Convergence studies for the temporal resolution: For $N=4$, (a) the loss probability and (b) excitation fraction obtained after the shuttling of $2.1$\,$\mu$m.}
\label{fig:convergence_temporal}

\end{figure}

The model consists of the Laplace solver to obtain the time-dependent potential in the device and the time-dependent Schr\"{o}dinger solver to obtain the evolution of the state forward in time. 

For this numerical model, appropriate spatial and temporal step sizes had to be chosen. The metric used to determine the convergence for the spatial resolution was the energy gap between the ground state and the first excited state for the initial Hamiltonian (t=0). The energy gap was first obtained by diagonalizing the Hamiltonian matrix, whose size is determined by the step sizes in x and y directions. These points were plotted in Figure~\ref{fig:convergence_spatial} with blue dots. Secondly, using the normalized initial and first excited states obtained by the diagonalization, the energy gap was once again obtained by evaluating the expectation values of the initial Hamiltonian. These points were plotted with orange dots in Figure~\ref{fig:convergence_spatial}. 

By fixing the step y to be 0.125, the convergence tests of step x were performed, whose results are shown in Figure~\ref{fig:convergence_step_x}. When step x changes from 0.125 (second point) to 0.0625 (last point), energy gap changes by 0.038 \%. Thus, 0.125 was chosen to be the step size in the x direction.

Similarly, the convergence tests of step y were performed by fixing the step x to be 0.125, and the results are shown in Figure~\ref{fig:convergence_step_y}. When step y changes from 0.125 (second to the last point) to 0.0625 (last point), energy gap changes by 0.063 \%. Thus, 0.125 was chosen to be the step size in the y direction.

Given the spatial resolutions of x and y directions, the convergence tests were performed for the temporal step size. The test was performed by shuttling the electron $2.1$\,$\mu$m from its initial position and calculating the loss probability and excitation fraction at the end of the shuttling. The results are shown in Figure~\ref{fig:convergence_temporal} with the y-axis in log scale. Even though the excitation fraction in Figure~\ref{fig:converegence_step_t_energy_excitation} converges at the step size of 0.00625 (second to the last point), the corresponding loss probability in Figure~\ref{fig:converegence_step_t_loss} only starts to converge at the step size of 0.003125 (last point). There is an order of magnitude change in the loss probability when step size changes from 0.00625 to 0.003125. Thus, 0.003125 was chosen to be the temporal step size.

\subsection{Numerical Precision}\label{appendix:subsec:numerical_precision}

\subsubsection{Results of Stationary Evolution}\label{appendix:subsubsec:stationary_evol}

To benchmark the results of simulations of conveyor-belt shuttling, we performed stationary evolution of the initial state with initial potential for the same time duration as the duration of shuttling $140$\,nm with the speed of $10$\,m/s and amplitude of $A=100$\,mV, for $N=3$ without noise. The loss probability was $4.43 \times 10^{-11}$, and the excitation fraction was $5.52 \times 10^{-8}$, which is smaller than the loss probability and excitation fraction for the corresponding shuttling scenario.

When there was a noise with the same parameters as the ones used in section \ref{sec:johnson-nyquist}, for one random run, the loss probability was $4.82 \times 10^{-8}$, and the excitation fraction was $5.12 \times 10^{-8}$. While the loss came out to be slightly bigger, the excitation fraction resulted in a smaller value. This shows that the noise changes the overall shape of the potential energy, changing the energy value of the ground state and characteristic energy gap in such a way that the excitation fraction came out to be slightly smaller.

\subsubsection{Normalization Drift and Energy Oscillation}

The sources of error are the trotterization error in the split operator method and the error below the tolerance threshold in the SOR. We observed two numerical artefacts: the normalization drift and the energy oscillation in the stationary potential. The simulation of shuttling with the target distance of $2.1$\,$\mu$m and the target speed $10$\,m/s was performed, and the normalization of the state during the entire process was recorded. At the end of the shuttling, there are additional 5000 time steps to do the stationary evolution with the final potential. During this additional 5000 time steps, the expectation value of energy was noted. In contrast to the reality, the normalization drifts from one by $10^{-8}$ during around $8 \times 10^{7}$ time steps. Furthermore, the energy expectation value after the shuttling oscillates by the scale of $10^{-8}$\,meV.

Such artefacts set the guideline on how small a number should be to be considered as \textit{numerical error}. The normalization drift is relevant to the loss probability as it calculates the probabilty of loss, i.e. the fraction of normalization outside of the potential well. The energy oscillation is relevant to the excitation fraction as the final energy is precise only up to the amplitude of oscillation. We claim that any number below these artefacts can be considered as a negligible quantity. One example is the probability of loss in shuttling of any distance in Figure~\ref{fig:loss_probability_target_dist} with no noise and no digitization.

\subsection{Lumped Element Model of a Voltage Source connected to Clavier Gates} \label{appendix:subsec:lumped_element_model}

\begin{figure*}
  \centering
  \includegraphics[width=\linewidth,trim={0cm 0cm 0cm 0cm},clip]{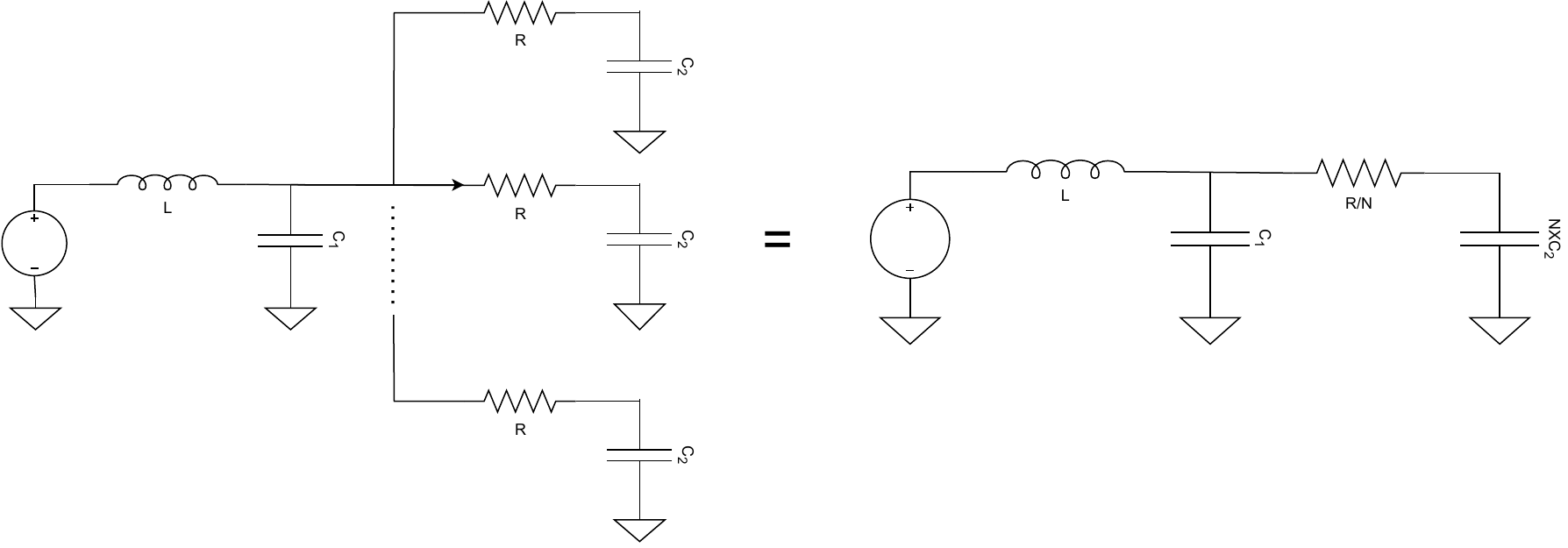}
\caption{(a) and (b) are equivalent circuits for lumped element model of a voltage source connected to $N$ clavier gate(s) via a single bondwire.
$L$ is the inductance of the bondwire. $C_{1}$ is the capacitance of the bondpad, and $C_{2}$ is the capacitance of the clavier gates. $R$ is the resistance of the metal connection from the bondpad to the gate.}
\label{fig:circuit}
\end{figure*}

Figure~\ref{fig:circuit} shows the lumped element model, i.e. a simplified model, of a voltage source connected to the clavier gates for shuttling. Multiple clavier gates of capacitance, $C_{2}$, are connected to a bondpad of capacitance, $C_{1}$ through metal connections of resistance, $R$. The bondpad is then connected to the voltage source via a bondwire of inductance $L$. The voltage source creates sinusoidal voltage pulses to the clavier gates as shown in Figure~\ref{fig:CB_mode_pulse_illustration_sinusoidal}. Due to thermal agitation, Johnson-Nyquist noise appear in the voltage pulse at the clavier gates, whose power spectral density is given by equation (\ref{eqn:PSD_johnson_nyquist_classical}) at room temperature and equation (\ref{eqn:PSD_johnson_nyquist_quantum}) at cryogenic temperatures.  For each element in Figure~\ref{fig:circuit}, we estimated typical values of the elements as $L=1$\,nH, $C_{1}=5$\,fF, and $C_{2}=100$\,aF. The resistance was varied in the range of $100\,\Omega$ to $2\,\mathrm{M}\Omega$ to vary the cut-off frequency, $\gamma$.

\subsection{\label{appendix:subsec:generation_of_noise}Generation of Johnson-Nyquist Noise}

Given the temporal step size, $dt$, in the time-dependent Schr\"{o}dinger solver, one can obtain a discretized power spectral density with the grid spacing of $2\pi/(N_{t} \cdot dt)$, where $N_{t}$ is the number of time steps of the entire shuttling process.

We first take the square root of the power spectral density to obtain the magnitudes of the modes at different frequencies. Then, we multiply each frequency mode by a random phase factor $e^{i\phi_{j}}$, where $\phi_{j} \in [0, 2\pi]$, and $j = 1,2, ..., N_{t}$.
Finally, we perform a Fast Fourier transform of the modes to produce a random time series $X(t)$, with the normalisation chosen to ensure Parseval's theorem is obeyed, i.e. $\int dt\, |X(t)|^{2}=\int d \omega \,S(\omega)$.  The integrals were approximated by Riemann sums $\int dt\, |X(t)|^{2} \approx \sum^{N_{t}}_{i=0} |X_{i}|^{2} \Delta t$.

\section{\label{appendix:} Additional Results of the Conveyor-Belt Mode Shuttling}

\subsection{\label{appendix:subsec:noise_free_speed} Noise-free Shuttling: Target Speed}

\begin{figure}
\subfloat[]{%
\includegraphics[width=\linewidth]{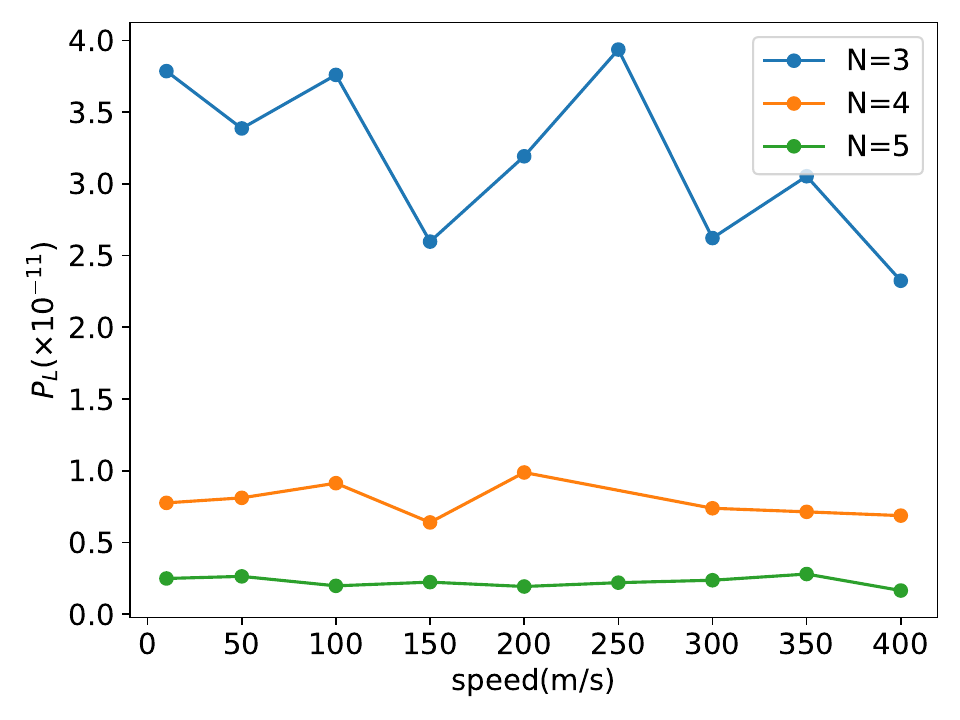}
  \label{fig:loss_probability_target_speed}
}%
\hfill
\subfloat[]{%
\includegraphics[width=\linewidth]{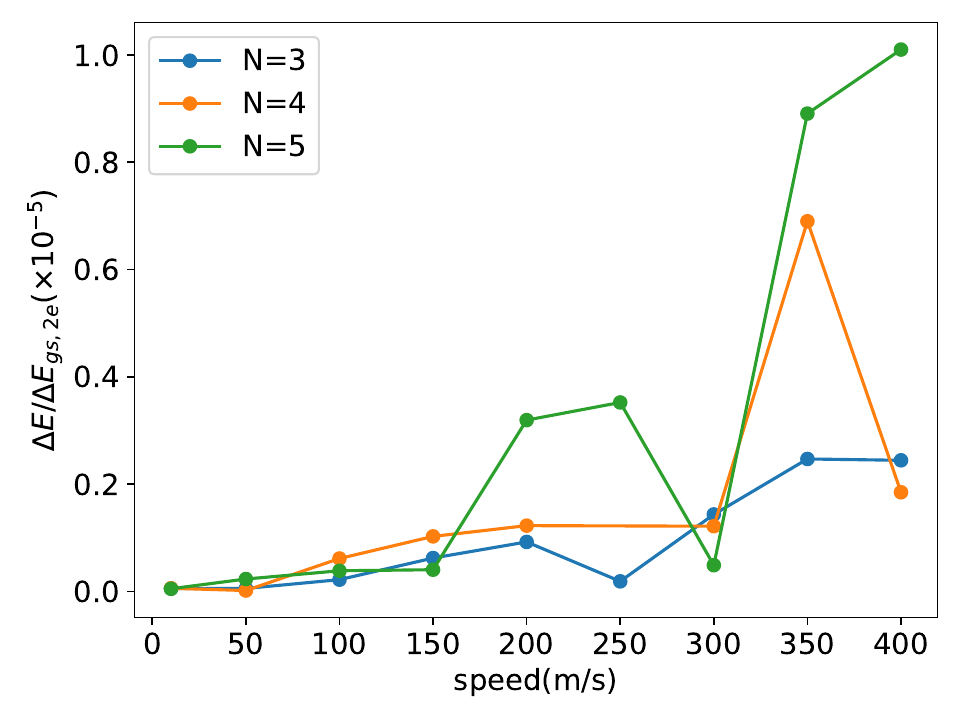}
  \label{fig:excitation_fraction_target_speed}
}%
\caption{(a) The loss probability and (b) excitation fraction with varying target speeds and the number of gates per unit cell.}
\label{fig:loss_and_excitation_target_speed}
\end{figure}

Figure~\ref{fig:loss_and_excitation_target_speed} shows the loss probability and excitation fraction with different target speeds. The shuttling distance and the amplitude of signals were fixed to $1.4$\,$\mu$m and $100$\,mV, respectively. While there is no clear trend of increase or decrease of loss probability with increasing target speed, the excitation fraction shows an upward trend with increasing target speeds. In all of the cases, the loss probability is bounded by $4 \times 10^{-11}$ and the excitation fraction is bounded by $1 \times 10^{-5}$.

\subsection{Sensitivity to step-changes in voltage control}\label{appendix:subsec:step_changes}

\begin{figure}
\centering
\subfloat[]{%
\includegraphics[width=\linewidth]{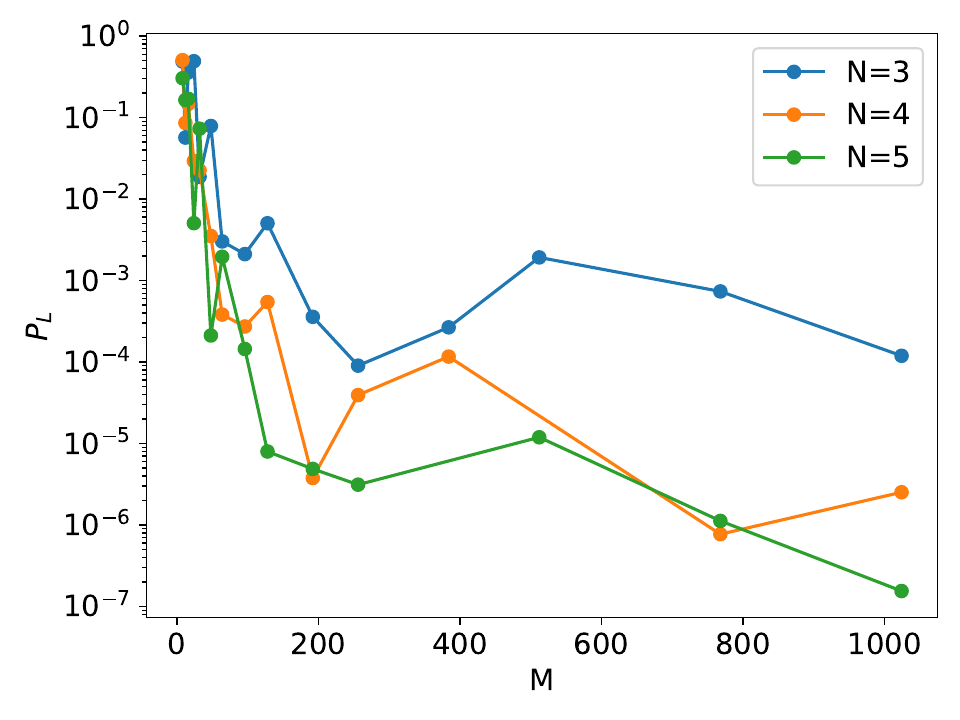}
  \label{fig:loss_probability_n_digits_V}}%
\hfill
\subfloat[]{%
\includegraphics[width=\linewidth]{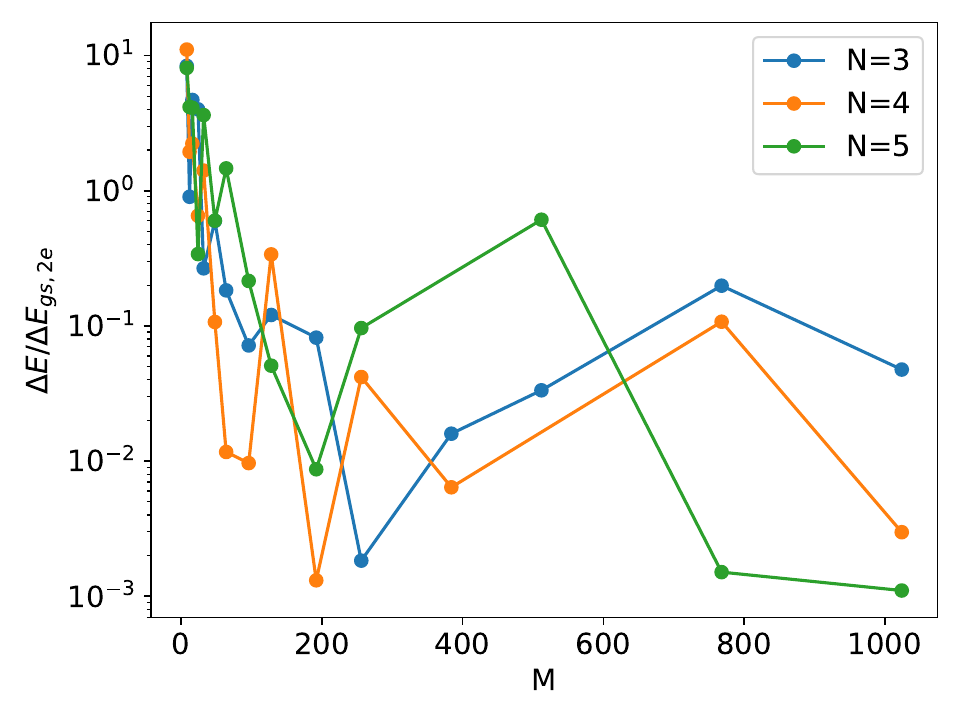}
  \label{fig:excitation_fraction_n_digits_V}
}%
\caption{(a) The loss probability and (b) excitation fraction with varying number of fixed potential settings and the number of gates per unit cell.}
\label{fig:loss_and_excitation_n_digits_V}
\end{figure}

\begin{figure}
\centering
\subfloat[]{%
\includegraphics[width=\linewidth]{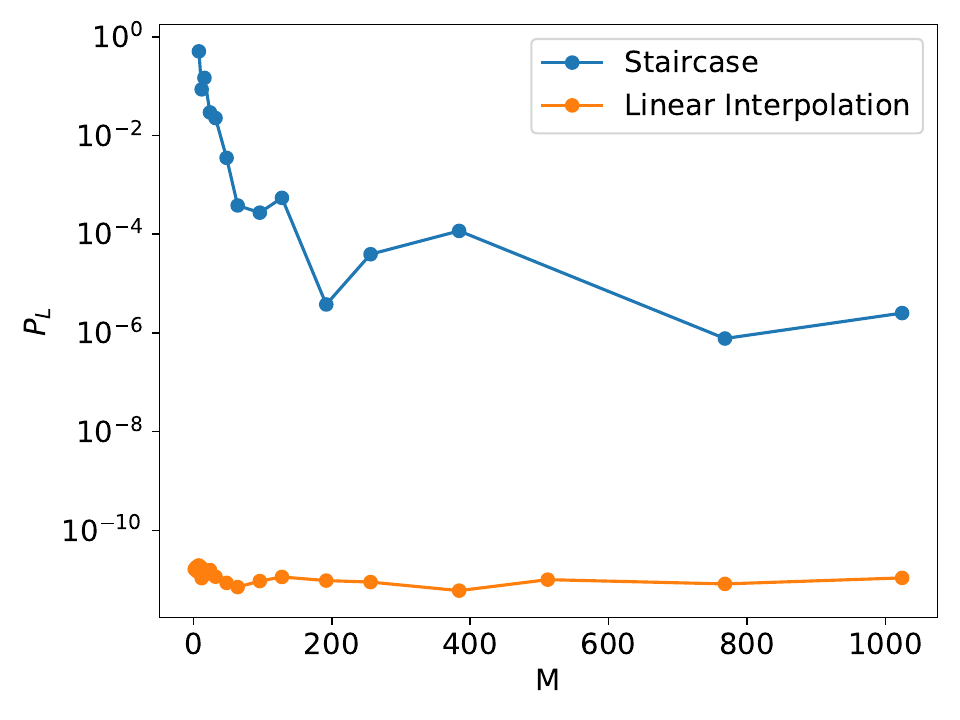}
  \label{fig:loss_probability_step_vs_linear}
}%
\hfill
\subfloat[]{%
\includegraphics[width=\linewidth]{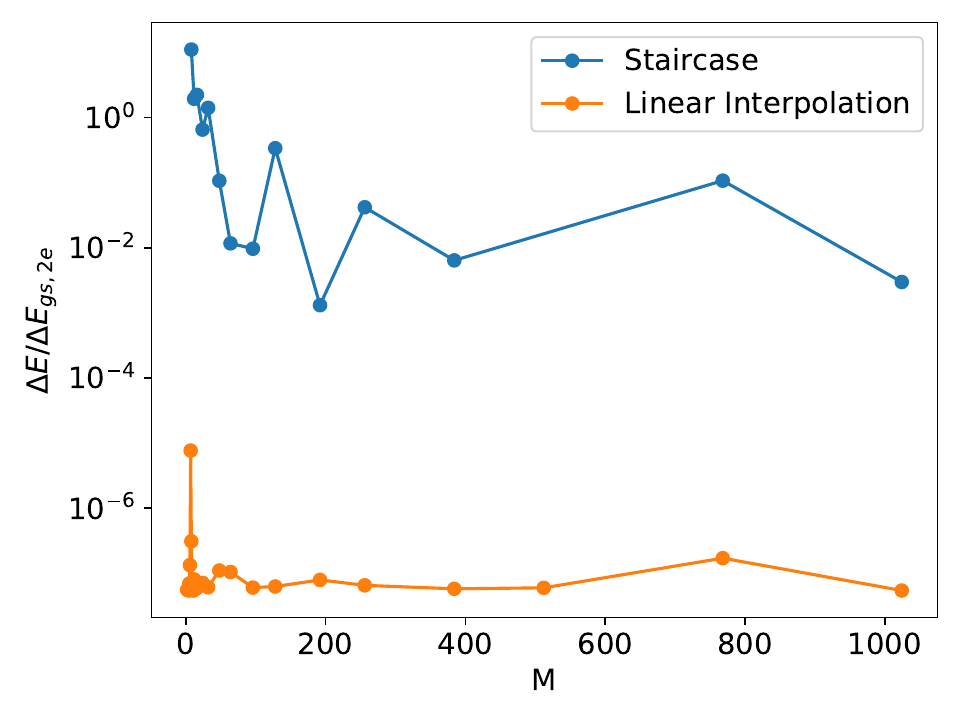}
  \label{fig:excitation_fraction_step_vs_linear}
}%
\caption{(a) The loss probability and (b) excitation fraction of digitization mode 1 (Staircase) and digitization mode 2 (Linear Interpolation)}
\label{fig:loss_and_excitation_step_vs_linear}
\end{figure}

Since the majority of the results with smoothly varying potential without any noise proved to be good shuttling scenarios, we investigated how the abrupt changes in voltage control affect the loss probability and excitation fraction. In particular, we considered the scenario where we have a finite number of voltage settings at hand as if the voltage signals are digitized. While the phase varies linearly like in section \ref{sec: noiseless_shuttling}, the sinusoidal voltage signals are mapped to the nearest voltage in the list of voltage settings:

\begin{figure}
    \centering
    \includegraphics[width=\textwidth]{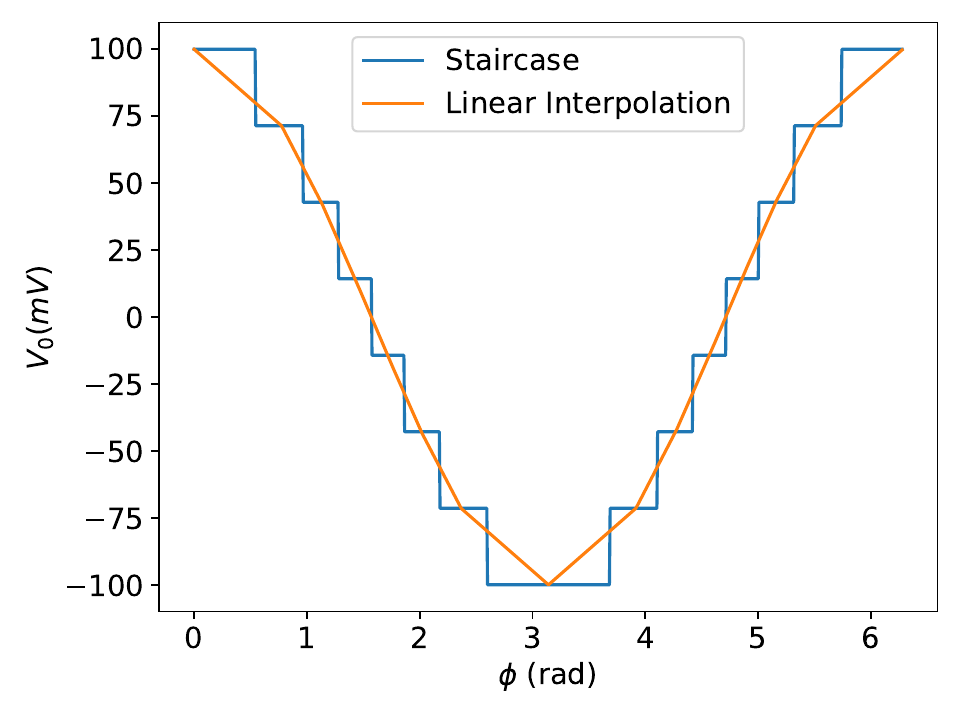}
    \caption{An example of voltage signal on the 1st gate in the unit cell for digitization mode 1(blue line) and mode 2(orange line).}
    \label{fig:voltage_signals_mode1_vs_mode2}
\end{figure}

\begin{equation}
\begin{split}
    m^{i}_{*}(t) &= \operatorname*{argmin}_{m}|A\cos(\phi(t) -\frac{2\pi i}{N}) - V_{m}| \\
    V^{i}(t) &= V_{m_{*}^{i}(t)},
\end{split}
\end{equation}where $N$ is the number of gates per unit cell, $\{V_{m}\}_{n=0,...,M-1}$ is the set of voltage settings, the superscript denotes the gate to which the voltage signal is applied and the subscript represents the voltage in the set of voltage settings. Thus, the voltage signals are step functions in time, which have step-changes in the voltage control. The blue line in figure\ref{fig:voltage_signals_mode1_vs_mode2} shows an example of such voltage signal.

On the other hand, we investigated another way of using the finite number of voltage settings such that the voltage is linearly interpolated between the two nearest voltage settings. To be specific, the voltage signals are made by linearly interpolating the mid points of the steps in the prior case as the orange line shows in figure\ref{fig:voltage_signals_mode1_vs_mode2}. Such signal is the opposite extreme from the prior case as there is no step-change in the voltage signal. Let's call the two methods digitization-mode 1 and 2, respectively.

Figure~\ref{fig:loss_and_excitation_step_vs_linear} shows the loss probability and excitation fraction of digitization mode 1 and 2 with different number of settings, $M$. This shows that the abrupt change in the voltage signal is detrimental to the quality of shuttling such that digitization-mode 2 with only two settings works as good as any other number of settings, i.e. $P_{L} \lesssim 2 \times 10^{-11}$. In contrast, the loss probability is $0.5$ when there are only two settings.

In this section, we conclude that discontinuities in voltage signals, such as the steps in staircase-like potential, significantly degrades the quality of shuttling, and this was observed by comparing the two digitization modes. Despite of this conclusion, we invented our new non-adiabatic shuttling method, which uses discontinuous updates of the potential (See section \ref{sec:snap}). Furthermore, this new method allows arbitrarily fast shuttling speeds, which can be controlled by the number of updates per unit cell and the strength of the gate voltages. This suggests that, while discontinuities in the voltage signals should be avoided, they can be useful when they are made at the right timings like our new method.

\subsection{Sensitivity to Classical Johnson Nyquist Noise} \label{appendix:subsec:johnson-nyquist_classical}

In this section, we present simulation results of shuttling with classical Johnson-Nyquist noise, whose power spectral density is given by equation~\ref{eqn:PSD_johnson_nyquist_classical} without the quantum correction factor. Note that the classical Johnson-Nyquist noise formula is only valid when the cut-off frequency and temperature satisfies $\gamma \lesssim \frac{k_{B}T}{\hbar}$.

\begin{figure*}
	\centering
    \subfloat[]{%
    \includegraphics[width=0.495\textwidth]{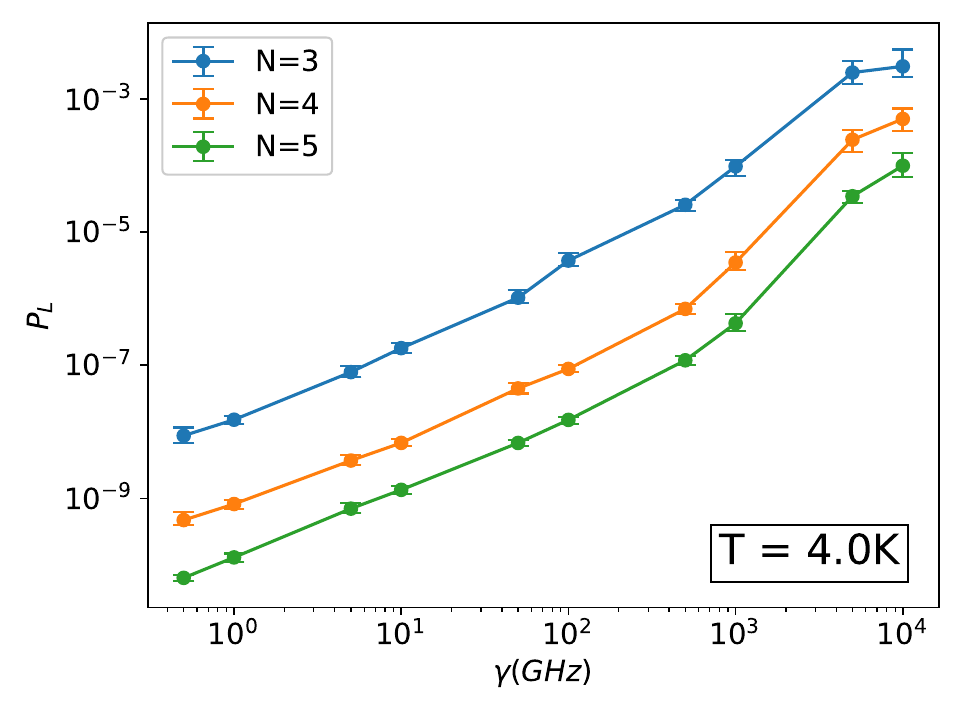}
      \label{fig:loss_probability_gamma_classical}}%
    ~
    \subfloat[]{%
    \includegraphics[width=0.495\textwidth]{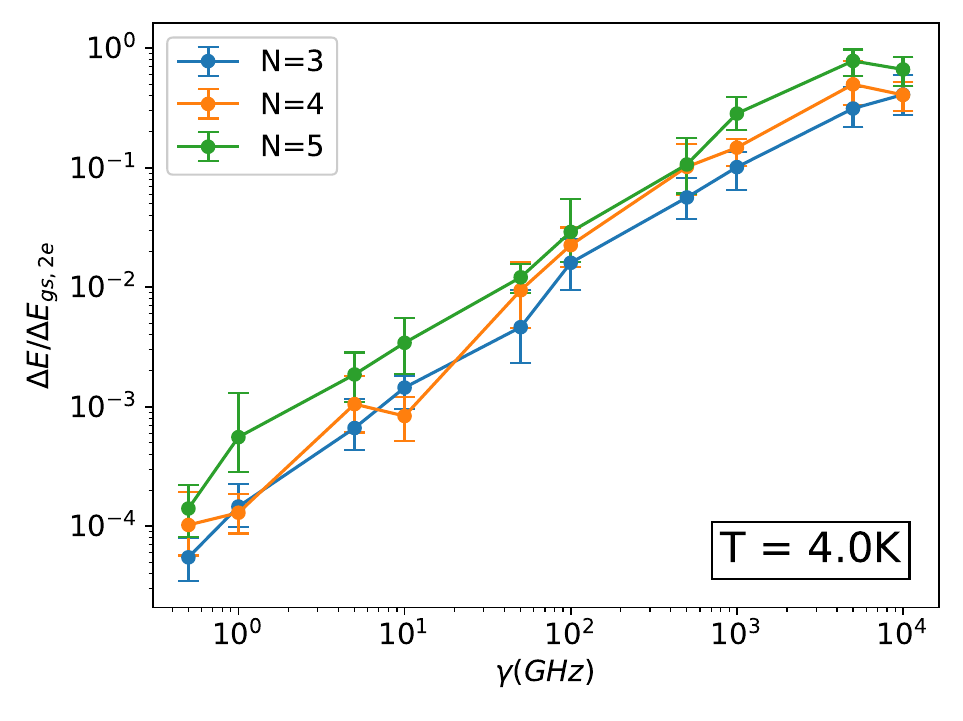}
          \label{fig:loss_probability_temp_classical}
    }%
    \hfill
    
    \subfloat[]{%
    \includegraphics[width=0.495\textwidth]{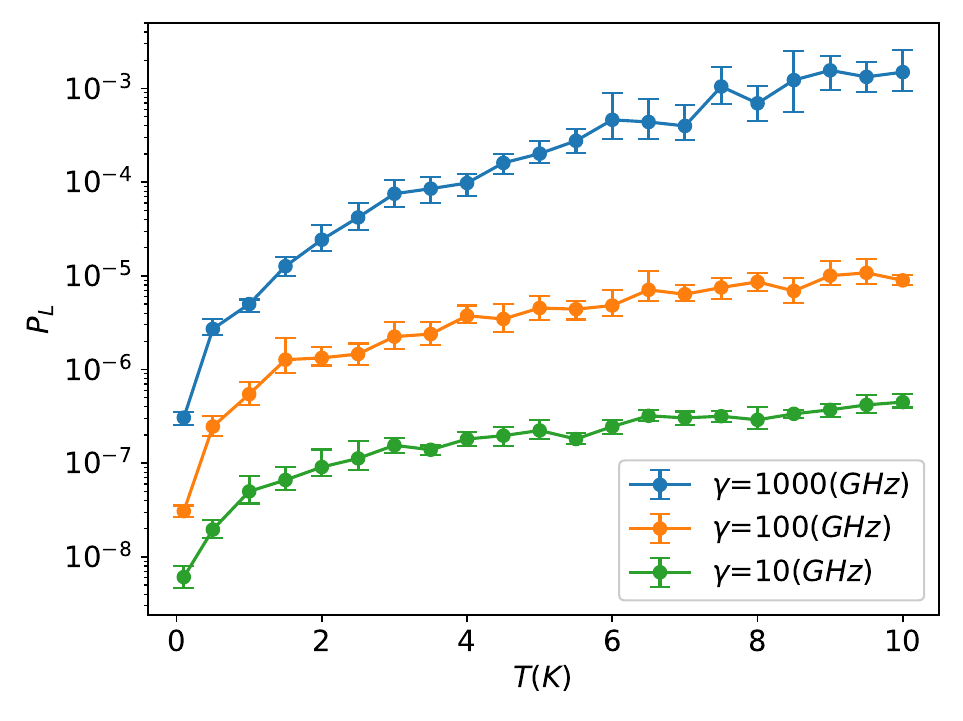}
    \label{fig:excitation_fraction_gamma_classical}
    }%
    ~
    \subfloat[]{%
    \includegraphics[width=0.495\textwidth]{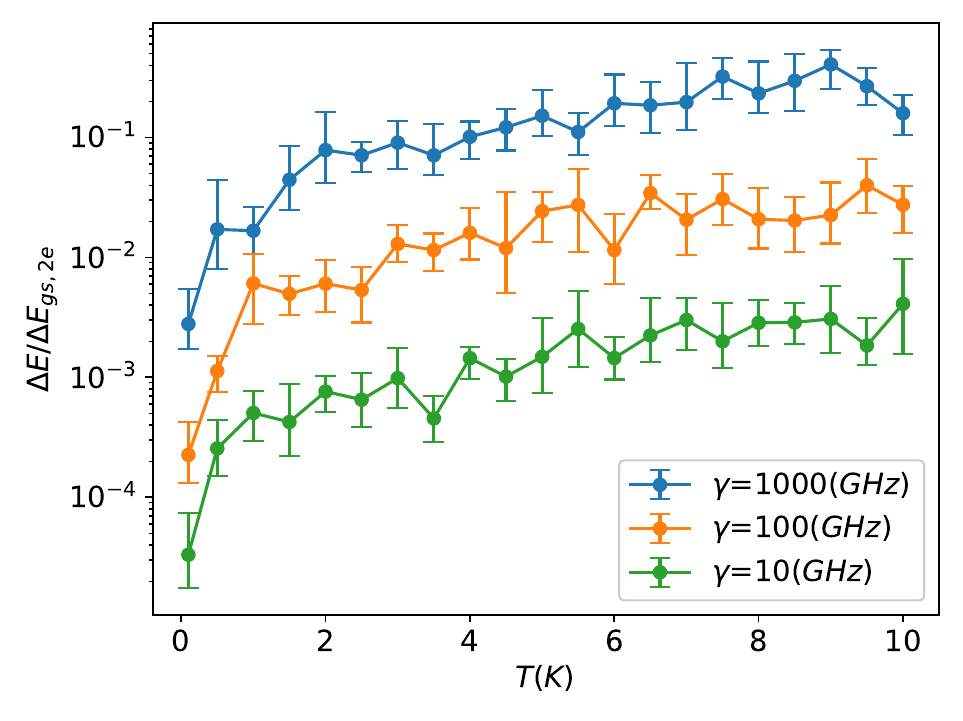}
          \label{fig:excitation_fraction_temp_classical}
    }%
	
    \caption{The loss probability and excitation fraction for classical Johnson-Nyquist noise: (a) loss probability and (b) excitation fraction as a function of cut-off frequency $\gamma$, at $T=4$\,K; (c) loss probability and (d) excitation fraction as a function of temperature with three different cut-off frequencies, $\gamma = 10, 100, 1000$\,GHz.}
    \label{fig:loss_and_excitation_gamma_temp_classical}
\end{figure*}

Figure~\ref{fig:loss_probability_gamma_classical} and Figure~\ref{fig:excitation_fraction_gamma_classical} show the loss probability and excitation fraction for different cut-off frequencies, $\gamma$, and different numbers $N$ of electrodes per unit cell. The temperature was chosen to be $4$\,K, which resulted in RMS noise of $0.118$\,meV. Figure~\ref{fig:loss_probability_gamma_classical} shows that shuttling is sensitive to high frequency noise ($\omega / 2\pi > 1$\,THz) as the loss probability reaches $10^{-4}$ for both $N=4$ and $N=5$ when the cut-off frequency reaches $10$\,THz, i.e. $\gamma =10$\,THz. Note that the characteristic energy gap is $6.02$\,meV, which corresponds to the frequency of $1.46$\,THz, i.e. $\omega_{c}/ 2\pi = \Delta E_{gs,2e}/h=1.46$\,THz. Thus, we conclude that the effect of noise becomes severe as the cut-off frequency becomes comparable to the characteristic energy gap. Furthermore, the shuttling process is more resilient to the noise when there are more gates per unit cell. This is because the QD becomes deeper, and the inter-dot distance becomes longer, for more number of electrodes per unit cell. The inter-dot distance is equal to the length of unit cell, which is $(35 \times N)$\,nm in our case. The depths of QD for $N=3,4,5$ are $49$\,mV, $69.5$\,mV, and $71.5$\,mV, respectively. Thus, given the same cut-off frequency, the loss probability goes down by couple of orders of magnitude as $N$ increases from 3 to 5. The excitation fraction does not reduce as dramatically as the loss probability when $N$ increases; hence, the primary motivation to use more of electrodes per unit cell is to reduce the loss probability.

Figures~\ref{fig:loss_probability_temp_classical} and \ref{fig:excitation_fraction_temp_classical} show the loss probability and excitation fraction as a function of temperature for different cut-off frequencies, i.e. $\gamma= 10, 100, 1000$\,GHz. As the temperature increases, the RMS noise increases (equation \ref{eqn:rms_noise_temperature}); to avoid growth in the loss probability and excitation fraction, it is therefore beneficial to perform shuttling at low temperature with large number of electrodes.

\begin{figure}
	\centering
	\subfloat[]{%
	\includegraphics[width=\linewidth]{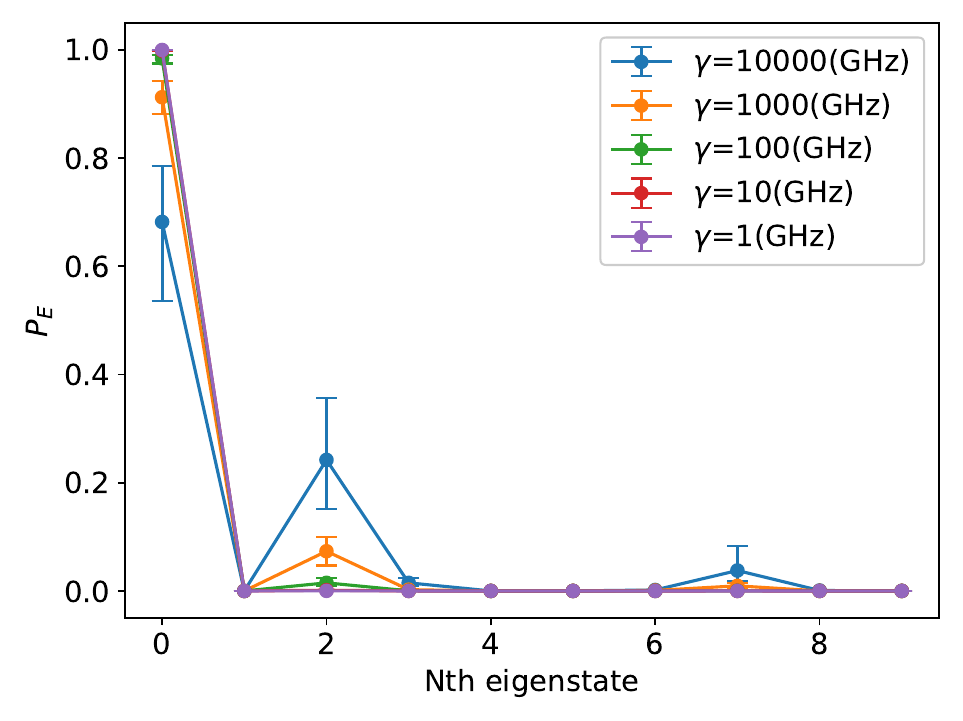}
      	\label{fig:fidelity_gamma}
	}%
    \hfill
    \subfloat[]{%
   \includegraphics[width=\linewidth]{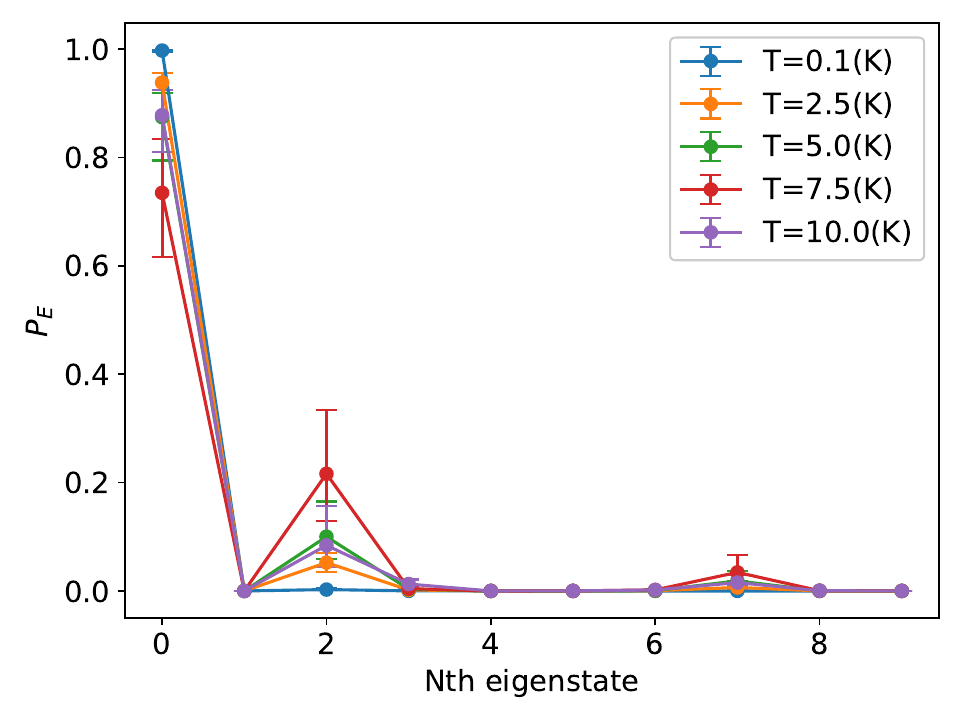}
      \label{fig:fidelity_temperature} 
    }%
    \caption{The probability of excitation to eigenstates of the instantaneous Hamiltonian for various (a) noise fractions and (b) cut-off frequency.}
    \label{fig:fidelity_gamma_temperature}
\end{figure}

Figure~\ref{fig:fidelity_gamma_temperature} shows the probability of excitation to the eigenstates of the instantaneous Hamiltonian at the end of the shuttling process for the case of three gates per unit cell. The dominant excitation are to the second excited state and seventh excited states, which are the first and second excited states in the direction of shuttling (See Figure~\ref{fig:excited_states_2d}). With increasing cut-off frequencies and higher temperatures, we can see that the probability to remain in the ground state decreases while the probability of excitation to the second excited increases. 

\subsection{Speed Vs. Adiabaticity}\label{appendix:subsec:speed_adiabaticity}

\begin{figure}
	\centering
	\subfloat[]{%
	\includegraphics[width=\linewidth]{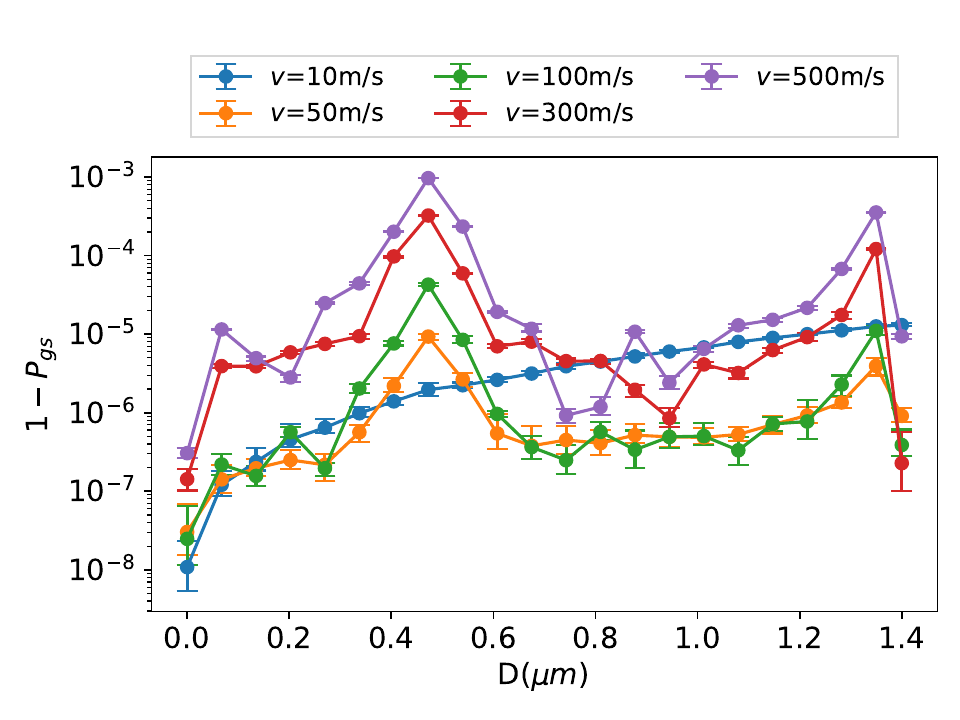}
  	\label{fig:infidelity_against_gs_along_distance_all_speeds}	
	}%
	\hfill
	\subfloat[]{%
	\includegraphics[width=\linewidth]{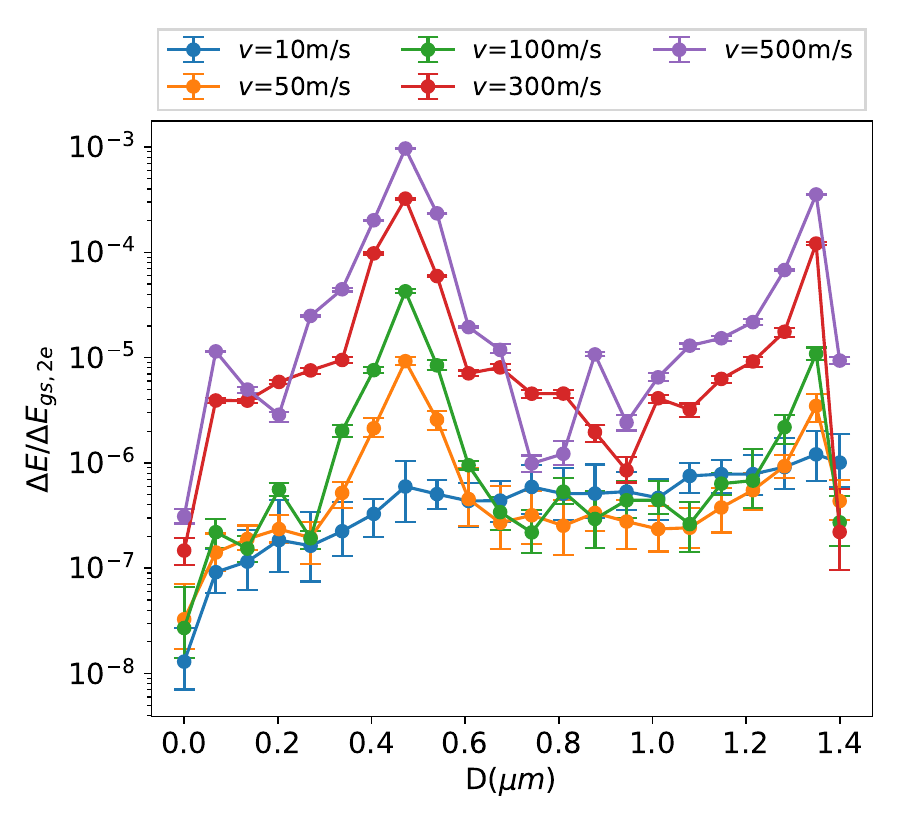}
  	\label{fig:excitation_fraction_along_distance_all_speeds}	
	}%
\caption{(a) Probability of excitation outside of the ground state and (b) excitation fraction during the shuttling for a target distance of $1.4$\,$\mu$m at 5 different speeds, i.e. $v=10, 50, 100, 300, 500$\,m/s. The data of $v=10$\,m/s (blue line) in these figures are the same as $1-P_{gs}$ data in Figure~\ref{fig:infidelity_exc_frac_along_distance}. Other parameters were set to $A=50$\,mV, $T=2$\,K, and $N=3$.}
\label{fig:infidelity_exc_frac_against_gs_along_distance_all_speeds}
\end{figure}

Figure~\ref{fig:infidelity_exc_frac_against_gs_along_distance_all_speeds} shows the probability of excitation outside of the ground state and the excitation fraction with 5 different shuttling speeds, i.e. $10, 50, 100, 300, 500$\,m/s. Other parameters were chosen realistically, $A=50$\,mV, $T=2$\,K, and $N=3$. Data for $10$\,m/s are identical to those of Figure~\ref{fig:infidelity_exc_frac_along_distance} in section \ref{sec: implication_spin}: The probability of excitation and excitation fraction at $10$\,m/s shows a regular behaviour of increase from 0 to $1.3 \times 10^{-5}$ for the excitation probability and around $10^{-6}$. In contrast, we observed irregular behaviours, which could be a sign of non-adabaticity, such that there are sharp peaks at $D = 0.47$\,$\mu$m and $D=1.35$\,$\mu$m for speeds greater than or equal to $50$\,m/s. The height of the peak increase as the shuttling speed increases, and the excitation fraction goes up to $10^{-3}$ for $300$\,m/s. This suggests that the impact of orbital excitation to the spin grows with the shuttling speed as more orbital excitation happens with higher excitation speed resulting in higher change of random phonon relaxation happening during the shuttling. The causes of the peaks at those particular positions could be studied in future works.

\subsection{Sensitivity to Charge Defects: Two Defects of Varying Separations.} \label{appendix:subsec:defects_varying_dist}

\begin{figure*}
\centering
    \subfloat[]{%
    \centering
    \includegraphics[width=0.495\textwidth]{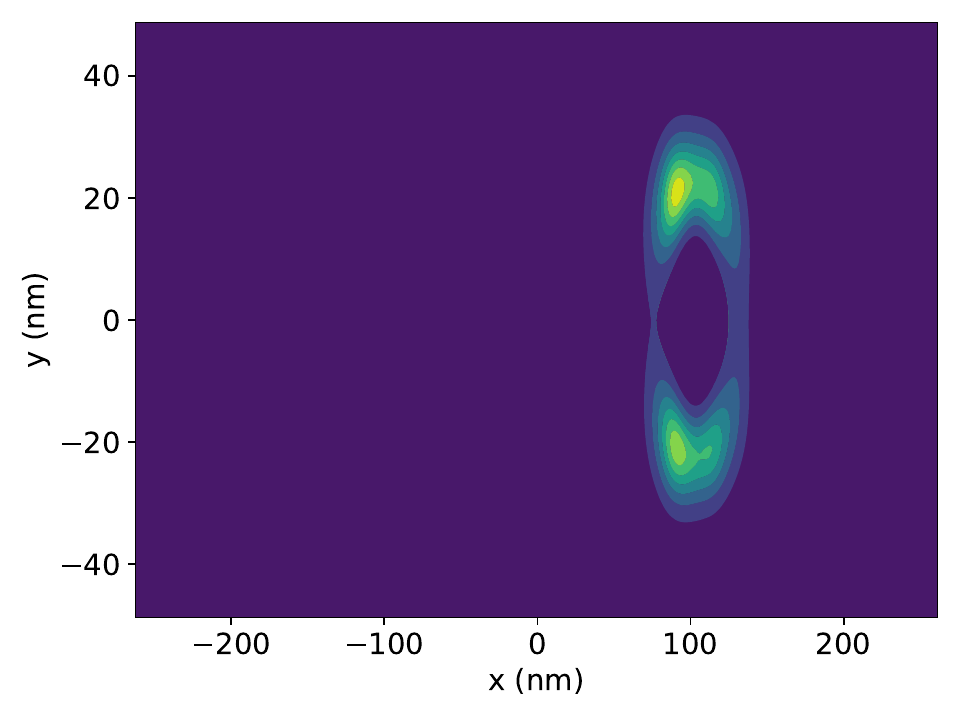}
      \label{fig:two_charge_defects_varying_distances_states_2nm}}%
    ~        
    \subfloat[]{%
    \centering
    \includegraphics[width=0.495\textwidth]{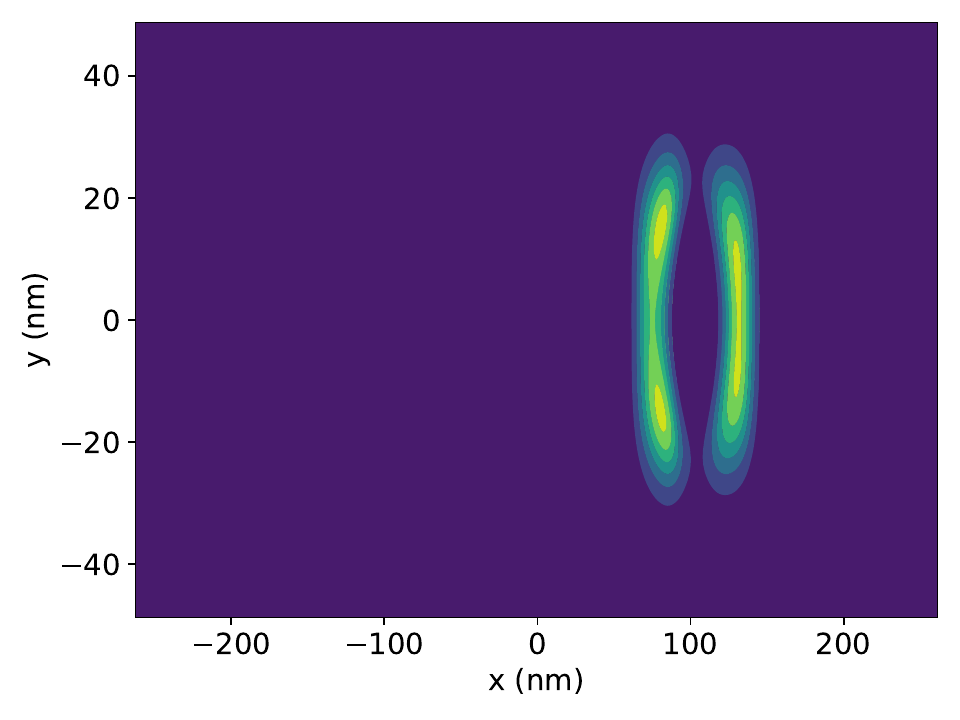}
    \label{fig:two_charge_defects_varying_distances_states_12nm}}%
    \hfill
    \subfloat[]{%
    \centering
    \includegraphics[width=0.495\textwidth]{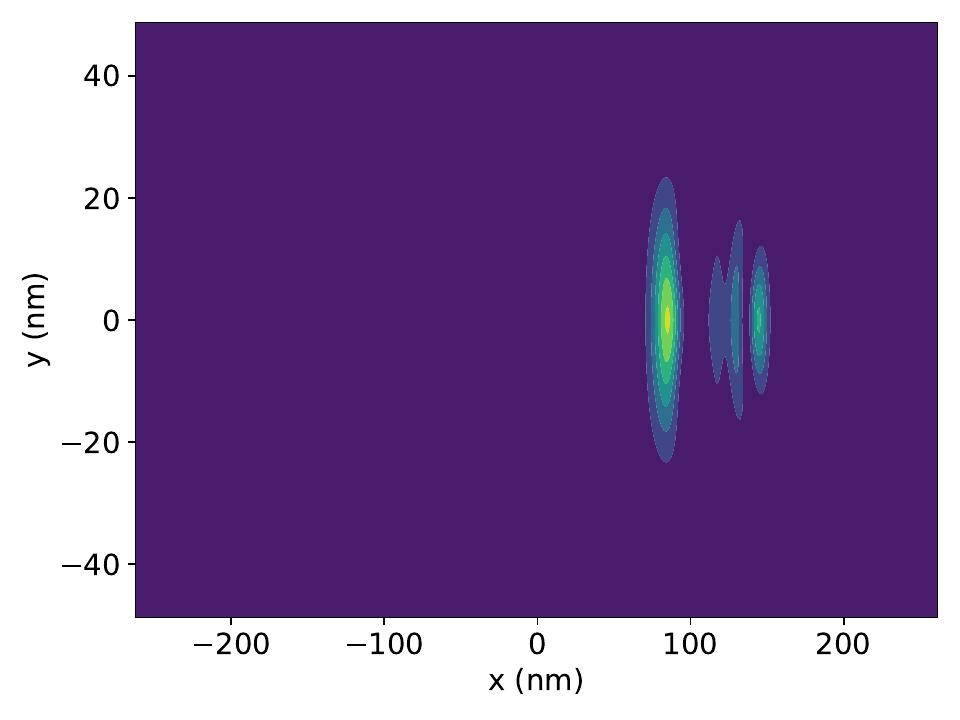}
    \label{fig:two_charge_defects_varying_distances_states_25nm}
    }%
    ~
    \subfloat[]{%
    \centering
    \includegraphics[width=0.495\textwidth]{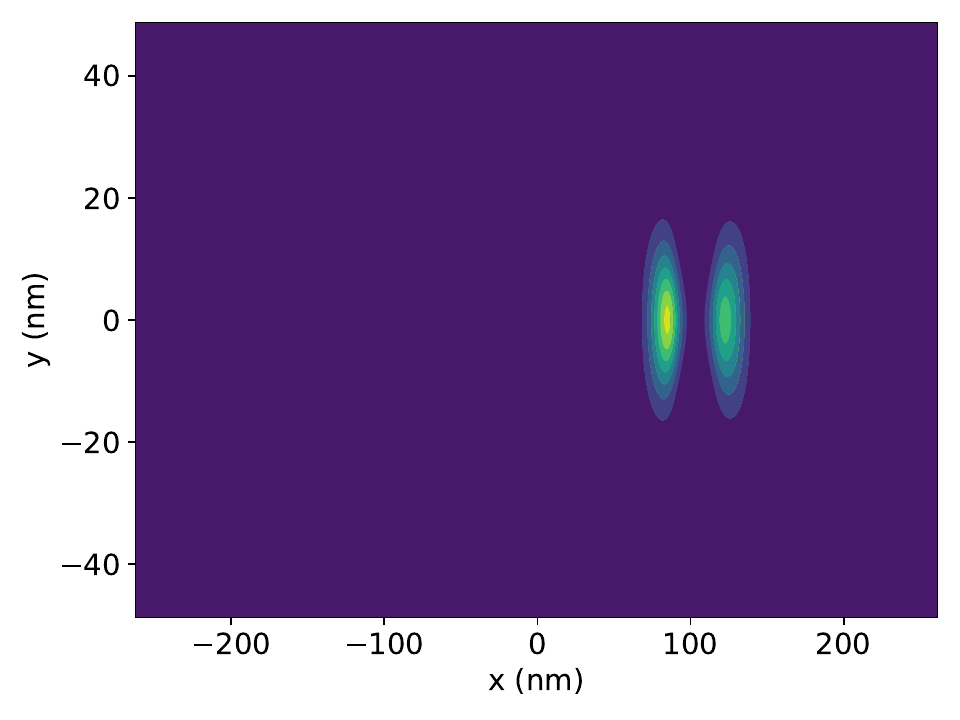}
          \label{fig:two_charge_defects_varying_distances_states_30nm}
    }%
	
    \caption{Contour plots of probability amplitudes of the electron moving through two charge defects for (a) $\Delta y=2$\,nm, (b) $\Delta y=12$\,nm, (c) $\Delta y=25$\,nm, (d) $\Delta y=30$\,nm. When the separation of two charge defects is small, ($\Delta y = 2, 12$\,nm), the electron moves around the defects as in (a) and (b). When the separation is large ($\Delta y = 30$\,nm) the electron moves through the middle without significant excitation as in (d). When neither of these actions is easy (e.g. $\Delta y = 25$\,nm), the tunneling through the barrier produces significant excitation in the state, as in (c).}
    \label{fig:two_charge_defects_varying_distances_states}
\end{figure*}

Figure~\ref{fig:two_charge_defects_varying_distances_states} shows the probability amplitude of the electron when it tunnels through the barrier created by two charge defects of varying separations $\Delta y = 2, 12, 25, \text{and } 30$\,nm. The images were taken at the time step when the expectation value of position in x-axis coincides with the x-axis coordinate of the defects. When two charge defects are close to each other, there is enough room for the electron to move around the central barrier at the sides of the channel. When two charge defects are far enough from each other, the central barrier is low enough for the electron to pass through the middle of the channel. However, when the distance between the defects makes both of these options hard, passage over the barrier produces significant excitation in the electron state.

\section{Results of Advanced Non-Adiabatic Ultra-fast shuttling} \label{appendix:sec:snap}

In this section, we present the results and analysis of simulation of the new non-adiabatic shuttling method, namely the snap method, proposed in section~\ref{sec:snap}. This scheme depends on how closely the trough of the potential energy can be approximated by an SHO potential in the range $x \in [x_{0} - \Delta x, x_{0} + \Delta x )$, where $x_{0}$ is the current position of the minimum of the potential energy: if the potential is perfectly harmonic, the displaced ground state forms a coherent state that moves in the potential without changing its spatial form, and step (3) exactly recovers the ground state of the final potential.

As the channel is placed deeper (i.e. more negative $z$), the shape of the potential becomes more harmonic, while the amplitude of the signal at the channel decreases. While the approach of the potential to the SHO potential limit benefits the shuttling, the smaller amplitude makes the loss probability bigger. Thus, these two factors compete with each other as the channel is placed deeper. 

\begin{figure}
    \centering
    \includegraphics[width=\textwidth]{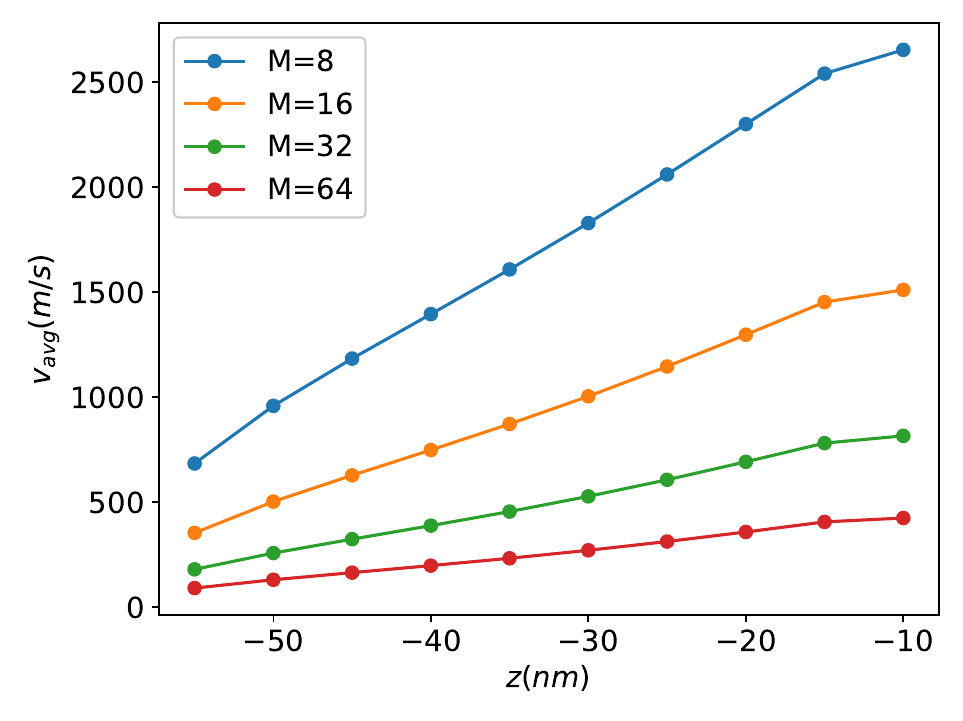}
    \caption{The average speed of adiabatic shuttling at varying depths with different number of instantaneous updates of the potential(M).}
    \label{fig:average_speed_snap}
\end{figure}

If the target distance is a multiple of the length of one unit cell, we can have a finite set of $\Delta t$ and $ \Delta x$ such that it can be repeatedly used after shuttling the electron by the length of one unit cell. Let $M$ be the number of instantaneous changes in potential while traversing one unit cell; as $M$ increases, the average speed of shuttling decreases with both increasing $M$ and increasing depth, as shown in Figure~\ref{fig:average_speed_snap}, because the interval between changes is set by the curvature of the potential minimum.

\begin{figure}
\subfloat[]{%
\includegraphics[width=\linewidth]{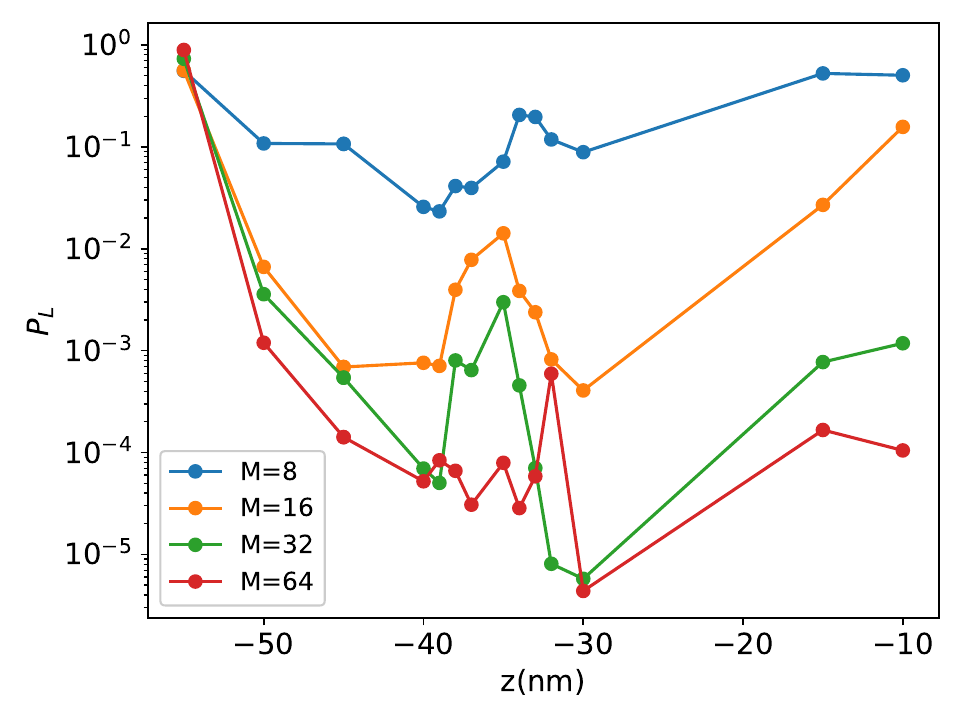}
  \label{fig:loss_snap_sample_z_multiple_M}
}%
\hfill
\subfloat[]{%
\includegraphics[width=\linewidth]{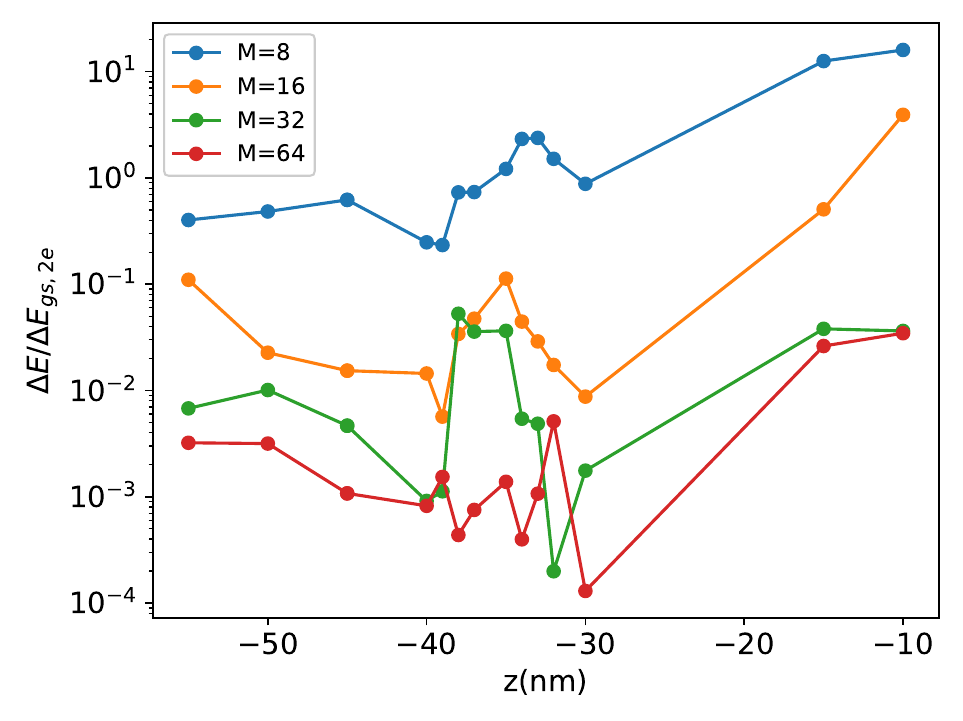}
  \label{fig:excitation_fraction_snap_sample_z_multiple_M}
}%
\caption{(a) Loss probability and (b) excitation fraction of the non-adiabatic shuttling at the varying depth(z) with different number of instantaneous updates of the potential(M).}
\label{fig:loss_excitation_fraction_snap_sample_z_multiple_M}
\end{figure}

\begin{figure*}
\centering
\begin{minipage}[t]{0.495\columnwidth}

\subfloat[]{%
\includegraphics[width=\linewidth]{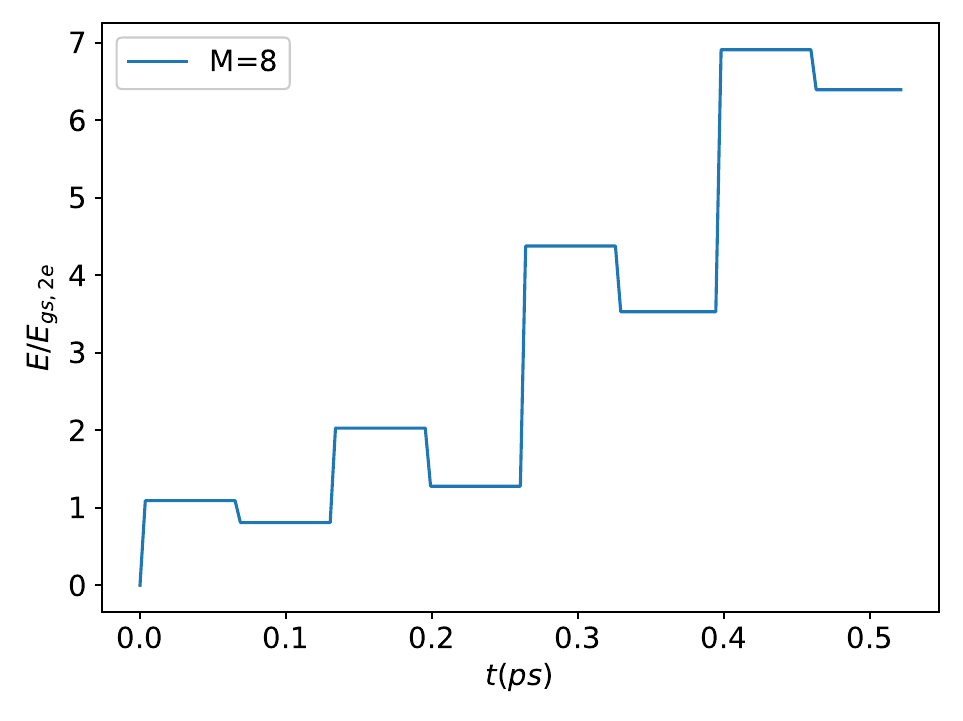}
  \label{fig:energy_fraction_snap_M_8}
}%
\hfill
\subfloat[]{%
	\includegraphics[width=\linewidth]{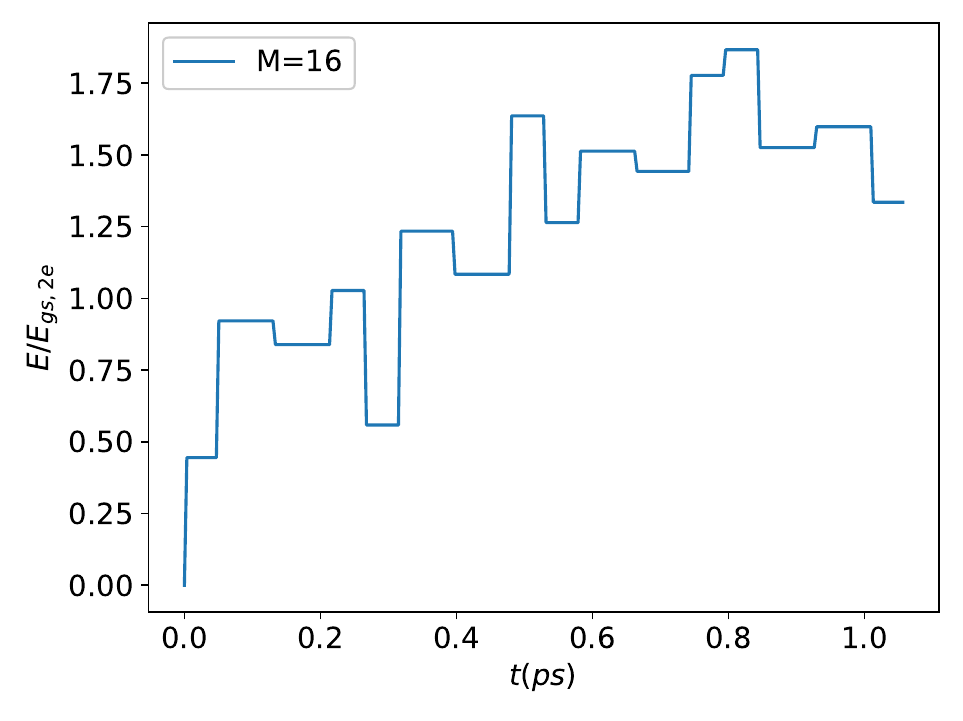}
  \label{fig:energy_fraction_snap_M_16}
}%
\end{minipage}
\hfill
\begin{minipage}[t]{0.495\columnwidth}
\subfloat[]{%
\includegraphics[width=\linewidth]{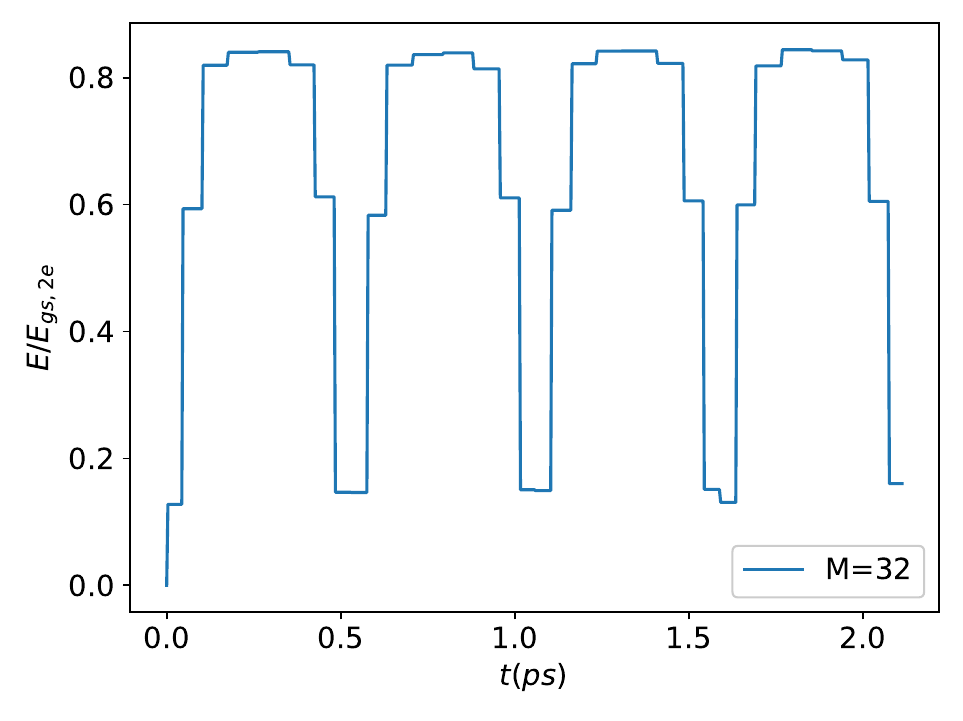}
  \label{fig:energy_fraction_snap_M_32}
}%
\hfill
\subfloat[]{%
\includegraphics[width=\linewidth]{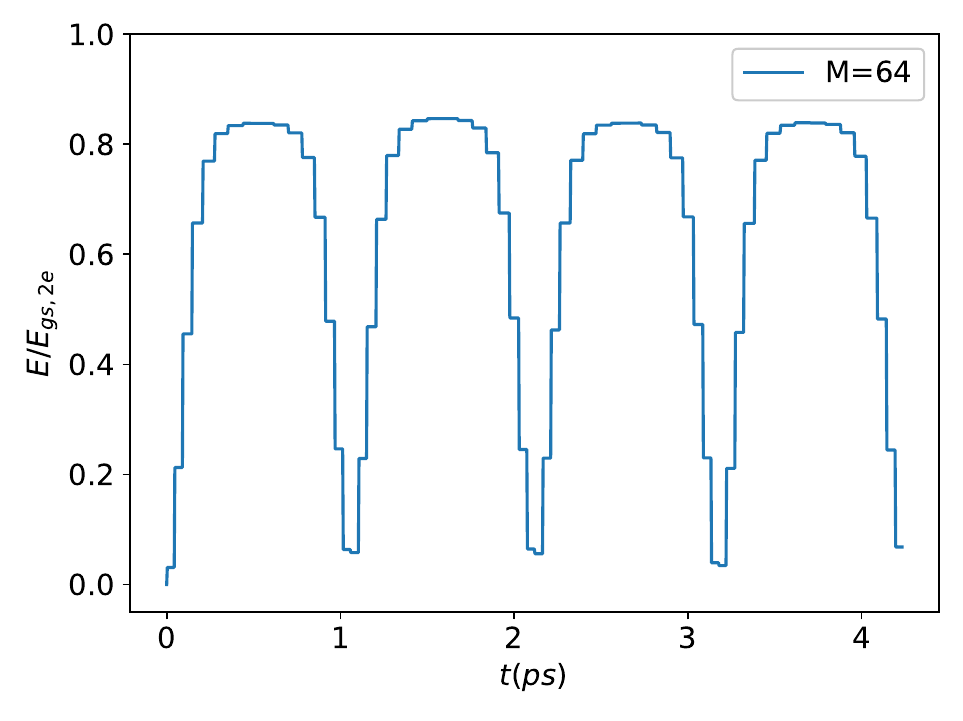}
  \label{fig:energy_fraction_snap_M_64}
}%
\end{minipage}
\caption{Fractional excitation energy in the fast non-adiabatic shuttling method for different values of $M$: (a) $M=8$, (b) $M=16$, (c) $M=32$, (d) $M=64$. The method becomes more stable and have a periodic behaviour as the number of instantaneous changes, $M$, per unit cell increases.}
\label{fig:energy_fraction_snap_M}
\end{figure*}

\begin{figure}
\centering
	\subfloat[]{%
	\includegraphics[width=\linewidth]{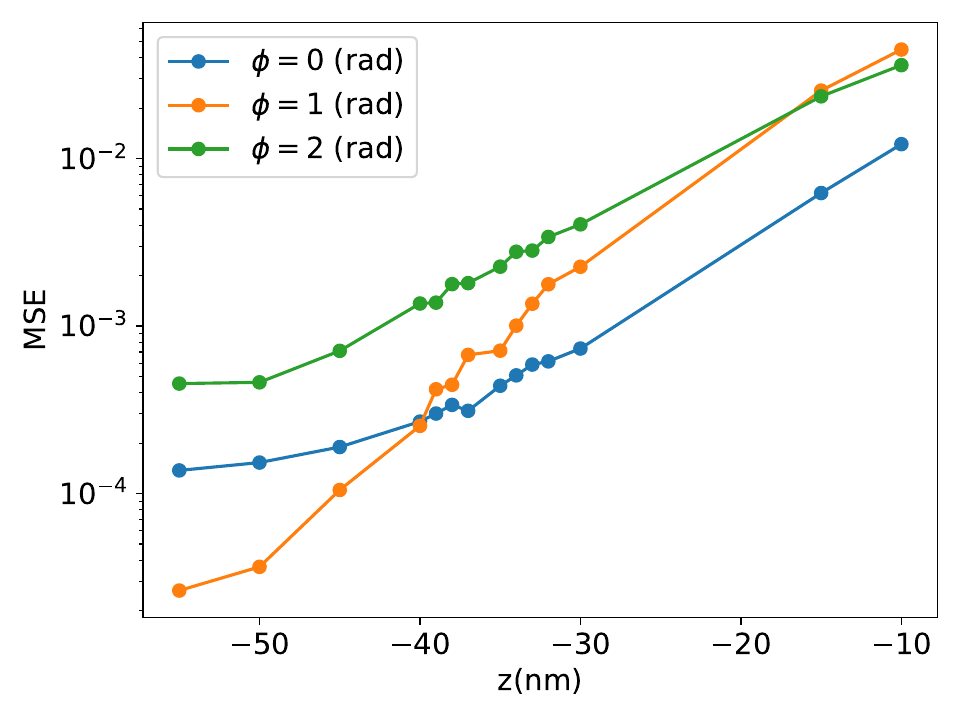}
        \label{fig:mse_potential_sinusoidal}
	}%
    \hfill
    \subfloat[]{%
    \includegraphics[width=\linewidth]{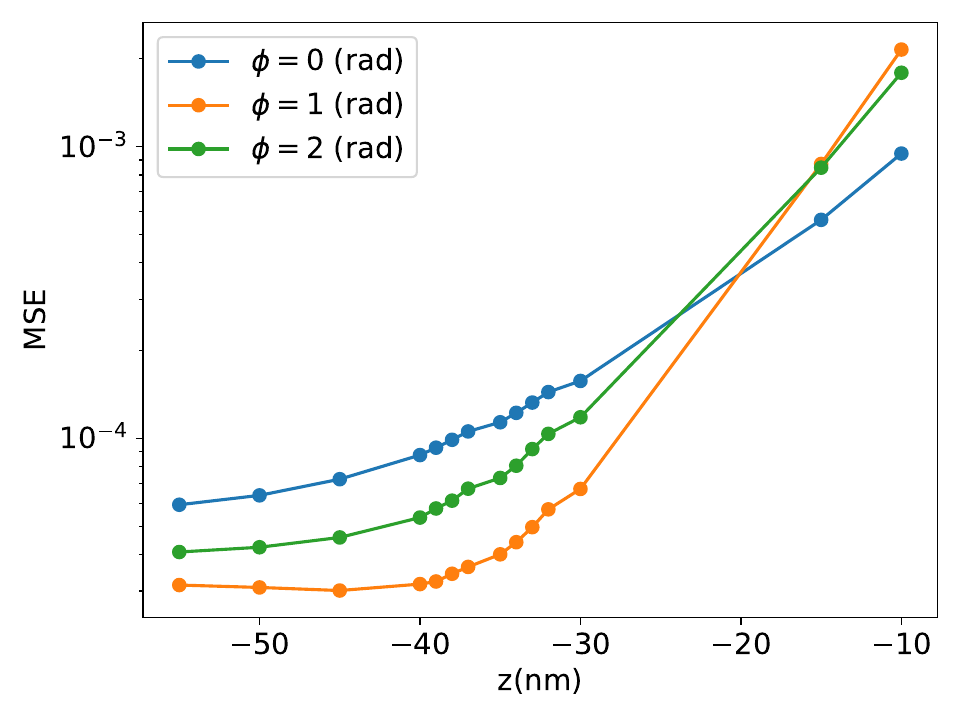}
        \label{fig:mse_potential_quadratic}
    }%
    \caption{The mean squared error(MSE) between the normalized potential, i.e. max$\abs{(V_{\phi}(x)}) = 1$ and two different functions: (a) $\cos(x)$ and (b) the quadratic fit at the bottom at three different phases, $\phi= 0, 1, 2$\,rad. This shows that the potential becomes more like a sinusoidal function and that quadratic fit becomes better when the potential is sampled at deeper part of the channel.}
    \label{fig:mse_potential}
\end{figure}

Figure~\ref{fig:loss_snap_sample_z_multiple_M} shows the competition of the two factors: closeness to the SHO potential (See Figure~\ref{fig:mse_potential}) and the amplitude of the voltage signal. The amplitude of the gate voltage was fixed at $100$\,mV. As $z$ becomes more negative, the loss probability and excitation fraction initially decrease, then increase beyond a certain point, e.g. at $z=-30$\,nm for $M=64$ (red line).  In the range $z=-30$\,nm to $z=-40$\,nm, there is a local maximum in both loss and excitation for all $M$; for $M=64$, this maximum takes the form of a sharp peak at $z=32$\,nm, followed by oscillations in the loss probability. We can see similar local maxima and oscillation of excitation fraction in Figure~\ref{fig:excitation_fraction_snap_sample_z_multiple_M}.

While this could be understood as another manifestation of the trade-off between the two factors mentioned earlier, it means that small errors in the timings of instantaneous changes to the potential (the `snaps') can lead to large changes in the quality of the shuttling. Errors in the update timings can lead to a de-synchronising between the electron's motion and the updates to the potential, and consequent excitation out of the desired mode.

As one might expect, it is possible to systematically optimise beyond the initial timings obtained from the idealised analytic model. We explored this using Limited-memory BFGS (L-BFGS)\cite{Liu_1989}, and setting the final energy of the shuttling as a target function. Optimisation led to improvement in both the loss probability and excitation fraction. In the chosen scenario, the number of electrodes used was $N=4$, the number of instantaneous changes within one unit cell was set to $M=8$, the amplitude of voltage at the gate was set to $A=100$\,mV, while the position of the channel was $z=-30$\,nm. With the default convergence criteria of the Scipy implementation\cite{2020SciPy-NMeth} adopted, we observed a reduction of the final loss probability by about $25$\,\% (from $0.089$ to $0.066$) and a reduction of $40$\,\% in the excitation fraction (from $0.88$ to $0.53$). No doubt further improvements could be made via other methods or other cost functions, e.g. excitation fraction or final loss probability.

Aside from optimising the `snap' event time intervals, a basic challenge for this non-adiabatic shuttling scheme is that the voltage changes are faster than the limits with current technology (around $14$\,mV/ps). In our model, the minimum rate of voltage change for $M=64$ at $z=-10$\,nm wass $5$\,V/ps, which is around $400$ times higher the $14$\,mV/ps. Further investigation can be made with slower voltage change: However, we expect that it would deteriorate the overall performance as the timings of instantaneous changes should be exact to seamlessly transport the electron. Furthermore, the presence of charge defects will make this method worse as the timings of instantaneous changes are affected by them. In particular, when the electron is near the charge defect and repelled by the Coulomb repulsion, the electron will linger longer to tunnel through the potential barrier.

Given these limitations, this `snap' method seems unrealistic for implementation with foreseeable technologies. However, in an era where silicon-based quantum devices are mature it might perhaps be an exploitable concept.

\vfill

\end{document}